\documentclass[a4, 11pt]{article}
\pdfoutput=1

\usepackage{amsmath, amssymb, bm}
\usepackage[top=1in, bottom=1.2in, left=0.5in, right=0.5in]{geometry}
\usepackage[]{graphicx}
\usepackage[font=scriptsize,labelfont=bf]{caption}
\usepackage[caption=false]{subfig}
\usepackage{textcomp}
\usepackage{authblk}
\usepackage{hyperref}

\newcommand{\bs}{\boldsymbol}

\begin{document}

\title{\Large{\bf{On the CMB circular polarization. I. \\ The Cotton-Mouton effect}}}
\author{Damian Ejlli}

\affil{\emph{\normalsize{Department of Physics, Novosibirsk State University, Novosibirsk 630090 Russia}}}

\date{}

\maketitle

\begin{abstract}
Generation of cosmic microwave background (CMB) elliptic polarization due to the Cotton-Mouton (CM) effect in a cosmic magnetic field is studied. We concentrate on the generation of CMB circular polarization and on the rotation angle of the CMB polarization plane from the decoupling time until at present. For the first time, a rather detailed analysis of the CM effect for an arbitrary direction of the cosmic magnetic field with respect to photon direction of propagation is done. Considering the CMB linearly polarized at the decoupling time, it is shown that the CM effect is one of the most substantial effects in generating circular polarization especially in the low part of the CMB spectrum. It is shown that in the frequency range $10^8$ Hz $\leq \nu_0\leq 10^9$ Hz, the degree of circular polarization of the CMB at present for perpendicular propagation with respect to the cosmic magnetic field is in the range $ 10^{-13}\lesssim P_C(t_0)\lesssim 7.65\times 10^{-7}$ or Stokes circular polarization parameter $2.7 \times 10^{-13}$ K $\lesssim |V(t_0)|\lesssim 2 \times 10^{-6}$ K for values of the cosmic magnetic field amplitude at present in the range $10^{-9}$ G $\lesssim B\lesssim 8\times 10^{-8}$ G.  On the other hand, for not perpendicular propagation with respect to the cosmic magnetic field we find $10^{-15}\lesssim P_C(t_0)\lesssim 6\times 10^{-12}$ or $2.72 \times 10^{-15}$ K $\lesssim |V(t_0)| \lesssim 10^{-11}$ K, for the same values of the cosmic magnetic field amplitude and same frequency range. Estimates on the rotation angle of the CMB polarization plane $\delta\psi_0$ due to the CM effect and constraints on the cosmic magnetic field amplitude from current constraints on $\delta\psi_0$ due to a combination of the CM and Faraday effects are found. 

\end{abstract}

\vspace{+1cm}

\section{Introduction}
\label{sec:1}

In the last two decades, there have been many established observational facts about the nature and properties of the CMB and their possible implications in cosmology. Among these, it has already been established the fact that the CMB has a linear polarization with a degree of polarization at present of the order $P_L(t_0)\simeq 10^{-6}$. This linear polarization is believed to have been generated at the decoupling time mostly due to the Thomson scattering of the CMB photons on electrons. In general, if the incident electromagnetic radiation has an isotropic intensity distribution, Thomson scattering does not generate a net linear polarization. In the specific case of the CMB the fact that linear polarization has been initially observed by DASI, WMAP and BOOMERANG collaborations \cite{Kovac:2002fg} and then re-confirmed by other collaborations, implies that at the decoupling time the CMB intensity did not have an isotropic distribution, a fact which is widely confirmed from the observation of the CMB temperature anisotropy. Another important consequence of the Thomson scattering is that it does not generates circular polarization in the case when electrons are assumed to be unpolarized. Based on this fact, during these years it has been erroneously assumed, at least from the theoretical point of view, that the CMB does not have a circular polarization at all even though there have been initial studies that might support its existence \cite{negroponte:80} and also initial experimental efforts to detect it \cite{smooth:83}.

In the recent years there have been several other theoretical studies exploring the possibility of CMB circular polarization from standard and non-standard effects and also new experiments such as MIPOL \cite{Mainini:2013mja} and SPIDER \cite{Nagy:2017csq} aiming to detect it. The MIPOL \cite{Mainini:2013mja} collaborations reported an upper limit on the degree of circular polarization at present of $P_C(t_0)\lesssim 7\times 10^{-5}-5\times 10^{-4}$ at the frequency 33 GHz and at angular scales between $8^\circ$ and $24^\circ$. On the other hand, the SPIDER collaboration reported an upper limit on the CMB circular polarization power spectrum $\ell (\ell+1) C_\ell^{VV}/(2\pi)< 255 {(\mu\text{K})}^2$ for multipole momenta $33<\ell<307$ at the CMB frequencies $\nu_0=95$ GHz and $\nu_0=150$ GHz. From the theoretical point of view, studies based on non-standard effects that generate circular polarization include; the interaction of the CMB with a vector field via a Chern-Simons term \cite{Alexander:2008fp}, non commutative geometry \cite{Bavarsad:2009hm} and free photon-photon scattering due to the Euler-Heisenberg Lagrangian term \cite{Sawyer:2012gn}.  On the other hand, some theoretical studies of standard effects include; the electron-positron scattering in magnetized plasma at the decoupling time \cite{Giovannini:2010ar}, the propagation of the CMB photons in magnetic field of supernova remnants of the first stars \cite{De:2014qza}, the scattering of the CMB photons with cosmic neutrino background \cite{Mohammadi:2013dea} and also the alignment of the cosmological matter particles in the post-decoupling epoch which results in an anisotropic susceptibility matter tensor \cite{Montero-Camacho:2018vgs}. For a recent and not complete review of the CMB circular polarization see Ref. \cite{King:2016exc}.

Apart from the circular polarization generation effects mentioned above, there is a class of effects called magneto-optic effects which generate CMB circular polarization as well. In Ref.  \cite{Ejlli:2016avx} and Ref. \cite{Ejlli:2017uli}, I studied the most important magneto-optic effects which can generate CMB circular polarization when the CMB interacts with large-scale cosmic magnetic fields. Among the effects which I studied one of them is a standard effect, namely the CM effect, and the other effects are non-standard and include the vacuum polarization in an external magnetic field due to one loop electron-positron, one loop millicharged fermion-antifermion and the photon-pseudoscalar mixing in a magnetic field. For all these effects to occur it is necessary the presence of a magnetic field which gives rise to birefringence effects due to the fact that each of the photon states acquires different indexes of refraction in the presence of the magnetized plasma. 

While it is well known that it does exist a magnetic field in galaxies and galaxy clusters with an order of magnitude of few $\mu$G, it is still not known if such a field is present also in the intergalactic space. The only information that we have about intergalactic magnetic fields are only in forms of upper and lower limits on the field magnitude at the present epoch. The upper limits on the magnetic field amplitude are found from observations of the CMB temperature anisotropy and from the rotation angle of the CMB polarization plane due to the Faraday effect. The temperature anisotropy upper limit is usually stronger than the Faraday effect limit, as reported by the Planck collaboration \cite{Ade:2015cva},  where the limit from CMB temperature anisotropy is $B_{e0}\lesssim 3$ nG at a scale $\lambda_B=1$ Mpc, while the limit from the Faraday effect is $B_{e0}\lesssim 1380$ nG at $\lambda_B=1$ Mpc. One important aspect of these limits is that they differ from each other roughly speaking by three orders of magnitude and most importantly the stronger limit on the magnetic field amplitude from the CMB temperature anisotropy does not exclude the weaker limit from the Faraday effect because these upper limits depend on how the magnetic field is modelled and other assumptions, see Ref. \cite{Ade:2015cva} for details. For simplicity, in this work we assume that the cosmic magnetic field amplitude changes in time $t$ and it is an almost constant function of the position $\bs x$. For a general review on large-scale cosmic magnetic field see Ref. \cite{Grasso:2000wj}.

One key aspect which distinguishes the CMB linear polarization with the CMB circular polarization, is that the former being generated at the decoupling time due to the Thomson scattering does not depend on the CMB frequency because of the nature of Thomson scattering which is frequency independent at lower energies, while the latter in most cases strongly depends on the CMB frequency. Because of this frequency dependence of the circular polarization, there is in some sense a kind of uncertainty on how to use and interpret the current limits obtained by experiments such as MIPOL and SPIDER since their limits are usually derived by observing the CMB in a specific frequency and it is not known how much substantial could be the signal at other frequencies.

In order to study and detect the CMB circular polarization, it is very important to first identify the circular polarization (possibly standard) effects that generate substantial CMB circular polarization and identify their frequency band where the signal is the strongest. So far, there has been a tendency in the literature to study the circular polarization in the high-frequency range, namely for frequencies above ten or few hundred GHz. This tendency has been partially influenced by the fact that most important CMB experiments such as WMAP and Planck operates at these frequencies where the CMB intensity is the highest and therefore their data at these frequencies might be useful in some way. In addition, there are some effects such as the photon-photon scattering in a cosmic magnetic field  \cite{Ejlli:2016avx} and the free photon-photon scattering \cite{Sawyer:2012gn}, \cite{Montero-Camacho:2018vgs} which are linearly proportional to the CMB frequency and one might hope that the higher is the frequency, the stronger is the circular polarization signal. Even though this is true, the signal for such effects is still very weak even at very high frequencies to be detected in the near future. 

Based on the facts discussed above, it is rather logical to explore the CMB circular polarization at low frequencies and study the magnitude of the signal. In this work, I study such possibility and concentrate on the CM effect in a large-scale cosmic magnetic field. As we will see, the CM effect is proportional to the square of the magnetic field amplitude, $B^2$, and inversely proportional to the third power of the CMB frequency, namely $\nu^{-3}$ in the case of perpendicular propagation with respect to the cosmic magnetic field. It is especially the scaling law with the frequency of $\nu^{-3}$ which makes the CM effect the most important effect in generating CMB circular polarization at low frequencies. I partially studied this effect in a previous work \cite{Ejlli:2016avx} where some estimates of the degree of circular polarization were made for a specific configuration of the cosmic magnetic field with the respect to the photon direction of propagation. In this work, I study the CM effect in details for an arbitrary configuration of the cosmic magnetic field direction. By generalizing the CM effect to an arbitrary direction of the cosmic magnetic field with respect to the observer's direction, the system of differential equations for the Stokes parameters has additional terms with respect to the case studied in Ref. \cite{Ejlli:2016avx}. In addition, I also study in details the impact that the CM effect has on the rotation angle of the CMB polarization plane and its interaction with the Faraday effect.

This paper is organized in the following way: in Sec. \ref{sec:2}, I discuss in a concise way the propagation of the electromagnetic radiation in a magnetized plasma and derive the elements of the photon polarization tensor in the cold magnetized plasma approximation. In Sec. \ref{sec:3}, I derive the system of differential equations for the Stokes parameters in an expanding universe. In Sec. \ref{sec:4}, I find perturbative solutions of the equations of motion in various regimes. In Sec. \ref{sec:5}, I calculate in details the generation of the CMB circular polarization due to the CM effect at present. In Sec. \ref{sec:6}, I study the rotation angle of the CMB polarization plane due to the CM effect alone and also due to a combination of the CM and Faraday effects. In Sec. \ref{sec:7}, I conclude. In this work I use the metric with signature $\eta_{\mu\nu}=\text{diag}[1, -1, -1, -1]$ and work with the rationalized Lorentz-Heaviside natural units ($k_B=\hbar=c=\varepsilon_0=\mu_0=1$) with $e^2=4\pi \alpha$. In addition in this work I use the values of the cosmological parameters found by the Planck collaboration \cite{Aghanim:2018eyx} with $\Omega_\Lambda \simeq 0.68, \Omega_\text{M}\simeq 0.31, h_0\simeq 0.67$ with zero spatial curvature where $\Omega_\kappa=0$.

\section{Propagation of the electromagnetic waves in a magnetized plasma}
\label{sec:2}

In this section we give a compact description of propagation of the electromagnetic waves in the cold magnetized plasma approximation. This description is useful because it would allow us to understand how electromagnetic waves propagate in a cold magnetized plasma and which are the most common effects which give rise to birefringence effects in the medium. In this section we use the same notation as in Ref. \cite{Ejlli:2016asd} where basics of propagation of the electromagnetic waves in a cold magnetized plasma are presented in the appendix.

When electromagnetic waves (photons) propagate in a medium several effects manifest which include dispersion, absorption and scattering of the electromagnetic radiation. In connection with the dispersion phenomena, the effects of the medium on the incident electromagnetic wave are usually described in terms of the photon polarization tensor $\Pi_{ij}$ ($i, j=x, y, z$) with components in a given cartesian coordinate system where the medium is at rest. Consequently, in a medium, the free Maxwell equations in momentum space, in absence of external currents, get modified to 
\begin{equation}\label{mod-max}
[\omega^2(\delta_{ij} - k_{ij})-\Pi_{ij}]E_j=0,
\end{equation}
for a plane electromagnetic wave travelling into the medium with electric field components $E_j$. Here $\omega$ is the incident photon angular frequency or energy and we used the expression $k_{ij}=\omega^2 n^2(\delta_{ij}-\hat k_i\hat k_j)$ with $k_{ij}$ being the photon momentum tensor, $n=|\bs k|/\omega$ is the index of refraction and $\hat k_i$ are the components of a unit vector along the electromagnetic direction of propagation wave-vector $\bs k$. We may see that the role of $\Pi_{ij}$ in \eqref{mod-max} is to give to photons an ''effective mass'' in the medium. In the case when the medium is isotropic, we have that $k_{ij}$ is a diagonal tensor with diagonal entries corresponding to the photon indexes of refraction in medium where $k_{ii}\neq 1$. In the case when photons propagate in vacuum, we have that $k_{ij}=\omega^2 \delta_{ij}$ and we get the on-shell photon relation $\omega=\bs k^2$ where $\Pi_{ij}=0$.

The explicit expression of the photon polarization tensor $\Pi_{ij}$ depends on the induced currents that enter a given problem. In this work we are interested in a cold magnetized plasma which is quite common situation in astrophysics and cosmology. We assume that the magnetized plasma is with almost no collisions, globally neutral and homogeneous. In addition, there is not an external electric field, namely $\bs E_e=0$ and the presence of the external magnetic field $\bs B_e$ locally breaks the isotropy of the plasma since it singles out a preferred direction in a given region of space where the plasma is located. 

In the cold magnetized plasma approximation, consider now an incident electromagnetic wave propagating along the observer's $z$ axis which points to the East, in a magnetized plasma with external magnetic field vector $\bs B_e = B_e \hat{\bs n}$. Here $\hat{\bs n}=[\cos(\Theta), \sin(\Theta)\cos(\Phi), \sin(\Theta)\sin(\Phi)]$ is a unit vector in the direction of the external magnetic field $\bs B_e$ and $\Theta, \Phi$ are, respectively, the polar and azimutal angles between the magnetic field $\bs B_e$ and $x$ and $y$ axes. As shown in Ref. \cite{Ejlli:2016asd}, the medium polarization vector $\bs P$ satisfies the equation of motion
\begin{equation}\label{pol-eq}
\ddot{\bs P}= \omega_\text{pl}^2 \bs E-\omega_c\,\dot{\bs P}\times \hat{\bs n},
\end{equation}
where $\bs E$ is the electric field of the incident electromagnetic wave, $\omega_\text{pl}^2=4 \pi \alpha n_e/m_e$ is the plasma frequency, $n_e$ is the free electron number density, $m_e$ is the electron mass and $\omega_c=e B_e/m_e$ is the cyclotron frequency. In Eq. \eqref{pol-eq} the dot symbol ($\cdot$) above $\bs P$ denotes the derivative with respect to the time $t$.

Assume that the fields evolve in time harmonically at a given point $\bs x$ 
\begin{equation}\label{field-exp-1}
\bs P(\bs x, t)=\bs P(\bs x, \omega) e^{-i\omega t}, \qquad \bs E(\bs x, t)=\bs E(\bs x, \omega) e^{-i\omega t}, 
\end{equation}
By using the expressions in \eqref{field-exp-1} in Eq. \eqref{pol-eq} and then solving for the components of $\bs P$, after we get the following solution in terms of the incident electric field components $E_j$, in the case when $\omega\neq 0$ and $\omega\neq \pm\omega_c$
\begin{equation}
P_i(\bs x, \omega)=\chi_{ij}(\omega) E_j(\bs x, \omega), \qquad (i, j=x, y, z),
\end{equation}
where $\chi_{ij}(\omega)$ are the components of the electric susceptibility tensor 
\begin{equation}\label{susc-tens}
\begin{gathered}
\chi_{xx}=-\frac{\omega_\text{pl}^2}{\omega^2-\omega_c^2}+\frac{\omega_\text{pl}^2\omega_c^2 \cos^2(\Theta)}{\omega^2(\omega^2-\omega_c^2)}, \quad \chi_{xy}=\frac{\omega_\text{pl}^2\,\omega_c^2 \sin(2 \Theta)\cos(\Phi)}{2\, \omega^2(\omega^2-\omega_c^2)}+i \frac{\omega_\text{pl}^2\omega_c \sin(\Theta)\sin(\Phi)}{\omega(\omega^2-\omega_c^2)}, \\
\chi_{xz}=\frac{\omega_\text{pl}^2\,\omega_c^2 \sin(2 \Theta)\sin(\Phi)}{2\,\omega^2(\omega^2-\omega_c^2)}-i \frac{\omega_\text{pl}^2\omega_c \sin(\Theta)\cos(\Phi)}{\omega(\omega^2-\omega_c^2)}, \quad \chi_{yx} = \chi_{xy}^*,\\
\chi_{yy}=-\frac{\omega_\text{pl}^2}{\omega^2-\omega_c^2}+\frac{\omega_\text{pl}^2\omega_c^2 \sin^2(\Theta)\cos^2(\Phi)}{\omega^2(\omega^2-\omega_c^2)}, \quad \chi_{yz} = \frac{\omega_\text{pl}^2\,\omega_c^2 \sin(2\Phi)\sin^2(\Theta)}{2\,\omega^2(\omega^2-\omega_c^2)} + i \frac{\omega_\text{pl}^2\omega_c \cos(\Theta)}{\omega(\omega^2-\omega_c^2)},\\
\chi_{zx}=\chi_{xz}^*, \quad \chi_{zy}=\chi_{yz}^*, \quad \chi_{zz}=-\frac{\omega_\text{pl}^2}{\omega^2-\omega_c^2}+\frac{\omega_\text{pl}^2\omega_c^2 \sin^2(\Theta)\sin^2(\Phi)}{\omega^2(\omega^2-\omega_c^2)}.
\end{gathered}
\end{equation}

The expressions for the components of $\chi_{ij}$ in \eqref{susc-tens} are valid for an incident electromagnetic wave with an arbitrary direction of propagation with respect to $\bs B_e$. In addition, the components $\chi_{ij}$ do not explicitly depend on $\bs x$ but only implicitly through $B_e(\bs x, t)$ which enters in $\omega_c$. After these general comments about \eqref{susc-tens}, let us find the components of the photon polarization tensor in a cold magnetized plasma. In order to do that we have to relate the components of $\chi_{ij}$ with $\Pi_{ij}$. It is well known that the components of $\Pi_{ij}$ are related to the relative permittivity tensor\footnote{Here we are assuming that in most cases of magnetized plasmas we have that the magnetic permeability tensor $\mu_{ij}\simeq 1$.} $\varepsilon_{ij}$ through the relation $\Pi_{ij}=\omega^2(\delta_{ij} - \varepsilon_{ij})$. On the other hand, the relative permittivity tensor $\varepsilon_{ij}$ is related to the electric susceptibility tensor $\chi_{ij}$, through the relation $\varepsilon_{ij}=\chi_{ij} + \delta_{ij}$. By using these relations, we get 
\begin{equation}\label{pi-chi-rel}
\Pi_{ij}=-\omega^2\,\chi_{ij}.
\end{equation}
By using the expressions for $\chi_{ij}$ in \eqref{susc-tens} into \eqref{pi-chi-rel}, we get 
\begin{equation}\label{pola-tens}
\begin{gathered}
\Pi_{xx}=\frac{\omega^2\,\omega_\text{pl}^2}{\omega^2-\omega_c^2} - \frac{\omega_\text{pl}^2\omega_c^2 \cos^2(\Theta)}{\omega^2-\omega_c^2}, \quad \Pi_{xy}=-\frac{\omega_\text{pl}^2\,\omega_c^2 \sin(2 \Theta)\cos(\Phi)}{2\,(\omega^2-\omega_c^2)} - i \frac{\omega_\text{pl}^2\omega_c\,\omega \sin(\Theta)\sin(\Phi)}{\omega^2-\omega_c^2}, \\
\Pi_{xz}= - \frac{\omega_\text{pl}^2\,\omega_c^2 \sin(2 \Theta)\sin(\Phi)}{2\,(\omega^2-\omega_c^2)} + i \frac{\omega\,\omega_\text{pl}^2\omega_c \sin(\Theta)\cos(\Phi)}{\omega^2-\omega_c^2}, \quad \Pi_{yx} = \Pi_{xy}^*,\\
\Pi_{yy}= \frac{\omega^2\,\omega_\text{pl}^2}{\omega^2-\omega_c^2} - \frac{\omega_\text{pl}^2\omega_c^2 \sin^2(\Theta)\cos^2(\Phi)}{\omega^2-\omega_c^2}, \quad \Pi_{yz} = -\frac{\omega_\text{pl}^2\,\omega_c^2 \sin(2\Phi)\sin^2(\Theta)}{2\,(\omega^2-\omega_c^2)} - i \frac{\omega\,\omega_\text{pl}^2\omega_c \cos(\Theta)}{\omega^2-\omega_c^2},\\
\Pi_{zx}=\Pi_{xz}^*, \quad \Pi_{zy}=\Pi_{yz}^*, \quad \Pi_{zz}= \frac{\omega^2\,\omega_\text{pl}^2}{\omega^2-\omega_c^2} - \frac{\omega_\text{pl}^2\omega_c^2 \sin^2(\Theta)\sin^2(\Phi)}{\omega^2-\omega_c^2}.
\end{gathered}
\end{equation}

The expression for $\Pi_{ij}$ in \eqref{pola-tens} quantify the dispersive effects induced by the medium on the incident electromagnetic wave. One thing which is very well known, is that in the presence of a medium, appears also a longitudinal component for the incident electromagnetic wave. So, the difference with respect to vacuum propagation, is that in the presence of a plasma one has also to deal with the induced longitudinal electric field. Indeed, as it is evident from the expression of $\Pi_{ij}$, all components $\Pi_{zi}\neq 0$, which mean that also the longitudinal component of the electromagnetic wave manifest dispersive phenomena. However, one important fact about the longitudinal photon state is that it does not transport energy and it does not propagate in space. Moreover, the longitudinal component of the electromagnetic wave has no magnetic field associated to it but only an electric field. It essentially correspond to a density wave of electrons in the plasma in the presence of $\bs B_e$ that does not propagate in space.

Consider now a harmonic electromagnetic wave with electric field vector with components $\bs E_i(\bs x, t)=[E_x(\bs x, \omega),\\ E_y(\bs x, \omega), E_z(\bs x, \omega)]^{\text{T}}e^{-i\omega t}$ which propagates along the $z$ direction in a given coordinate system where the medium is at rest. Here the (T) symbol indicates the transpose of a row vector. For an electromagnetic wave propagating in the $z$ direction, we have that all electric field components depend only on $z$, namely $E_i(\bs x, \omega)=E_i(z, \omega)$. In the presence of a medium, the Maxwell equation for the electric field $\bs E_i(\bs x, t)$ in an isotropic medium is given by
\begin{equation}\label{max-eq-med}
\frac{1}{v_\text{ph}^2}\frac{\partial^2 \bs E_i}{\partial t^2}-\nabla^2 \bs E_i=0
\end{equation}
where $v_\text{ph}^2=1/\epsilon\mu=1/n^2$ is the phase velocity of the electromagnetic wave in the plasma. Here $\epsilon$ and $\mu$ are respectively the electric permittivity and magnetic permeability in the plasma. By assuming that the electromagnetic wave evolves harmonically in time at a given point $z$ in space and considering the propagation in an anisotropic plasma, we get from \eqref{max-eq-med} and \eqref{pi-chi-rel}
\begin{equation}\label{wave-eq-1}
\partial_z^2 E_i(z, \omega)=\left[\Pi_{ij}(\omega) -  \omega^2\delta_{ij} \right] E_j(z, \omega).
\end{equation}

One important aspect of the wave equation \eqref{wave-eq-1} is that for $i=z$, namely for the $z$ component of the electric field, we have that $\partial_z^2 E_z(z, \omega)=0$. In this case the equation \eqref{wave-eq-1} for $i=z$ gives a constraint on $E_z$ which implies that it depends linearly on the transverse components of the electric field through the relation
\begin{equation}\label{Ez-condition}
E_z=\frac{\Pi_{zx} E_x + \Pi_{zy} E_y}{\omega^2 - \Pi_{zz}}.
\end{equation}
Now by using the constraint \eqref{Ez-condition} into the ($i=x, y$) components of the electric field in \eqref{wave-eq-1}, we get the following equation for the transverse components of the electric field
\begin{equation}\label{wave-eq-2}
 \partial_z^2 E_i(z, \omega)=\left[\tilde\Pi_{ij} -\omega^2 \delta_{ij}\right]E_j(z, \omega), \quad \text{for}\quad (i, j=x, y),
\end{equation}
where $\tilde\Pi_{ij}$ for $(i, j=x, y)$ is the effective photon polarization tensor of the transverse electromagnetic field in the cold magnetized plasma
\begin{equation}\label{efe-pola}
\tilde\Pi_{ij}=\Pi_{ij}+\frac{\Pi_{iz}\Pi_{zj}}{\omega^2-\Pi_{zz}},
\end{equation}
where the components of $\Pi_{ij}$ are given in \eqref{pola-tens}. The effective expression for the polarization tensor in \eqref{efe-pola} takes into account the mixing of the longitudinal electromagnetic wave in plasma with the usual transverse electromagnetic waves. From expressions \eqref{efe-pola} and \eqref{pola-tens}, we find the following expressions for the components of $\tilde\Pi_{ij}$

\begin{equation}\label{efe-pola-comp}
\begin{gathered}
\tilde \Pi_{xx}=\frac{\omega^2\,\omega_\text{pl}^2}{\omega^2-\omega_c^2}\left[ 1 - \frac{\omega_c^2}{\omega^2} \cos^2(\Theta) +\frac{\omega_\text{pl}^2 \omega_c^2 \left( 4\, \omega^2\cos^2(\Phi)\sin^2(\Theta) + \omega_c^2 \sin^2(2\Theta) \sin^2(\Phi) \right)}{4 \,\omega^2 \left( \omega^4-\omega^2\,\omega_\text{pl}^2 -\omega^2\,\omega_c^2 +\omega_c^2\,\omega_\text{pl}^2\sin^2(\Theta)\sin^2(\Phi)\right)} \right], \\
 \tilde \Pi_{xy}=  -\frac{\omega_\text{pl}^2\,\omega_c^2}{2\,(\omega^2-\omega_c^2)} \left[\sin(2 \Theta)\cos(\Phi) + \frac{\omega_\text{pl}^2 \left( 2\, \omega^2\sin(2 \Theta)\cos(\Phi) - \omega_c^2 \sin^2(\Theta)\sin(2\Theta) \sin(\Phi) \sin(2 \Phi)\right)}{2\left( \omega^4-\omega^2\,\omega_\text{pl}^2 -\omega^2\,\omega_c^2 +\omega_c^2\,\omega_\text{pl}^2\sin^2(\Theta)\sin^2(\Phi)\right)} \right] \\  - i \frac{\omega_\text{pl}^2\omega_c\,\omega}{\omega^2-\omega_c^2} \left[ \sin(\Theta)\sin(\Phi) + \frac{\omega_\text{pl}^2 \omega_c^2 \left( \sin(\Phi)\cos(\Theta)\sin(2\Theta) +  \sin^3(\Theta)\sin(2\Phi) \cos(\Phi) \right)}{2\left( \omega^4-\omega^2\,\omega_\text{pl}^2 -\omega^2\,\omega_c^2 +\omega_c^2\,\omega_\text{pl}^2\sin^2(\Theta)\sin^2(\Phi)\right)} \right], \\  
 \tilde \Pi_{yx} = \tilde \Pi_{xy}^*,\\
\tilde\Pi_{yy}= \frac{\omega^2\,\omega_\text{pl}^2}{\omega^2-\omega_c^2}\left[ 1 - \frac{\omega_c^2}{\omega^2} \sin^2(\Theta)\cos^2(\Phi) +\frac{\omega_\text{pl}^2 \omega_c^2 \left( 4\, \omega^2\cos^2(\Theta) + \omega_c^2 \sin^4 (\Theta) \sin^2(2 \Phi) \right)}{4 \,\omega^2 \left( \omega^4-\omega^2\,\omega_\text{pl}^2 -\omega^2\,\omega_c^2 +\omega_c^2\,\omega_\text{pl}^2\sin^2(\Theta)\sin^2(\Phi)\right)} \right].
\end{gathered}
\end{equation}

The expressions given in \eqref{efe-pola-comp} are the most general form of the elements of $\tilde \Pi_{ij}$ in a cold magnatized plasma. As already mentioned above they take into account the mixing of the longitudinal electric field with the usual transverse electric field. We may note that this contribution in \eqref{efe-pola-comp} is inversely proportional to $ \omega^4-\omega^2\,\omega_\text{pl}^2 -\omega^2\,\omega_c^2 +\omega_c^2\,\omega_\text{pl}^2\sin^2(\Theta)\sin^2(\Phi)\neq 0$. The latter condition is satisfied as far as $\omega>0$ and
\begin{equation}
\omega\neq \sqrt{\frac{\omega_\text{pl}^2 + \omega_c^2}{2}} \left[ 1\pm \left( 1- \frac{4 \omega_c^2\omega_\text{pl}^2 \sin^2(\Phi)\sin^2(\Theta)}{\omega_\text{pl}^2 +\omega_c^2}\right)^{1/2} \right]^{1/2},
\end{equation}
where we must have $\omega_\text{pl}^2 +\omega_c^2\geq 2 \omega_c\, \omega_\text{pl} |\sin(\Phi)\sin(\Theta)|$ in order to have real and positive roots of the quadratic equation. Another important question to ask is for what minimum frequencies we have propagating transverse electromagnetic waves? This can be seen by requiring that all spatial derivatives on the left hand side in Eq. \eqref{wave-eq-2} are zero, namely a non propagating electric field in space. In that case we would have $\left[ \tilde \Pi_{ij}-\omega^2 \delta_{ij} \right] E_j=0$ where nontrivial solution exist only if det$(M_{ij})=0$ with $M_{ij}\equiv \tilde \Pi_{ij}-\omega^2 \delta_{ij}$. However, the solution of det$(M_{ij})=0$ in terms of $\omega$ would be quite complicated due to the presence of $\Theta$ and $\Phi$. Therefore, it is more convenient to rotate the coordinate system in  such a way that $\Phi=\pi/2$ and $\Theta=\pi/2$, namely $\bs B_e$ is along the direction of propagation of the electromagnetic wave. Under a rotation of the coordinate system we have that $M_{ij}^\prime$ in the new coordinate system is related to the old $M_{ij}$ through $M_{ij}^\prime=R_{il} R_{jm} M_{lm}$ and $E_j^\prime = R_{jk} E_k$ where $R_{il} $ is an orthogonal rotation matrix with unit determinant. In the rotated coordinate system the equation $M_{ij} E_j=0$ becomes $M_{ij}^\prime E_j^\prime=M_{ij}(\Phi=\pi/2, \Theta=\pi/2) E_j^\prime=0$. Consequently, the condition det$(M_{ij}^\prime)=0$ is equivalent to det$[M_{ij}(\Phi=\pi/2, \Theta=\pi/2)]=0$. Now by requiring that det$[M_{ij}(\Phi=\pi/2, \Theta=\pi/2)]=0$ and after doing some algebra we find that the lower bound on the frequencies for propagation are $\omega> \left(\pm\omega_c + \sqrt{\omega_c^2+4 \omega_\text{pl}^2}\right)/2$.

\section{Solutions of the equations of motion of the Stokes parameters}
\label{sec:3}

In the previous section we derived the most general form of the elements of the photon polarization tensor in a cold magnetized plasma for arbitrary direction of propagation of the electromagnetic waves with respect to the external magnetic field $\bs B_e$. In this section, we focus our attention in deriving the equations of motion of the Stokes parameters in an expanding universe and provide perturbative solutions of the equations of motion. As in the previous section, let us consider an electromagnetic wave propagating along the $z$ direction in a cartesian reference system with wave vector $\bs k=(0, 0, k)$ in a cold magnetized plasma with arbitrary direction of the external magnetic field $\bs B_e$. The linearized equations of motion in an unperturbed FRW metric for the CMB photons are given by
\cite{Ejlli:2016avx}
\begin{equation}\label{wave-eq}
i\partial_t \Psi(k, t)=\left[M(k, B_e, \Phi, \Theta) - \frac{3}{2} H(t) \bs I \right] \Psi(k, t),
\end{equation}
where $A_x$ and $A_y$ are respectively the transverse components of the vector potential $\bs A$ of the CMB photons with respect to the $x$ and $y$ axes, $\Psi(k, t)=[A_x(k, t), A_y(k, t)]^\text{T}$ is a two component field, $H(t)$ is the Hubble parameter, $\bs I$ is a $2\times 2$ identity matrix and $M$ is the mixing matrix which is given by 
\begin{equation}
M(k, B_e, \Phi, \Theta)=
\left[
\begin{matrix}
k-M_x & -M_\text{CF}\\
-M_\text{CF}^* & k-M_y
\end{matrix}
\right],
\end{equation}
where $M_x=-\tilde\Pi_{xx}/(2\omega)$, $M_y=-\tilde\Pi_{yy}/(2\omega)$ and $M_\text{CF}=-\tilde\Pi_{xy}/(2\omega)$. The term $M_\text{CF}=M_\text{C}+iM_\text{F}$ takes into account the combination of the CM and Faraday effects in a magnetized plasma.

In order to describe the polarization of the light and more precisely in our case of the CMB photons, it is better to work with the Stokes parameters rather than the wave equation \eqref{wave-eq}. The procedure in obtaining the equations of motion of the Stokes parameters has been presented in \cite{Ejlli:2016avx} and it consists on two steps; first write the equations of motion for the polarization density matrix $\rho$ based on the wave equation \eqref{wave-eq} and second, express the polarization density matrix in terms of the Stokes parameters in order to get the equations of motion of the latter quantities. The equations of motion of the polarization density matrix in an unperturbed FRW metric are given by \cite{Ejlli:2016avx}
\begin{equation}
\frac{\partial \rho}{\partial t}=-i [M, \rho] - \{D, \rho\},
\end{equation}
where $D=(3/2) H(t) \bs I$ is the damping matrix which takes into account the damping of the electromagnetic waves in an expanding universe due to the Hubble friction. In our case the field mixing matrix $M$ is Hermitian, namely $M=M^\dagger$ since in our case we do not include any process which might change the number of photons due to decay or absorption in the medium\footnote{ If there is any process that may change the number of the CMB photons, its magnitude is a very small quantity at post decoupling epoch.}.  Now by using the connection between the Stokes parameters and the polarization density matrix elements as shown in Ref. \cite{Fano:1954zza}, see also the appendix of Ref. \cite{Ejlli:2016avx}, we get the following equations of motion of the effective Stokes parameters
\begin{align}\label{Stokes-eq}
\dot I (k, \hat{\bs n}, t) & =-3 H(t) I (k, \hat{\bs n}, t),\nonumber\\
\dot Q (k, \hat{\bs n}, t) & =-2M_\text{F}(k) U (k, \hat{\bs n}, t) - 2 M_\text{C}(k) V (k, \hat{\bs n}, t) - 3 H(t)Q (k, \hat{\bs n}, t),\\
\dot U(k, \hat{\bs n}, t)  & =2M_\text{F}(k) Q(k, \hat{\bs n}, t)  + \Delta M(k) V(k, \hat{\bs n}, t) -3 H(t)U(k, \hat{\bs n}, t) ,\nonumber\\
\dot V (k, \hat{\bs n}, t) &= 2M_\text{C}(k)Q(k, \hat{\bs n}, t) -\Delta M(k) U(k, \hat{\bs n}, t) -3 H(t) V(k, \hat{\bs n}, t) ,\nonumber
\end{align}
where we have defined $\Delta M \equiv  M_{x}- M_{y}$ with the dot sign above Stokes parameters indicating the time derivative with respect to the cosmological time $t$. For simplicity, in \eqref{Stokes-eq} we have dropped the symbols $B_e, \Phi$ and $\Theta$ which do appear in the elements of $M$.

The system of linear differential equations \eqref{Stokes-eq} can be written in a more compact form as $\dot S(k, \hat{\bs n}, t)=A(k, t)S(k, \hat{\bs n}, t)$ where $S=(I, Q, U, V)^\text{T}$ is the Stokes vector formed with the Stokes parameters and $A(k, t)$ is the time dependent coefficient matrix which is given by

\begin{equation}
A(k, t)=
\begin{pmatrix}
- 3 H & 0 & 0 & 0\\
0 & -3 H & -2 M_\text{F} & -2 M_\text{C}\\
0 & 2 M_\text{F} & -3 H & \Delta M\\
0 & 2 M_\text{C} & -\Delta M & -3 H
\end{pmatrix}.
\end{equation}
In most cases it is more convenient to express the quantities in $A$ as a function of the photon temperature $T$ rather than the cosmological time $t$, so, in this case one needs to express the time derivative in an expanding universe as $\partial_t= - H T\partial_T$ in the equations of motion of the Stokes vector, namely $S^\prime(k, \hat{\bs n}, T)= \tilde A(k, T) S(k, \hat{\bs n}, T)$. At this stage it is more convenient to write the matrix $\tilde A(k, T)$ as the sum of $\tilde A(k, T)=B(k, T) + (3/T) \bs I_{4\times 4}$, where the matrix $B(k, T)$ is given by
\begin{equation}\label{B-matrix}
B(k, T)=\frac{1}{HT}
\begin{pmatrix}
0 & 0 & 0 & 0\\
0 & 0 & 2 M_\text{F}(T) & 2 M_\text{C}(T)\\
0 & -2 M_\text{F}(T) & 0 & -\Delta M(T)\\
0 & -2 M_\text{C}(T) & \Delta M(T) & 0
\end{pmatrix},
\end{equation}
where in an expanding universe the wave-vector $k=k(T)$ is a function of the temperature $T$. We may note that with respect to the case when the direction of $\bs B_e$ is in the $xz$ plane as studied in Ref. \cite{Ejlli:2016avx}, for arbitrary magnetic field direction also appear the terms $2 M_\text{C}$ in the matrix $B$. The appearance of these terms which make possible the mixing of the $Q$ parameter with $U$ and $V$ parameters, complicate the situation with respect to the case when $M_\text{C}=0$.

\section{Series solution of the polarization equations of motion}
\label{sec:4}

In the previous section, Sec. \ref{sec:3}, we found the equations of motion of the Stokes parameters in an expanding universe for an arbitrary direction of the external magnetic field $\bs B_e$ with respect to the electromagnetic wave direction of propagation. In this section we focus our attention on perturbative solutions of the equations of motion in some limiting cases. Before aiming to find these solutions, it is very important to calculate first each term that enters the matrix $B(k, T)$. Let us recall the definitions of $M_\text{F} \equiv -\text{Im}\{\tilde \Pi_{xy}\}/(2\omega)$, $M_\text{C} \equiv -\text{Re}\{\tilde \Pi_{xy}\}/(2\omega)$ and $\Delta M \equiv M_x-M_y=(\tilde \Pi_{yy}-\tilde \Pi_{xx})/(2\omega)$. Now by using the expressions of the photon polarization tensor given in \eqref{efe-pola-comp} we get 
\begin{equation}\label{M-comp}
\begin{gathered}
M_\text{C}=\frac{\omega_\text{pl}^2\,\omega_c^2}{4\,\omega(\omega^2-\omega_c^2)} \left[\sin(2 \Theta)\cos(\Phi) + \frac{\omega_\text{pl}^2 \left( 2\, \omega^2\sin(2 \Theta)\cos(\Phi) - \omega_c^2 \sin^2(\Theta)\sin(2\Theta) \sin(\Phi) \sin(2 \Phi)\right)}{2\left( \omega^4-\omega^2\,\omega_\text{pl}^2 -\omega^2\,\omega_c^2 +\omega_c^2\,\omega_\text{pl}^2\sin^2(\Theta)\sin^2(\Phi)\right)} \right], \\
 M_\text{F}=  \frac{\omega_\text{pl}^2\omega_c}{2(\omega^2-\omega_c^2)} \left[ \sin(\Theta)\sin(\Phi) + \frac{\omega_\text{pl}^2 \omega_c^2 \left( \sin(\Phi)\cos(\Theta)\sin(2\Theta) +  \sin^3(\Theta)\sin(2\Phi) \cos(\Phi) \right)}{2\left( \omega^4-\omega^2\,\omega_\text{pl}^2 -\omega^2\,\omega_c^2 +\omega_c^2\,\omega_\text{pl}^2\sin^2(\Theta)\sin^2(\Phi)\right)} \right], \\  
\Delta M= -\frac{\omega_c^2\,\omega_\text{pl}^2}{2 \omega\,(\omega^2-\omega_c^2)}\left[ \sin^2(\Theta)\cos^2(\Phi) -\cos^2(\Theta) \right. \\ \left.+ \frac{\omega_\text{pl}^2 \left[ 4\, \omega^2\left(\cos^2(\Phi)\sin^2(\Theta)-\cos^2(\Theta)\right) + \omega_c^2 \left(\sin^2(2\Theta)\sin^2(\Phi)-\sin^4 (\Theta) \sin^2(2 \Phi)\right) \right]}{4 \,\left( \omega^4-\omega^2\,\omega_\text{pl}^2 -\omega^2\,\omega_c^2 +\omega_c^2\,\omega_\text{pl}^2\sin^2(\Theta)\sin^2(\Phi)\right)} \right].
\end{gathered}
\end{equation}

We want to stress that from expressions \eqref{M-comp} one can immediately see the contribution of the CM and Faraday effects explicitly. To the leading order (see calculations below), the CM effect terms $\Delta M$ and $M_\text{C}$ are proportional to $\omega_c^2\propto B_e^2$ while the Faraday effect term $M_\text{F}$ is proportional to $\omega_c\propto B_e$. We may also note that the term corresponding to the plasma effects exactly cancel out in $\Delta M$ because the indexes of refraction of light in a cold plasma are the same for both propagating photon transverse states. Thus, to the polarization of light in a magnetized plasma contribute the CM effect through the terms $\Delta M, M_\text{C}$ and the Faraday effect through the term $M_\text{F}$.

The expression for the elements of the matrix $B$ given in \eqref{M-comp}, which are the most general ones for arbitrary magnetic field direction and magnitude, can be further simplified by making some reasonable assumptions on the parameters. Since in this work we concentrate on the CMB frequency spectrum we have that $\omega\gg \omega_\text{pl}$ and  $\omega\gg \omega_c$. In order to see this, let us calculate explicitly the numerical values of the parameters. The numerical value of the angular plasma frequency which enters the expressions in \eqref{M-comp} can be written as $\omega_\text{pl}=5.64\times 10^4 \sqrt{n_e/\text{cm}^3}$ (rad/s) or $\nu_\text{pl}=\omega_\text{pl}/(2\pi)=8976.33 \sqrt{n_e/\text{cm}^3}$ (Hz) for the frequency. On the other hand the numerical value of the cyclotron angular frequency is given by $\omega_c=1.76\times 10^7( B_{e0}/\text{G})$ (rad/s) or $\nu_c=2.8\times 10^6( B_{e0}/\text{G})$ (Hz). However, in the case of CMB photons propagating in an expanding universe, we can express the time $t$ in terms of the cosmological temperature $T$ as $t=t(T)$ as we did in the previous section. Therefore, the conditions $\omega\gg \omega_\text{pl}$ and  $\omega\gg \omega_c$, in an expanding universe, are respectively satisfied when
\begin{equation}\label{nu-con-1}
\left(\frac{\nu_0}{\text{Hz}}\right)\gg 8976.33 \left(\frac{0.76\,n_B(T_0) X_e(T)}{\text{cm}^3}\right)^{1/2} \left(\frac{T}{T_0}\right)^{1/2} \quad \text{and} \quad \left(\frac{\nu_0}{\text{Hz}}\right)\gg 2.8\times 10^6 \left(\frac{ B_{e0}}{\text{G}}\right)\left(\frac{T}{T_0}\right),
\end{equation}
where we expressed $\nu(t)=\nu_0 [a(t_0)/a(t)]=\nu_0(T/T_0)$ with $\nu_0$ being the frequency of the electromagnetic radiation at the present time $t=t_0$ at the temperature $T=T_0$, $a(t)$ being the universe expansion scale factor and $ B_{e0}= B_e(t_0)=B_e(T_0)$ is the magnetic field strength at the present time. Here we expressed the number density of free electrons as $n_e(t)=n_e(T)\simeq 0.76\, n_B(T_0) X_e(T)(T/T_0)^3$ where $n_B(T_0)$ is the total baryon number density at the present time and $X_e(T)$ is the ionization function of the free electrons. The factor of $0.76$ takes into account the contribution of hydrogen atoms to the free electrons at the post decoupling time. By taking for example $n_B(T_0)\simeq 2.47\times 10^{-7}$ cm$^{-3}$ as given by the Planck collaboration \cite{Aghanim:2018eyx}, and expressing $a(t_0)/a(t)=T/T_0$, we can write the conditions \eqref{nu-con-1} as
\begin{equation}\label{nu-con-2}
\left(\frac{\nu_0}{\text{Hz}}\right)\gg 3.88 X_e^{1/2}(T)\left(T/T_0 \right)^{1/2} \quad \text{and} \quad \left(\frac{\nu_0}{\text{Hz}}\right)\gg 2.8 \times 10^6 \left(\frac{ B_{e0}}{\text{G}}\right) \left( T/T_0 \right).
\end{equation}

Given the fact that the present day CMB photon frequencies are in the frequency part above $\nu_0\geq 10^8$ Hz, the conditions given in \eqref{nu-con-2} are well satisfied for physically reasonable values of $X_e(T)$ and $B_{e0}$. With these considerations in mind, we can simplify $\omega^4-\omega^2\,\omega_\text{pl}^2 -\omega^2\,\omega_c^2 +\omega_c^2\,\omega_\text{pl}^2\sin^2(\Theta)\sin^2(\Phi)\simeq \omega^4$ in \eqref{M-comp} for $\omega\gg \omega_\text{pl}$ and  $\omega\gg \omega_c$. 
After this simplification, from the expressions \eqref{M-comp} we may note that each expression within the square brackets is composed of a first term of trigonometric functions and a second term which is the product of trigonometric functions with terms $\omega_\text{pl}^2/\omega^2$ or $\omega_\text{pl}^2\,\omega_c^2/\omega^4$. However, since we are in the regime when $\omega\gg \omega_\text{pl}$ and  $\omega\gg \omega_c$, we also have $\omega_\text{pl}^2/\omega^2\ll 1$ and $\omega_\text{pl}^2\,\omega_c^2/\omega^4 \ll 1$. This fact tells us that in the case when the trigonometric functions in the first and second terms within the square brackets in \eqref{M-comp} are different from zero, the second term is usually much smaller than the first term.  In order to see this, let us consider the case when $\Theta=0$, namely when the magnetic field has components only along the $x$. In this case $M_\text{C}=M_\text{F}=0$ and $\Delta M=[\omega_\text{pl}^2\omega_c^2/(2\omega^3)][1+(\omega_\text{pl}/\omega)^2]$, where $(\omega_\text{pl}/\omega)^2\ll 1$, so the contribution coming from the second term can be completely neglected. One can see that by making similar examples, the contribution of the second terms within the square brackets in \eqref{M-comp}, that arise due to the mixing of the longitudinal electromagnetic wave with the transverse waves, can be neglected with respect to the first terms. Consequently, in the regime studied in this work $\omega\gg \omega_\text{pl}$ and  $\omega\gg \omega_c$, we have that 

\begin{equation}\label{M-comp-2}
\begin{gathered}
M_\text{C} \simeq \frac{\omega_\text{pl}^2\,\omega_c^2}{4\,\omega^3} \sin(2 \Theta)\cos(\Phi), \quad
 M_\text{F} \simeq  \frac{\omega_\text{pl}^2\omega_c}{2\,\omega^2} \sin(\Theta)\sin(\Phi), \quad  
\Delta M \simeq - \frac{\omega_c^2\,\omega_\text{pl}^2}{2 \omega^3} \left[\sin^2(\Theta)\cos^2(\Phi) -\cos^2(\Theta)\right].
\end{gathered}
\end{equation}

\subsection{Neumann series solutions}
\label{subsec:4.1}

Here we present a Neumann series solutions of the equations of motion by making use of the perturbation theory. Let us concentrate on the full equation $S^\prime(k, \hat{\bs n}, T)= \left[B(k, T) + (3/T) \bs I_{4\times 4}\right] S(k, \hat{\bs n}, T)$ and omit from now on the dependence of the Stokes vector  on $\hat{\bs{n}}$ and $k$ and also that of matrix $B$ on $k$. Starting from now, in what follows in this work, we use the expressions found in \eqref{M-comp-2} for the elements of the matrix $B$ in \eqref{B-matrix}. From the equation of motion of the Stokes vector, the term $3/T$ is a term which takes into account the damping of fields in an expanding universe. In the case when there is not a magnetic field, the solution of the equation $S^\prime(T)= \left[B(T) + (3/T) \bs I_{4\times 4}\right] S(T)$ is $S(T)=\exp[3 \bs I_{4\times 4}\int_{T_i}^T dT^\prime/T^\prime] S(T_i)= (T/T_i)^3 \bs I_{4\times 4} S(T_i)$ for $B(T)=0$. It is worth to stress since now that the effective scaling of the Stokes vector in an expanding universe is not $(T/T_i)^3$ but $(T/T_i)^2$ as discussed in details in Ref. \cite{Ejlli:2016avx}. In the case when the magnetic field is present, namely when $B(T) \neq 0$, it is convenient to define $ S(T)\equiv (T/T_i)^3 \bs I_{4\times 4} \tilde S(T)$. In this case, the equation of motion for $S^\prime(T)= \left[B(T) + (3/T) \bs I_{4\times 4}\right] S(T)$ in components becomes
\begin{equation}\label{equ-0}
\tilde S_i^\prime(T)=B_{ij}(T) \tilde S_j(T) \quad (i, j=1, 2, 3, 4).
\end{equation}

The system of first order of linear differential equations given in \eqref{equ-0} cannot be solved exactly except in some particular cases. However, one of the main characteristic of a linear system of first order of differential equations is that its general solution is given by $\tilde S(T)=\tilde M(T) \tilde S(T_i)$, where $\tilde M(T)$ is the solution matrix. Consequently, if we put the general solution $\tilde S(T)=\tilde M(T) \tilde S(T_i)$ in \eqref{equ-0}, we get that the solution matrix $\tilde M$ satisfies the equation
\begin{equation}\label{equ-1}
\tilde M_{ij}^\prime(T)=B_{il}(T) \tilde M_{lj}(T) \quad (i, j, l=1, 2, 3, 4),
\end{equation}
with the initial conditions $\tilde M_{lj}(T_i)=\bs I_{4\times 4}$. Therefore the solution of the system \eqref{equ-0} is reduced to the solution of the differential equations for the matrix $\tilde M_{lj}$ in \eqref{equ-1}. The system of differential equations given in \eqref{equ-1} can formally be solved as a convergent Neumann series in the case when the non zero elements of the matrix $B_{lm}(T)$ satisfy $\left|\int_{T_i}^T dT^\prime B_{lm}(T^\prime) \right| < 1$
\begin{equation}\label{Series}
\tilde M_{ij}(T)= \bs I_{4\times 4}+ \int_{T_i}^T dT^\prime B_{ij}(T^\prime) +\int_{T_i}^T \int_{T_i}^{T^\prime} dT^{\prime} dT^{\prime\prime} B_{il}(T^{\prime}) B_{lj}(T^{\prime\prime}) +  \int_{T_i}^T \int_{T_i}^{T^\prime} \int_{T_i}^{T^{\prime\prime}} dT^{\prime} dT^{\prime\prime} dT^{\prime\prime\prime} B_{il}(T^{\prime}) B_{lm}(T^{\prime\prime}) B_{mj}(T^{\prime\prime\prime})+ ...
\end{equation}
where we have that $...<T^{\prime\prime\prime}<T^{\prime\prime}<T^{\prime}<T$.

In order to find the parameter space arising from the conditions $\left|\int_{T_i}^T dT^\prime B_{lm}(T^\prime) \right| < 1$, we need to evaluate explicitly each element in the matrix $B_{ij}(T)$. In each element in $B_{ij}(T)$ enters the product $H(T)T$, where the Hubble parameter in the case of zero spatial curvature is given by
\begin{equation}
H(T)=H_0\left[ \Omega_\Lambda + \Omega_\text{M}(T/T_0)^3 + \Omega_\text{R}(T/T_0)^4\right]^{1/2},
\end{equation}
where after the decoupling epoch the contribution of relativistic particles to the total energy density and consequently to the Hubble parameter can be safely neglected. In addition, since the contribution of the cosmological parameter to the Hubble parameter is important only for low redshifts, we may approximate the Hubble parameter in our calculations as $H(T) \simeq H_0 \sqrt{\Omega_\text{M}}(T/T_0)^{3/2}$. Now let us define
\begin{equation}
\mathcal M_\text{F}(T) \equiv \int_{T}^{T_i} dT^\prime\, \frac{2M_\text{F}(T^\prime)}{H(T^\prime) T^\prime}, \quad  \mathcal M_\text{C}(T) \equiv \int_{T}^{T_i} dT^\prime\, \frac{2M_\text{C}(T^\prime)}{H(T^\prime) T^\prime}, \quad \Delta\mathcal M(T) \equiv \int_{T}^{T_i} dT^\prime\, \frac{\Delta M(T^\prime)}{H(T^\prime) T^\prime}.
\end{equation}
The conditions that $\left|\int_{T_i}^T dT^\prime B_{lm}(T^\prime) \right| < 1$ imply that $|\mathcal M_\text{F}(T)|<1, |\mathcal M_\text{C}(T)|<1$, and $|\Delta \mathcal M(T)|<1$. The condition $|\mathcal M_\text{F}(T)|<1$ is also satisfied by the following stronger condition
\begin{equation}\label{F-cond}
8.71\times 10^{25} \left(\frac{\text{Hz}}{\nu_0}\right)^2\left( \frac{B_{e0}}{\text{G}}\right) T_0^{-1/2} \left|\int_{T}^{T_i} dT^\prime X_e(T^\prime) T^{\prime 1/2} \right|  \quad (\text{K}^{-1}) < 1,
\end{equation}
where we used the fact that $\left| \sin(\Theta)\sin(\Phi) \right|\leq 1$ in $|\mathcal M_\text{F}(T)|<1$. So, the condition \eqref{F-cond} is a stronger condition on the parameter space with respect to the case when the term $\left| \sin(\Theta)\sin(\Phi) \right|$ is taken into account. In case when $\left| \sin(\Theta)\sin(\Phi) \right|\rightarrow 0$, the condition $|\mathcal M_\text{F}(T)|<1$ is in principle satisfied for any finite value of the parameters $B_{e0}, \nu_0$ and $T$.
On the other hand, the conditions  $|\mathcal M_\text{C}(T)|<1$ and $|\Delta \mathcal M(T)|<1$ are respectively satisfied by the stronger conditions 
\begin{eqnarray}\label{C-M-cond}
6.05\times 10^{31}  \left(\frac{\text{Hz}}{\nu_0}\right)^3\left( \frac{B_{e0}}{\text{G}}\right)^2 T_0^{-3/2}\left|\int_{T}^{T_i} dT^\prime X_e(T^\prime) T^{\prime 3/2} \right|\,  \quad (\text{K}^{-1}) < 1 \quad \text{and} \\  1.21\times 10^{32}  \left(\frac{\text{Hz}}{\nu_0}\right)^3\left( \frac{B_{e0}}{\text{G}}\right)^2 T_0^{-3/2}\left|\int_{T}^{T_i} dT^\prime X_e(T^\prime) T^{\prime 3/2} \right|\,  \quad (\text{K}^{-1}) < 1.\nonumber
\end{eqnarray}
where we used the fact $| \sin(2 \Theta)\cos(\Phi) |\leq 1$ in $|\mathcal M_\text{C}(T)|<1$ and that $\left| \left[\sin^2(\Theta)\cos^2(\Phi) -\cos^2(\Theta)\right] \right|\leq 1$ in $|\Delta M(T)|<1$. Again, in the cases when $| \sin(2 \Theta)\cos(\Phi) |\rightarrow 0$ and $\left| \left[\sin^2(\Theta)\cos^2(\Phi) -\cos^2(\Theta)\right] \right|\rightarrow 0$, the conditions $|\mathcal M_\text{C}(T)|<1$ and $|\Delta \mathcal M(T)|<1$ are in principle satisfied for any finite values of the parameters $B_{e0}, \nu_0$ and $T$.

\begin{figure*}[h!]
\centering
\mbox{
\subfloat[\label{fig:Fig1}]{\includegraphics[scale=0.63]{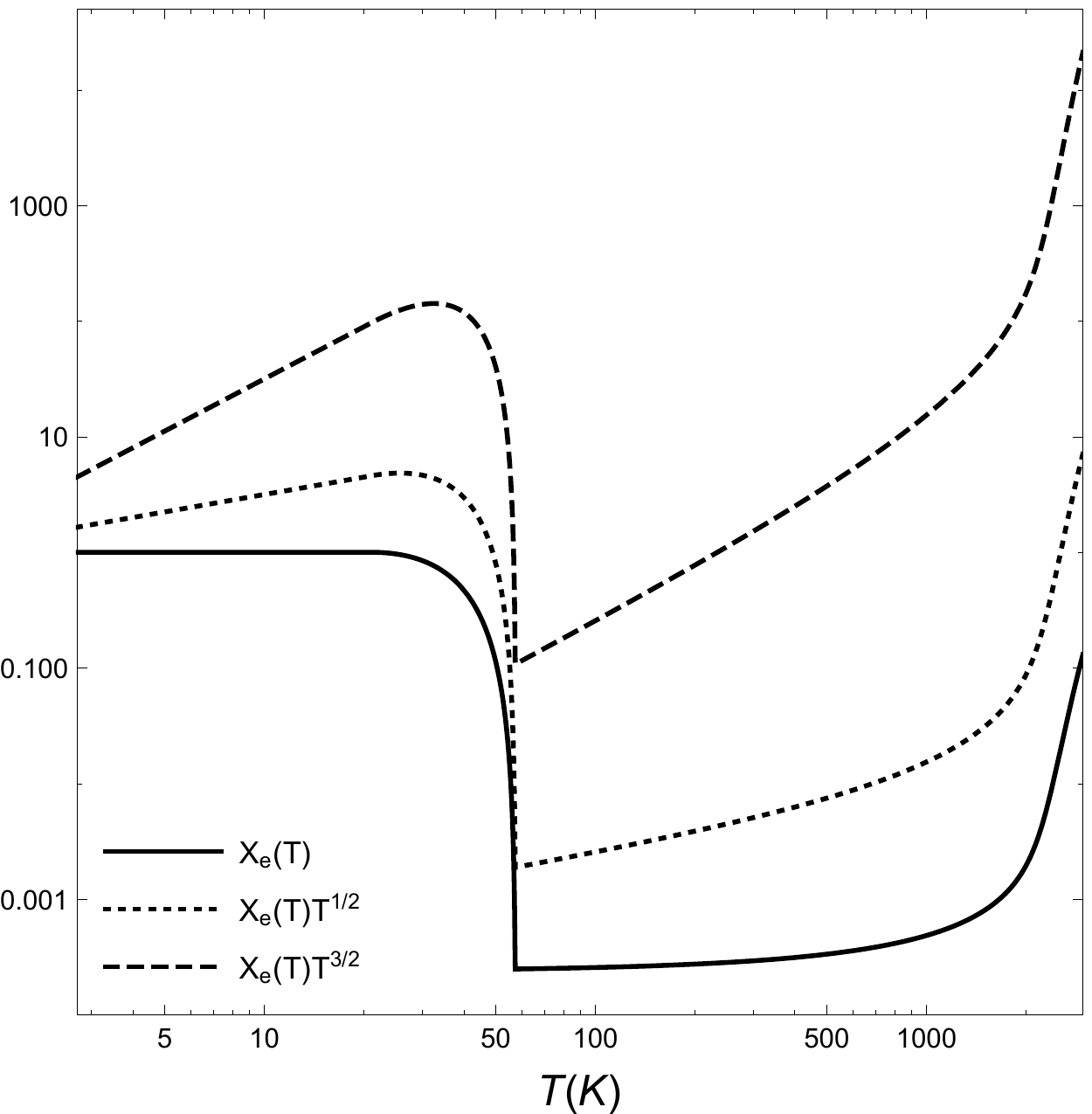}}\qquad
\subfloat[\label{fig:Fig2}]{\includegraphics[scale=0.67]{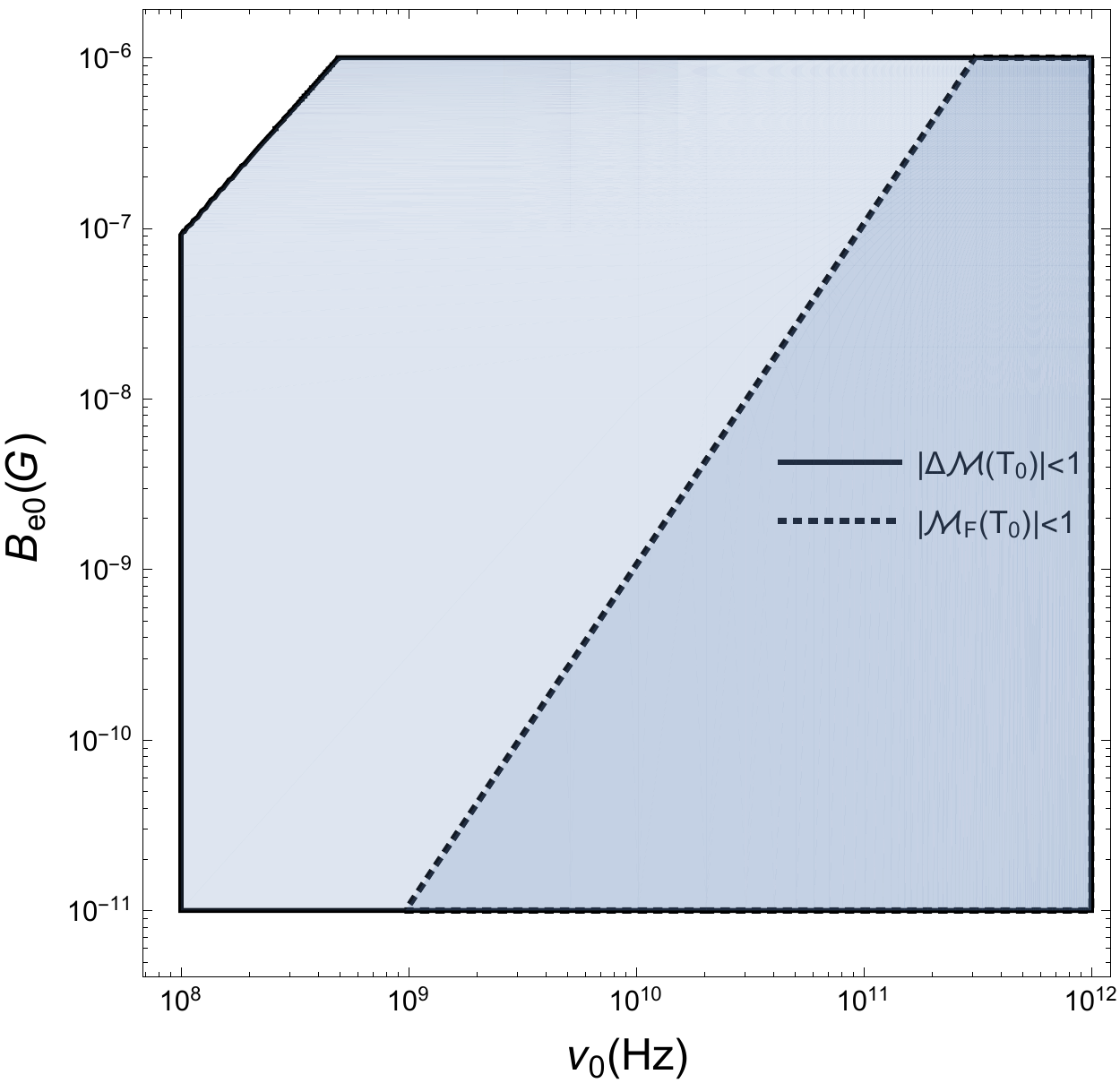}}}
\caption{(a) Logarithmic scale plots of the ionization function $X_e(T)$ (dimensionless), $X_e(T)T^{1/2}$ (in units of K$^{1/2}$) and $X_e(T)T^{3/2}$ (in units of K$^{3/2}$) as a function of the CMB temperature  $T\in [2.725, 2970]$ (K) are shown. (b) Logarithmic scale region plots of the stronger conditions \eqref{F-cond} and \eqref{C-M-cond} of $| \mathcal M_\text{F}(T_0)|<1$ and $|\Delta \mathcal M(T_0)|<1$ only as functions of $\nu_0\in [10^8, 10^{12}]$ (Hz) and $B_{e0}\in [10^{-11}, 10^{-6}]$ (G) are shown. The region within the black line is the allowed region for stronger condition \eqref{C-M-cond} of $|\Delta \mathcal M(T_0)|<1$ while the region within the black dotted line is the allowed region for the stronger condition \eqref{F-cond} of $|\mathcal M_\text{F}(T_0)|<1$. }
\label{fig:Fig1a}
\end{figure*}

In order to find the parameter space for the conditions $\left|\int_{T_i}^T dT^\prime B_{lm}(T^\prime) \right| < 1$ it is necessary to know the expression for the free electron ionization function $X_e(T)$. This function satisfies a complicated differential equations as shown in Ref. \cite{Weinberg:2008zzc} and in general it is calculated by solving the differential equation numerically. In Fig. \ref{fig:Fig1} the plots of  $X_e(T)$, $X_e(T)T^{1/2}$ and $X_e(T)T^{3/2}$ as a function of the CMB temperature  $T$ are shown. In the temperature interval 57.22 K$\leq T \leq 2970$ K the curve of the ionization function $X_e(T)$ is obtained by solving the differential equation for $X_e(T)$ as given in Ref. \cite{Weinberg:2008zzc}, where the lower limit $T = 57.22$ K corresponds to the start of reionization epoch at redshift $z_\text{ion}\sim 20$ and the upper limit corresponds to the CMB decoupling temperature $T_i=2970$ K for redshift $1+z\simeq 1090$. The complete re-ionization is reached approximately at $z_\text{ion}\simeq 7$. The evolution of $X_e(T)$ in the temperature interval 21.8 K $\leq T \leq $ 57.22 K has been obtained by a smooth interpolation of the curve $X_e(T)$ in the interval 57.22 K $\leq T \leq$ 2970 K with $X_e(T) = 1$ in the interval 2.725 K $\leq T \leq$ 21.8 K. By using the numerical solutions found for $X_e(T)$ as descibed above and plotted in Fig. \ref{fig:Fig1}, we get the following values for $\left|\int_{T_0}^{T_i} dT^\prime X_e(T^\prime) T^{\prime 3/2} \right| \simeq 4.45\times 10^6$ (K$^{5/2}$) and $\left|\int_{T_0}^{T_i} dT^\prime X_e(T^\prime) T^{\prime 1/2} \right| \simeq 1790.3$ (K$^{3/2}$). With these values of the integrals, the stronger conditions \eqref{F-cond} and \eqref{C-M-cond} are respectively satisfied when
\begin{equation}\label{F-C-M-cond}
\left(\frac{\text{Hz}}{\nu_0}\right)^2\left( \frac{B_{e0}}{\text{G}}\right) < 1.05\times 10^{-29}, \quad \left(\frac{\text{Hz}}{\nu_0}\right)^3\left( \frac{B_{e0}}{\text{G}}\right)^2 < 1.67 \times 10^{-38} \quad \text{and} \quad \left(\frac{\text{Hz}}{\nu_0}\right)^3\left( \frac{B_{e0}}{\text{G}}\right)^2 < 8.35 \times 10^{-39}.
\end{equation}
In Fig. \ref{fig:Fig2} the region plots of the stronger conditions \eqref{F-cond} and \eqref{C-M-cond} of $|\Delta \mathcal M(T_0)|<1$ and $|\mathcal M_\text{F}(T_0)|<1$ only as functions of the parameters $\nu_0$ and $B_{e0}$ are shown. We may observe that the second stronger condition in \eqref{C-M-cond} of $|\Delta \mathcal M(T_0)|<1$ is satisfied for a wide range of the parameters $\nu_0$ and $B_{e0}$, while the stronger condition \eqref{F-cond} of $|\mathcal M_\text{F}(T_0)|<1$ is satisfied for more stringent values of the parameters. 

Having found the parameter space where the condition for the convergence of the Neumann series certainly holds, we are at the position now to calculate the Stokes parameters. Since the Stokes vector is given by $\tilde S(T)=\tilde M(T) \tilde S(T_i)$, the only thing that we need to calculate to obtain $\tilde S(T)$ is to calculate the elements of the matrix $\tilde M(T)$ given in \eqref{Series}. In the regime where $\left|\int_{T_i}^T dT^\prime B_{lm}(T^\prime) \right| < 1$, it will be sufficient for our purposes to truncate the Neumann series \eqref{Series} at the second order. Now by looking at the structure of the matrix $B(T)$ in \eqref{B-matrix}, we may note that $B_{1j}=B_{j1}=B_{jj}=0$ with the rest of the elements different from zero. Let us define for commodity 
\begin{equation}
G_\text{F}(T) \equiv \frac{2 M_\text{F}(T)}{H(T) T}, \quad G_\text{C}(T) \equiv \frac{2 M_\text{C}(T)}{H(T) T}, \quad 
\Delta G(T) \equiv \frac{ \Delta M(T)}{H(T) T}.
\end{equation}
Then the vanishing elements of $\tilde M$ are $\tilde M_{1j}=\tilde M_{j1}=0$ while the non zero elements of $\tilde M$ are given by
\begin{equation}\label{M-ele}
\begin{gathered}
\tilde M_{11}(T)=1, \quad \tilde M_{22}(T)= 1- \int_T^{T_i} dT^\prime G_\text{F}(T^\prime)\int_{T^\prime}^{T_i} dT^{\prime\prime} G_\text{F}(T^{\prime\prime}) - \int_T^{T_i} dT^\prime G_\text{C}(T^\prime)\int_{T^\prime}^{T_i} dT^{\prime\prime} G_\text{C}(T^{\prime\prime}), \\
\tilde M_{23}(T)= -\mathcal M_\text{F}(T) + \int_{T}^{T_i} dT^\prime G_\text{C}(T^\prime) \int_{T^\prime}^{T_i} dT^{\prime\prime} \Delta G(T^{\prime\prime}), \quad \tilde M_{24}(T)= -\mathcal M_\text{C}(T) - \int_{T}^{T_i} dT^\prime G_\text{F}(T^\prime) \int_{T^\prime}^{T_i} dT^{\prime\prime} \Delta G(T^{\prime\prime}), \\
\tilde M_{32}(T)= \mathcal M_\text{F}(T) + \int_{T}^{T_i} dT^\prime \Delta G(T^\prime) \int_{T^\prime}^{T_i} dT^{\prime\prime} G_\text{C}(T^{\prime\prime}),  \\ \tilde M_{33}(T)= 1- \int_T^{T_i} dT^\prime G_\text{F}(T^\prime)\int_{T^\prime}^{T_i} dT^{\prime\prime} G_\text{F}(T^{\prime\prime}) - \int_T^{T_i} dT^\prime \Delta G(T^\prime)\int_{T^\prime}^{T_i} dT^{\prime\prime} \Delta G(T^{\prime\prime}),  \\
\tilde M_{34}(T)= \Delta \mathcal M(T) - \int_{T}^{T_i} dT^\prime G_\text{F}(T^\prime) \int_{T^\prime}^{T_i} dT^{\prime\prime} G_\text{C}(T^{\prime\prime}), \quad \tilde M_{42}(T)= \mathcal M_\text{C}(T) - \int_{T}^{T_i} dT^\prime \Delta G(T^\prime) \int_{T^\prime}^{T_i} dT^{\prime\prime} G_\text{F}(T^{\prime\prime}), \\
\tilde M_{43}(T)= -\Delta \mathcal M(T) - \int_{T}^{T_i} dT^\prime G_\text{C}(T^\prime) \int_{T^\prime}^{T_i} dT^{\prime\prime} G_\text{F}(T^{\prime\prime}),  \\
\tilde M_{44}(T)= 1- \int_T^{T_i} dT^\prime G_\text{C}(T^\prime)\int_{T^\prime}^{T_i} dT^{\prime\prime} G_\text{C}(T^{\prime\prime}) - \int_T^{T_i} dT^\prime \Delta G(T^\prime)\int_{T^\prime}^{T_i} dT^{\prime\prime} \Delta G(T^{\prime\prime}).
\end{gathered}
\end{equation}

The matrix elements found in \eqref{M-ele} allow us to find explicit expression for the Stokes vector $\tilde S(T)$ up to second order in the perturbation theory. Consequently, the expression of the Stokes parameters are given by 
\begin{equation}\label{Stokes-1}
\begin{gathered}
\tilde I(T)=\tilde I_i, \quad \tilde Q(T)=\tilde M_{22}(T) \tilde Q_i + \tilde M_{23}(T)\tilde U_i+\tilde M_{24}(T)\tilde V_i, \\
 \tilde U(T)=\tilde M_{32}(T) \tilde Q_i + \tilde M_{33}(T)\tilde U_i+\tilde M_{34}(T)\tilde V_i, \quad  \tilde V(T)=\tilde M_{42}(T) \tilde Q_i + \tilde M_{43}(T)\tilde U_i+\tilde M_{44}(T)\tilde V_i,
 \end{gathered}
 \end{equation}
where $\tilde I_i\equiv \tilde I(T_i), \tilde Q_i\equiv \tilde Q(T_i), \tilde U_i\equiv \tilde U(T_i),$ and $\tilde V_i\equiv V(T_i)$. It is very important to stress that the expressions found of the Stokes parameters in \eqref{Stokes-1} are valid for arbitrary direction of the external magnetic field with respect to the photon propagation, namely for arbitrary $\Theta, \Phi$ and for arbitrary profile of $B_{e}(T)$ and $n_e(T)$.

\subsection{Power series solution for dominant Faraday effect}
\label{subsec:4.2}

In the previous section we found Neumann series solutions in the case when the conditions $|\mathcal M_\text{F}(T_0)|<1$, $|\mathcal M_\text{C}(T_0)|<1$ and $|\Delta \mathcal M(T_0)|<1$ are satisfied. However, we did not make any specific assumptions on the relative magnitude among $|\mathcal M_\text{F}(T_0)|$, $|\mathcal M_\text{C}(T_0)|$ and $|\Delta \mathcal M(T_0)|$, namely we did not specify which of these terms is bigger than  the others. 
 As we can see from Fig. \ref{fig:Fig2}, the stronger condition of $|\Delta \mathcal M(T_0)|<1$ is satisfied for a wide range of values that $\nu_0$ and $B_{e0}$ can have. On the other hand, it may happen that the stronger condition of $|\mathcal M_\text{F}(T_0)|<1$ is not satisfied at all. This case would correspond to pairs of points outside the dotted region in Fig. \ref{fig:Fig2}. For example if $\nu_0=10^9$ Hz and $B_{e0}=10^{-9}$ G, the stronger condition of $|\mathcal M_\text{F}(T_0)|<1$ is not satisfied. So, it may happen that in some cases the term corresponding to the Faraday effect is the dominating one for fixed values of $\Theta$ and $\Phi$. Therefore, in this section we study the case when $|2M_\text{F}(T)| \gg |2M_\text{C}(T)|, |\Delta M(T)|$ in the matrix $B(T)$ and require that $|\mathcal M_\text{C}(T_0)|<1$ and $|\Delta \mathcal M(T_0)|<1$. In addition, the conditions $|2M_\text{F}(T)| \gg |2M_\text{C}(T)|, |\Delta M(T)|$ explicitly require that $|\sin(\Theta)\sin(\Phi)|\neq 0$. The conditions $|2M_\text{F}(T)| \gg |2M_\text{C}(T)|$ and $|2M_\text{F}(T)| \gg  |\Delta M(T)|$, for fixed values of the angles $\Theta$ and $\Phi$, are respectively satisfied for any temperature in the interval $T_0\leq T\leq T_i$, only when $T=T_i$
 \begin{equation}\label{angle-cond}
 \begin{gathered}
\left(\frac{\nu_0}{\text{Hz}}\right)\left( \frac{\text{G}}{B_{e0}}\right)\gg 6.94\times 10^5 \left| \frac{\sin(2\Theta)\cos(\Phi)}{\sin(\Theta)\sin(\Phi)}\right| \left( \frac{T_i}{T_0}\right) \quad \text{and} \\  \left(\frac{\nu_0}{\text{Hz}}\right)\left( \frac{\text{G}}{B_{e0}}\right)\gg 3.47\times 10^5 \left| \frac{\sin^2(\Theta)\cos^2(\Phi) -\cos^2(\Theta)}{\sin(\Theta)\sin(\Phi)}\right| \left( \frac{T_i}{T_0}\right).
\end{gathered}
\end{equation}
The conditions in \eqref{angle-cond} are almost always satisfied for realistic values of $\nu_0$ and $B_{e0}$ that we consider in this work unless $|\sin(\Theta)\sin(\Phi)|\rightarrow  0$.

Let us write the matrix $B(T)=B_1(T)+\epsilon B_2(T)$ where $B_1(T)$ is the matrix which contains only the Faraday effect term $2M_\text{F}(T)$ while the matrix $\epsilon B_2(T)$ contains the terms $2M_\text{C}$ and $\Delta M(T)$ only. Here $\epsilon \ll 1$ is a small parameter that can be commonly extracted from $G_\text{C}(T)$ and $\Delta G(T)$ and which depends on $\nu_0$ and $B_{e0}$ only. Let us look for solutions of the equation \eqref{equ-1} in the form 
\begin{equation}\label{M-expansion}
\tilde M= \tilde M^{(0)} + \epsilon \tilde M^{(1)} + \epsilon^2 \tilde M^{(2)} +.... \epsilon^n \tilde M^{(n)}.
\end{equation}
After by inserting the expansion \eqref{M-expansion} and $B(T)=B_1(T)+\epsilon B_2(T)$  in the equation \eqref{equ-1} and collecting the terms with the appropriate power in $\epsilon$, we get the following matrix system of equations
\begin{equation} 
\begin{gathered}\label{M-system-exp}
\tilde M^{\prime (0)}(T) = B_1(T) \tilde M^{(0)}(T), \\
 \epsilon \tilde M^{\prime (1)}(T) = B_1(T) \epsilon \tilde M^{(1)} + \epsilon B_2(T) \tilde M^{(0)}(T),\\
  \epsilon^2 \tilde M^{\prime (2)}(T) = B_1(T) \epsilon^2 \tilde M^{(2)} +\epsilon  B_2(T) \epsilon \tilde M^{(1)}(T),\\
 \vdots \\
\epsilon^n \tilde M^{\prime (n)}(T) = B_1(T) \epsilon^n \tilde M^{(n)} + \epsilon B_2(T) \epsilon^{n-1} \tilde M^{(n-1)}(T),
 \end{gathered}
 \end{equation}
where for simplicity we suppressed the matrix elements indexes in $B_1, B_2$ and $\tilde M$. The system of equations \ref{M-system-exp} has to be solved with the initial conditions $\tilde M^{(0)}(T_i)=\bs I_{4\times 4}$ and $\epsilon^n \tilde M^{n}(T_i)=0$ for $n\geq 1$.

The zero order matrix equation in \eqref{M-system-exp} can immediately be solved by noticing that the different temperature commutator of $B_1(T)$ is zero, namely $[B_1(T), B_1(T^\prime)]=0$, and consequently the solution of the zero order matrix equation in \eqref{M-system-exp} is simply given by taking the matrix exponential\footnote{Here we are assuming that the reader is familiar on how to find the matrix exponential.} of $B_1(T)$ with the initial condition $\tilde M^{(0)}(T_i)=\bs I_{4\times 4}$. After doing several calculations we get

\begin{equation}\label{M0-matrix}
\tilde M_{ij}^{(0)}(T)=
\begin{pmatrix}
1 & 0 & 0 & 0\\
0 & \cos[\mathcal M_\text{F}(T)] & -\sin[\mathcal M_\text{F}(T)] & 0\\
0 & \sin[\mathcal M_\text{F}(T)] & \cos[\mathcal M_\text{F}(T)] & 0\\
0 & 0 & 0 & 1
\end{pmatrix}.
\end{equation}
Now let us consider the $n$ order matrix equation in \eqref{M-system-exp} and let us define $n-1=m$. Next, let us note that the non zero elements of the matrix $B_1(T)$ are $B_{1, (23)}(T)=G_\text{F}(T)$ and $B_{1, (32)}(T)= -G_\text{F}(T)$. On the other hand the non zero elements of the matrix $\epsilon B_2(T)$
are $\epsilon B_{2, (24)}(T)=G_\text{C}(T)$, $\epsilon B_{2, (42)}(T)=-G_\text{C}(T)$, $\epsilon B_{2, (34)}(T)=-\Delta G(T)$ and $\epsilon B_{2, (43)}(T)=\Delta G(T)$. After these consideration, we obtain the following results for the components of the $n$ order matrix equation in \eqref{M-system-exp}
\begin{equation}\label{elem-sys}
\begin{gathered}
\epsilon^{m+1} \tilde M_{1j}^{\prime (m+1)}(T)=0, \\ 
 \epsilon^{m+1} \tilde M_{2j}^{\prime (m+1)}(T)= G_\text{F}(T) \epsilon^{m+1} \tilde M_{3j}^{(m+1)}(T) +  G_\text{C}(T) \epsilon^m \tilde M_{4j}^{(m)}(T),\\
\epsilon^{m+1} \tilde M_{3j}^{\prime (m+1)}(T)= -G_\text{F}(T) \epsilon^{m+1}  \tilde M_{2j}^{(m+1)}(T) - \Delta G(T)\epsilon^m  \tilde M_{4j}^{(m)}(T),\\
\epsilon^{m+1} \tilde M_{4j}^{\prime (m+1)}(T)= -G_\text{C}(T) \epsilon^{m}  \tilde M_{2j}^{(m)}(T) + \Delta G(T) \epsilon^m \tilde M_{3j}^{(m)}(T).
\end{gathered}
\end{equation}

In order to solve the system in \eqref{elem-sys} let us multiply the third equation with the imaginary unit ($i$) and after sum it with the second equation. Then we get
\begin{equation}\label{eq-3}
 \epsilon^{m+1} \left[\tilde M_{2j}^{\prime (m+1)}(T) + i \tilde M_{3j}^{\prime (m+1)}(T) \right]= -i G_\text{F}(T)\left[ \tilde M_{2j}^{ (m+1)}(T) + i \tilde M_{3j}^{ (m+1)}(T) \right] \epsilon^{m+1} + \left[ G_\text{C}(T) - i \Delta G(T) \right] \epsilon^m \tilde M_{4j}^{(m)}(T).
\end{equation}
We may observe that \eqref{eq-3} is a first order non homogeneous linear differential equation for the function\\ $\left(\tilde M_{2j}^{ (m+1)}(T) + i \tilde M_{3j}^{ (m+1)}(T)\right) \epsilon^{m+1} $ and it can be solved with the method of the variation of coefficients. Performing several operation and using the initial conditions $\tilde M_{ij}^{(n)}(T_i)=0$ for $n\geq 1$ we get the following solution
\begin{equation}\label{eq-4}
\left(\tilde M_{2j}^{ (m+1)}(T) + i \tilde M_{3j}^{ (m+1)}(T)\right) \epsilon^{m+1} =-\exp[i \mathcal M_\text{F}(T)] \int_{T}^{T_i} dT^\prime
\left[ G_\text{C}(T^\prime) - i \Delta G(T^\prime) \right] \epsilon^m \tilde M_{4j}^{(m)}(T^\prime)\exp[- i \mathcal M_\text{F}(T^\prime)]. 
\end{equation}
Now by equating the real and imaginary parts of the left hand side of \eqref{eq-4} with the those of the right hand side and directly integrating the first and the fourth equations in \eqref{elem-sys} we get 
\begin{equation}\label{m-order-eq}
\begin{gathered}
\epsilon^{m+1} \tilde M_{1j}^{(m+1)}(T)=0\\
\epsilon^{m+1} \tilde M_{2j}^{ (m+1)}(T) = - \cos[\mathcal M_\text{F}(T)] \int_T^{T_i} dT^\prime \left[ G_\text{C}(T^\prime) \cos[\mathcal M_\text{F}(T^\prime)] - \Delta G(T^\prime) \sin[\mathcal M_\text{F}(T^\prime)] \right] \epsilon^m \tilde M_{4j}^{(m)}(T^\prime)\\
- \sin[\mathcal M_\text{F}(T)] \int_T^{T_i} dT^\prime \left[ G_\text{C}(T^\prime) \sin[\mathcal M_\text{F}(T^\prime)] + \Delta G(T^\prime) \cos[\mathcal M_\text{F}(T^\prime)] \right] \epsilon^m \tilde M_{4j}^{(m)}(T^\prime),\\
\epsilon^{m+1} \tilde M_{3j}^{ (m+1)}(T) =\cos[\mathcal M_\text{F}(T)] \int_T^{T_i} dT^\prime \left[ G_\text{C}(T^\prime) \sin[\mathcal M_\text{F}(T^\prime)] + \Delta G(T^\prime) \cos[\mathcal M_\text{F}(T^\prime)] \right] \epsilon^m \tilde M_{4j}^{(m)}(T^\prime)\\
- \sin[\mathcal M_\text{F}(T)] \int_T^{T_i} dT^\prime \left[ G_\text{C}(T^\prime) \cos[\mathcal M_\text{F}(T^\prime)] - \Delta G(T^\prime) \sin[\mathcal M_\text{F}(T^\prime)] \right] \epsilon^m \tilde M_{4j}^{(m)}(T^\prime),\\
\epsilon^{m+1} \tilde M_{4j}^{ (m+1)}(T) = - \int_T^{T_i} dT^\prime \left[ -G_\text{C}(T^\prime) \epsilon^m \tilde M_{2j}^{(m)}(T^\prime) + \Delta G(T^\prime) \epsilon^m \tilde M_{3j}^{(m)}(T^\prime) \right].
\end{gathered}
\end{equation}

The expressions in \eqref{m-order-eq} allows us to find recursively the elements of $\tilde M_{ij}(T)$ to the order $m+1$ in the case when the elements at the order $m$ are known. Since we already know the elements of $\tilde M_{ij}$ at the order $m=0$ as given in \eqref{M0-matrix}, we can recursively calculate those at the order $m+1$. Let us define for simplicity 
\begin{equation}\label{L-M-def}
\begin{gathered}
L_j^{(m)}(T) \equiv \int_T^{T_i} dT^\prime \left[ G_\text{C}(T^\prime) \cos[\mathcal M_\text{F}(T^\prime)] - \Delta G(T^\prime) \sin[\mathcal M_\text{F}(T^\prime)] \right] \epsilon^m \tilde M_{4j}^{(m)}(T^\prime),\\
K_j^{(m)}(T) \equiv \int_T^{T_i} dT^\prime \left[ G_\text{C}(T^\prime) \sin[\mathcal M_\text{F}(T^\prime)] + \Delta G(T^\prime) \cos[\mathcal M_\text{F}(T^\prime)] \right] \epsilon^m \tilde M_{4j}^{(m)}(T^\prime).
\end{gathered}
\end{equation}
By using the definitions in \eqref{L-M-def} and the expressions in \eqref{m-order-eq} we get the following expressions for $m=0$ of the matrix elements of $\epsilon \tilde M_{ij}^{(1)}$
\begin{equation}\label{M1-matrix}
\epsilon \tilde M_{ij}^{(1)}(T)=
\begin{pmatrix}
0 & 0 & 0 & 0\\
0 & 0 & 0 & -\cos[\mathcal M_\text{F}(T)] L_4^{(0)}(T) - \sin[\mathcal M_\text{F}(T)] K_4^{(0)}(T)\\
0 & 0 & 0 & -\sin[\mathcal M_\text{F}(T)] L_4^{(0)}(T) + \cos[\mathcal M_\text{F}(T)] K_4^{(0)}(T)\\
0 & L_4^{(0)}(T) & -K_4^{(0)}(T) & 0
\end{pmatrix}.
\end{equation}
We can proceed in the same way to find the matrix elements of $\epsilon^2 \tilde M_{ij}^{(2)}$ starting from the elements of $\epsilon \tilde M_{ij}^{(1)}$ given in \eqref{M1-matrix}. However, for our purposes it will be sufficient to consider only the elements of $\tilde M_{ij}$ up to the first order in $\epsilon$. By using the expression $\tilde S_j(T)=\tilde M_{jl}(T) \tilde S_l(T_i)\simeq \left[ \tilde M_{jl}^{(0)}(T) + \epsilon \tilde M_{jl}^{(1)}(T) \right] \tilde S_l(T_i)$, we get for the elements of the Stokes vector the following expressions
\begin{equation}\label{M-ele-1}
\begin{gathered}
\tilde I(T)=\tilde I_i, \\
\tilde Q(T) = \cos[\mathcal M_\text{F}(T)] \tilde Q_i - \sin[\mathcal M_\text{F}(T)] \tilde U_i - \left( \cos[\mathcal M_\text{F}(T)] L_4^{(0)}(T) + \sin[\mathcal M_\text{F}(T)] K_4^{(0)}(T) \right) \tilde V_i, \\
\tilde U(T) = \sin[\mathcal M_\text{F}(T)] \tilde Q_i + \cos[\mathcal M_\text{F}(T)] \tilde U_i - \left( \sin[\mathcal M_\text{F}(T)] L_4^{(0)}(T) - \cos[\mathcal M_\text{F}(T)] K_4^{(0)}(T) \right) \tilde V_i,\\
\tilde V(T)= L_4^{(0)}(T) \tilde Q_i - K_4^{(0)}(T) \tilde U_i +\tilde V_i.
\end{gathered}
\end{equation}
It is worth to remind that the expressions found in \eqref{M-ele-1} are valid for $\mathcal M_\text{F}(T)\neq 0$ or when $\sin(\Theta)\sin(\Phi) \neq 0$.

\subsection{Another perturbative solution}
\label{subsec:4.3}

In the previous two sections we found perturbative solutions in the case when $|\mathcal M_\text{C}(T_0)|< 1, |\mathcal M_\text{F}(T_0)|<1$ and $\Delta \mathcal M(T_0)<1$, namely in sec. \ref{subsec:4.1} and when $|2 M_\text{F}|\gg |\Delta M|, |2 M_\text{C}|$ with $|\mathcal M_\text{C}(T_0)|< 1$ and $\Delta \mathcal M(T_0)<1$ in sec. \ref{subsec:4.2}. Especially in sec. \ref{subsec:4.2} we worked under the hypothesis that $|\sin(\Theta) \sin(\Phi)| \neq 0$ which essentially means non vanishing Faraday effect. In this section we show another perturbative approach based on the Neumann series where the condition $|\sin(\Theta) \sin(\Phi)| \neq 0$ does not necessarily has to hold and that it came out as a byproduct of use of the regular perturbation theory only. In order to do so, let us use the same notations as in sec. \ref{subsec:4.2} and split the matrix $B(T)=B_1(T)+ B_2(T)$ where here we do not specify which matrix is the perturbation matrix. In the case when the magnetic field is completely longitudinal, namely when it is present only the Faraday effect, we have that the solution of the equation $\tilde S^\prime(T)=B(T) \tilde S(T)$ is given by
$\tilde S(T)=\exp\left[-\int_T^{T_i} dT^\prime B_1(T^\prime) \right] \tilde S(T_i)$ since $B_1(T)$ commutes with itself for different temperatures. In the case when the magnetic field has also a transverse component, namely when $B_2(T)\neq 0$, let us define $\tilde S(T) \equiv \exp\left[-\int_T^{T_i} dT^\prime B_1(T^\prime) \right] \bar S(T)$. Then the equation $\tilde S^\prime(T)=[B_1(T) + B_2(T)] \tilde S(T)$ becomes
\begin{equation}\label{an-eq}
\bar S^\prime(T)=\exp\left[\int_T^{T_i} dT^\prime B_1(T^\prime) \right] B_2(T) \exp\left[-\int_T^{T_i} dT^\prime B_1(T^\prime) \right] \bar S(T).
\end{equation}

As we may note, so far we did not make any assumption on the matrix $B_1(T)$ and in principle it can be also a null matrix depending on the situation. After doing some lengthy calculations we get
\begin{equation}\label{bar-M}
\bar M(T)\equiv \exp\left[\int_T^{T_i} dT^\prime B_1(T^\prime) \right] B_2(T) \exp\left[-\int_T^{T_i} dT^\prime B_1(T^\prime)  \right] =
\begin{pmatrix}
0 & 0 & 0 & 0\\
0 & 0 & 0 & - L_4^{\prime (0)}(T) \\
0 & 0 & 0 &  K_4^{\prime (0)}(T)\\
0 & L_4^{\prime (0)}(T) & -K_4^{\prime (0)}(T) & 0
\end{pmatrix}.
\end{equation}
We can solve Eq. \eqref{an-eq} as a convergent Neumann series as we did in sec. \ref{subsec:4.2} as far as $\left|\int_{T_i}^T dT^\prime \bar M_{ij}(T^\prime) \right|<1$. By using the matrix expression \eqref{bar-M} in \eqref{an-eq}, we get the following solution for $\bar S(T)$ up to the first order 
\begin{equation}\label{bar-S-sol}
\bar S(T)= \left[ \bs I_{4\times 4} -\int_{T}^{T_i} dT^\prime \bar M(T^\prime) +...\right] \bar S(T_i) \simeq 
\begin{pmatrix}
1 & 0 & 0 & 0\\
0 & 1 & 0 &  -L_4^{ (0)}(T) \\
0 & 0 & 1 & K_4^{ (0)}(T)\\
0 & L_4^{ (0)}(T) & -K_4^{ (0)}(T) & 1
\end{pmatrix} \bar S(T_i).
\end{equation}
Now by returning back to the to the components of $\tilde S(T) \equiv \exp\left[-\int_T^{T_i} dT^\prime B_1(T^\prime) \right] \bar S(T)$ we get the following solution for $\tilde S(T)$ up to first order (for  solutions up to the second order in perturbation theory of $\tilde S(T)$ see expressions \eqref{second-order-sol} in appendix \ref{app:1})
\begin{equation}\label{tilde-S-sol}
\tilde S(T)\simeq 
\begin{pmatrix}
1 & 0 & 0 & 0\\
0 & \cos[\mathcal M_\text{F}(T)] & -\sin[\mathcal M_\text{F}(T)] &  -\cos[\mathcal M_\text{F}(T)] L_4^{(0)}(T) - \sin[\mathcal M_\text{F}(T)] K_4^{(0)}(T)  \\
0 & \sin[\mathcal M_\text{F}(T)] & \cos[\mathcal M_\text{F}(T)] & - \sin[\mathcal M_\text{F}(T)] L_4^{(0)}(T) + \cos[\mathcal M_\text{F}(T)] K_4^{(0)}(T) \\
0 & L_4^{ (0)}(T) & -K_4^{ (0)}(T) & 1
\end{pmatrix} \tilde S(T_i).
\end{equation}
We may note that the solution \eqref{tilde-S-sol} exactly coincides with the solutions found in \eqref{M-ele-1} which we found by using the regular perturbation theory.

The solution \eqref{tilde-S-sol} has been found without any restriction on the magnitude and sign of $\mathcal M_\text{F}(T)$ which is different from the result of Sec. \ref{subsec:4.2} where we worked under the assumption that $\mathcal M_\text{F}(T)\neq 0$, which for fixed and non zero values of $B_{e0}, \nu_0$ and $T$, is equivalent to $\sin(\Theta)\sin(\Phi)\neq 0$. This fact tells us that the condition on the Faraday effect term $\mathcal M_\text{F}(T)\neq 0$ is not necessary in order to find the solution \eqref{tilde-S-sol} and that the condition $\mathcal M_\text{F}(T)\neq 0$ comes out only in the regular perturbation theory. On the other hand, in this section in order to use the Neumann series expansion we required that $\left|\int_{T_i}^T dT^\prime \bar M_{ij}(T^\prime) \right|<1$ or equivalently that $| L_4^{(0)}(T)|<1$ and $| K_4^{(0)}(T)|<1$. For example we may note 
\begin{equation}\label{ineq-1}
\begin{gathered}
|L_4^{(0)}(T)|= \left|\int_T^{T_i} dT^\prime \left[ G_\text{C}(T^\prime) \cos[\mathcal M_\text{F}(T^\prime)] - \Delta G(T^\prime) \sin[\mathcal M_\text{F}(T^\prime)] \right] \right| \\
\leq \int_T^{T_i} dT^\prime \left| \left[ G_\text{C}(T^\prime) \cos[\mathcal M_\text{F}(T^\prime)] - \Delta G(T^\prime) \sin[\mathcal M_\text{F}(T^\prime)] \right] \right| \leq \int_T^{T_i} dT^\prime  \left[ \left| G_\text{C}(T^\prime) \cos[\mathcal M_\text{F}(T^\prime)] \right| + \left| \Delta G(T^\prime) \sin[\mathcal M_\text{F}(T^\prime)] \right|\right] \\
\leq \int_T^{T_i} dT^\prime  \left[ \left| G_\text{C}(T^\prime) \right| + \left| \Delta G(T^\prime) \right|\right].
\end{gathered}
\end{equation}
Thus if we require that $\int_T^{T_i} dT^\prime  \left[ \left| G_\text{C}(T^\prime) \right| + \left| \Delta G(T^\prime) \right|\right] <1$ by \eqref{ineq-1} we also must have that $|L_4^{(0)}(T)|<1$. Similar result can be also found from the condition $|K_4^{(0)}(T)|<1$. Obviously the requirement  $\int_T^{T_i} dT^\prime  \left[ \left| G_\text{C}(T^\prime) \right| + \left| \Delta G(T^\prime) \right|\right] <1$ is much stronger than $|L_4^{(0)}(T)|<1$. Without going into details, one can show based on \eqref{C-M-cond}, as far as the stronger condition on $|\Delta \mathcal M(T)|<2/3$ is satisfied, also $\int_T^{T_i} dT^\prime  \left[ \left| G_\text{C}(T^\prime) \right| + \left| \Delta G(T^\prime) \right|\right] <1$  is satisfied. Thus, in order for the last inequality in \eqref{ineq-1} to be less than unity, we must have the stronger condition on $|\Delta \mathcal M(T)|<2/3$ and not the stronger condition on $|\Delta \mathcal M(T)|<1$ as in \eqref{C-M-cond}. This implies that we must impose slightly stronger constraints on the parameters $\nu_0$ and $B_{e0}$ in order to satisfy the last inequality in \eqref{ineq-1}.


\section{Degree of circular polarization}
\label{sec:5}

In the previous section we found perturbative solutions of the equations of motion of the Stokes parameters in two different regimes by using perturbation theory. In this section, we focus our attention on the generation of circular polarization, where in specific, we study the expected degree of circular polarization at present time and the expected rotation angle of the CMB polarization plane. We separate our analysis by first studying the solutions found in Sec. \ref{subsec:4.1} and second, study those found in Sec. \ref{subsec:4.2}. In what follows, we consider the evolution of the CMB polarization and rotation angle of the polarization plane starting from the decoupling epoch at the temperature $T=T_i$ until at the present time at the temperature $T=T_0$. Moreover, we consider the CMB at the decoupling epoch partially polarized where it acquires only a linear polarization due to the Thomson scattering off the CMB photons on electrons with no initial circular polarization, namely $Q_i\neq 0, U_i\neq 0$ and $V_i=0$ as studied in Ref. \cite{Ejlli:2016avx}.

\subsection{Case when $|\mathcal M_\text{F}(T_0|< 1, |\mathcal M_\text{C}(T_0)|< 1,|\Delta \mathcal M(T_0)|< 1$.}
\label{subsec:5.1}

Let us consider first the generation of the circular polarization and calculate its degree of polarization at present time where $T=T_0$ for  $|\mathcal M_\text{F}(T_0|< 1, |\mathcal M_\text{C}(T_0|< 1$ and $|\Delta \mathcal M(T_0)|< 1$. By using the expression for the Stokes parameters $\tilde I(T)$ and $\tilde V(T)$ found in \eqref{Stokes-1}, the degree of circular polarization of the CMB at present is defined
\begin{equation}\label{circ-pol-1}
P_C(T_0) \equiv \frac{\left|\tilde V(T_0)\right|}{\tilde I(T_0)}= \frac{\left | \tilde M_{42}(T_0) \tilde Q_i + \tilde M_{43}(T_0)\tilde U_i+\tilde M_{44}(T_0)\tilde V_i \right |}{\tilde I_i}.
\end{equation}
It is quite convenient at this stage to normalize the CMB intensity at the decoupling time to unity $\tilde I_i=1$. In addition, we have that $\tilde I_i=I_i, \tilde Q_i=Q_i, \tilde U_i=U_i$ and $\tilde V_i=V_i$. In what follows, we assume that $V_i=0$ if not specified otherwise. To calculate $P_C(T_0)$ we need to calculate explicitly the matrix elements $\tilde M_{42}(T_0)$ and $\tilde M_{43}(T_0)$. For the first term entering in $\tilde M_{42}(T_0)$ we have
\begin{equation}\label{A}
\mathcal M_\text{C}(T_0)=2.69\times 10^{38} \left(\frac{\text{Hz}}{\nu_0}\right)^3\left( \frac{B_{e0}}{\text{G}}\right)^2 T_0^{-3/2} \, \sin(2\Theta)\cos(\Phi) \quad (\text{K}^{3/2}),
\end{equation}
while for the second term we have
\begin{equation}
\begin{gathered}
\int_{T_0}^{T_i} dT^\prime \Delta G(T^\prime) \int_{T^\prime}^{T_i} dT^{\prime\prime} G_\text{F}(T^{\prime\prime})= -1.05\times 10^{58} \left[ \sin^3(\Theta)\sin(\Phi)\cos^2(\Phi)-\sin(\Theta)\cos^2(\Theta)\sin(\Phi) \right] \left(\frac{\text{Hz}}{\nu_0}\right)^5\left( \frac{B_{e0}}{\text{G}}\right)^3 T_0^{-2}\\
\times \int_{T_0}^{T_i} dT^\prime X_e(T^\prime) T^{\prime 3/2} \int_{T^\prime}^{T_i} dT^{\prime\prime} X_e(T^{\prime\prime}) T^{\prime\prime 1/2} \quad (\text{K}^{-2}) \simeq -3.63\times 10^{67}\left[ \sin^3(\Theta)\sin(\Phi)\cos^2(\Phi)-\sin(\Theta)\cos^2(\Theta)\sin(\Phi) \right] \\ \times \left(\frac{\text{Hz}}{\nu_0}\right)^5\left( \frac{B_{e0}}{\text{G}}\right)^3 T_0^{-2} \quad (\text{K}^{2}), \end{gathered}
\end{equation}
where we used the numerical value of $\int_{T_0}^{T_i} dT^\prime X_e(T^\prime) T^{\prime 3/2} \int_{T^\prime}^{T_i} dT^{\prime\prime} X_e(T^{\prime\prime}) T^{\prime\prime 1/2}\simeq 3.46\times 10^{9}$ (K$^4$) for $T_0=2.725$ K and $T_i=2970$ K. On the other hand, for the first term in the matrix element $\tilde M_{43}(T_0)$ we get
\begin{equation}
\Delta \mathcal M(T_0)= - 5.38\times 10^{38} \left(\frac{\text{Hz}}{\nu_0}\right)^3\left( \frac{B_{e0}}{\text{G}}\right)^2 T_0^{-3/2} \,\left[\sin^2(\Theta)\cos^2(\Phi) -\cos^2(\Theta)\right] \quad (\text{K}^{3/2}),
\end{equation}
while the second term in $\tilde M_{43}(T_0)$ is given by
\begin{equation}\label{D}
\begin{gathered}
\int_{T}^{T_i} dT^\prime G_\text{C}(T^\prime) \int_{T^\prime}^{T_i} dT^{\prime\prime} G_\text{F}(T^{\prime\prime}) \simeq 1.82\times 10^{67}\left[ \sin(2\Theta)\sin(\Theta) \sin(\Phi)\cos(\Phi) \right] \left(\frac{\text{Hz}}{\nu_0}\right)^5\left( \frac{B_{e0}}{\text{G}}\right)^3 T_0^{-2} \quad (\text{K}^{2}).
\end{gathered}
\end{equation}

Since all terms in $\tilde M_{42}$ and $\tilde M_{43}$ depend on the angles $\Theta$ and $\Phi$ and because some terms in $\tilde M_{42}$ and $\tilde M_{43}$ are equal to zero when averaged over the angles $\Theta$ and $\Phi$, it is more convenient to calculate the root mean square of the degree of circular polarization instead of the mean value. By using the expressions \eqref{A}-\eqref{D} in $\tilde M_{42}$ and $\tilde M_{43}$ we get (by suppressing for the moment the units)
\begin{equation}\label{av-deg-pol}
\begin{gathered}
P_C^\text{rms}(T_0) \equiv \left\langle P_C^{2}(T_0) \right \rangle^{1/2}= \left\langle \tilde M_{42}^2(T_0) Q_i^2 + \tilde M_{43}^2(T_0) U_i^2 + 2 \tilde M_{42}(T_0) \tilde M_{43}(T_0) Q_i U_i \right\rangle^{1/2}=\\
\left[ \left( 7.23\times 10^{-14}\left( \frac{1}{4} \right) \left(\frac{\text{GHz}}{\nu_0}\right)^6 \left( \frac{B_{e0}}{\text{nG}}\right)^4 T_0^{-3} + 1.31\times 10^{-9} \left( \frac{9}{256} \right) \left(\frac{\text{GHz}}{\nu_0}\right)^{10}\left( \frac{B_{e0}}{\text{nG}}\right)^6 T_0^{-4} \right) Q_i^2  + \right. \\ \left.  \left( 2.89\times 10^{-13}\left( \frac{25}{64} \right) \left(\frac{\text{GHz}}{\nu_0}\right)^6 \left( \frac{B_{e0}}{\text{nG}}\right)^4 T_0^{-3} + 3.31\times 10^{-10} \left( \frac{1}{32} \right) \left(\frac{\text{GHz}}{\nu_0}\right)^{10}\left( \frac{B_{e0}}{\text{nG}}\right)^6 T_0^{-4} \right) U_i^2 \right]^{1/2},
\end{gathered}
\end{equation}
where $\left \langle \tilde M_{42}(T_0) \tilde M_{43}(T_0) \right \rangle =0$ and the average value on the angles of the mixed terms in $\tilde M_{42}^2(T_0)$ and $\tilde M_{43}^2(T_0)$ are identically equal to zero. In obtaining the expression \eqref{av-deg-pol} we used the following results of integrals of trigonometric functions
\begin{equation}\nonumber
\begin{gathered}
\int_0^\pi \int_0^{2\pi} d\Theta d\Phi \left[ \sin(2\Theta) \cos(\Phi) \right]^2=\frac{\pi^2}{2}, \quad \int_0^\pi \int_0^{2\pi} d\Theta d\Phi \left[ \sin(2\Theta) \sin(\Theta) \sin(\Phi)\cos(\Phi) \right]^2=\frac{\pi^2}{16}, \\
\int_0^\pi \int_0^{2\pi} d\Theta d\Phi \left[ \sin^3(\Theta)\sin(\Phi) \cos^2(\Phi) -\sin(\Theta)\cos^2(\Theta) \sin(\Phi) \right]^2=\frac{9 \pi^2}{128}, \\
  \int_0^\pi \int_0^{2\pi} d\Theta d\Phi \left[ \sin^2(\Theta) \cos^2(\Phi) -\cos^2(\Theta) \right]^2=\frac{25 \pi^2}{32},\\
\int_0^\pi \int_0^{2\pi} d\Theta d\Phi \left[ \sin(2\Theta) \cos(\Phi) \right] \left[ \sin^3(\Theta)\sin(\Phi) \cos^2(\Phi) -\sin(\Theta)\cos^2(\Theta) \sin(\Phi) \right]=0,\\
 \int_0^\pi \int_0^{2\pi} d\Theta d\Phi \left[ \sin^2(\Theta) \cos^2(\Phi) -\cos^2(\Theta) \right] \left[ \sin(2\Theta) \sin(\Theta) \sin(\Phi)\cos(\Phi) \right] =0,\\
 \int_0^\pi \int_0^{2\pi} d\Theta d\Phi \left[ \sin(2\Theta) \cos(\Phi)\right] \left[ \sin^2(\Theta) \cos^2(\Phi) - \cos^2(\Theta)\right]=0,\\
 \int_0^\pi \int_0^{2\pi} d\Theta d\Phi \left[ \sin(2\Theta) \cos(\Phi)\right] \left[ \sin(2\Theta) \sin(\Theta) \sin(\Phi)\cos(\Phi) \right]=0,\\
 \int_0^\pi \int_0^{2\pi} d\Theta d\Phi \left[ \sin^3(\Theta)\sin(\Phi) \cos^2(\Phi) - \sin(\Theta)\cos^2(\Theta) \sin(\Phi) \right] \left[ \sin(2\Theta) \sin(\Theta) \sin(\Phi)\cos(\Phi) \right]=0,\\
 \int_0^\pi \int_0^{2\pi} d\Theta d\Phi \left[ \sin^3(\Theta)\sin(\Phi) \cos^2(\Phi) - \sin(\Theta)\cos^2(\Theta) \sin(\Phi) \right] \left[ \sin^2(\Theta) \cos^2(\Phi) - \cos^2(\Theta) \right]=0.
       \end{gathered}
\end{equation}
One important thing about expression \eqref{av-deg-pol} is that the second terms proportional to $Q_i^2$ and $U_i^2$ must be in magnitude smaller that the first terms. The reason is because these terms correspond to second order terms in perturbation theory where their magnitudes must be smaller than the first order terms in order to have a convergent series. This fact implies that care must be used in order to choose the values of $B_{e0}$ and $\nu_0$ in order to evaluate $P_C^\text{rms}(T_0)$. However, since we are in the regime where the constraints \eqref{F-C-M-cond} must be satisfied, usually there is not reason to worry about since the values of the parameters $\nu_0$ and $B_{e0}$ that satisfy \eqref{F-C-M-cond}  automatically keep the magnitudes of the second order terms smaller than the first ones.

\begin{figure*}[h!]
\centering
\mbox{
\subfloat[\label{fig:Fig3}]{\includegraphics[scale=0.65]{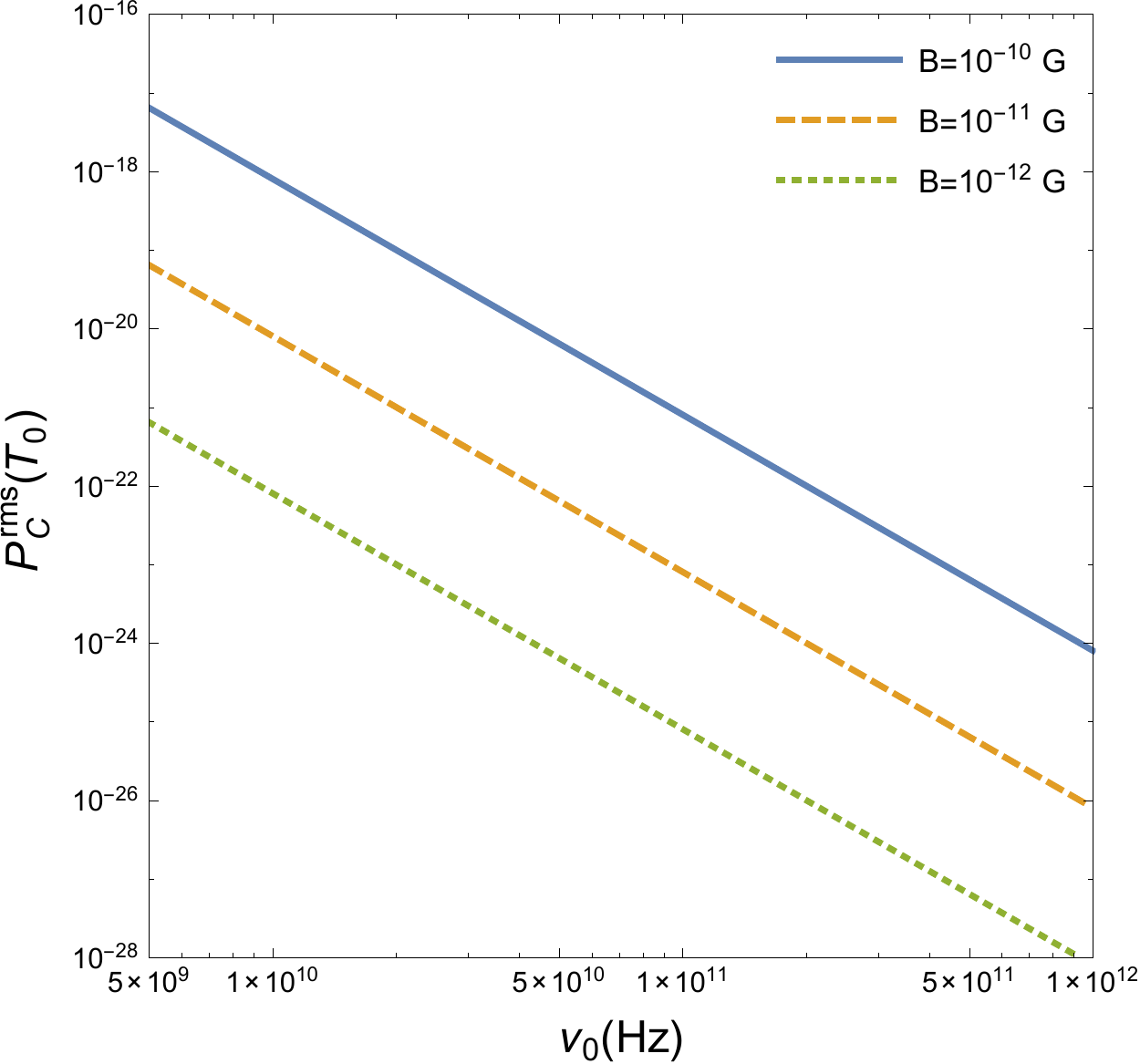}}\qquad
\subfloat[\label{fig:Fig4}]{\includegraphics[scale=0.65]{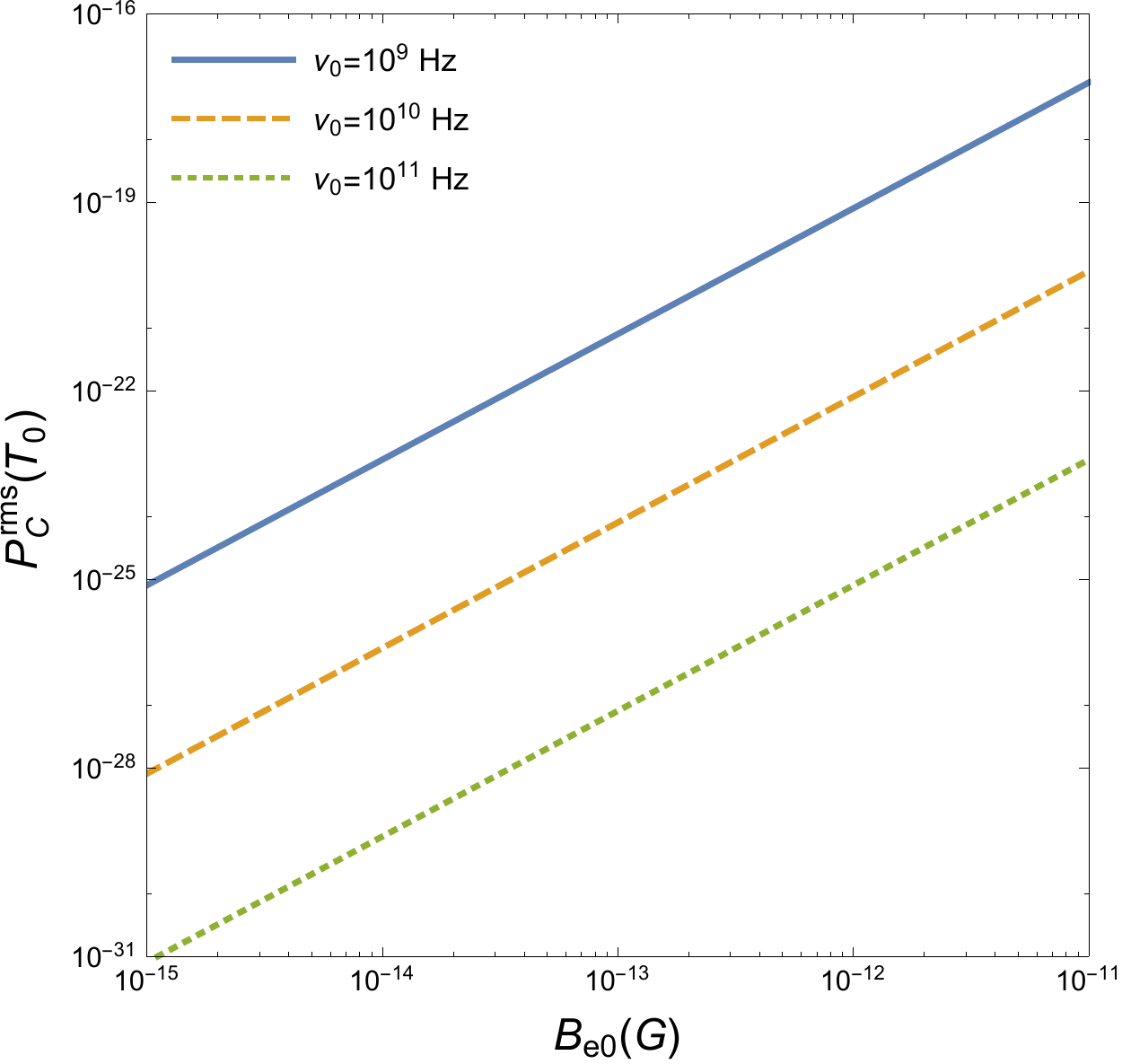}}}
\caption{(a) Logarithmic scale plots of the root mean square of the degree of circular polarization  $P_C^\text{rms}(T_0)$ as a function of the CMB frequency $\nu_0$ for values of $B_{e0}=10^{-10}$ G, $B_{e0}=10^{-11}$ G and $B_{e0}=10^{-12}$ G are shown.  (b) Logarithmic scale plots of the root mean square of the degree of circular polarization  $P_C^\text{rms}(T_0)$ as a function of the magnetic field amplitude $B_{e0}$ for values of the CMB frequencies $\nu_{0}=10^{9}$ Hz, $\nu_{0}=10^{10}$ Hz and $\nu_{0}=10^{11}$ Hz are shown. In both plots (a) and (b) we used a value of $|r|=1$ and $|Q_i |= 10^{-6}$ where $r \equiv U_i/Q_i$.}
\label{fig:Fig2a}
\end{figure*}

\begin{figure*}[h!]
\centering
\mbox{
\subfloat[\label{fig:Fig5}]{\includegraphics[scale=0.65]{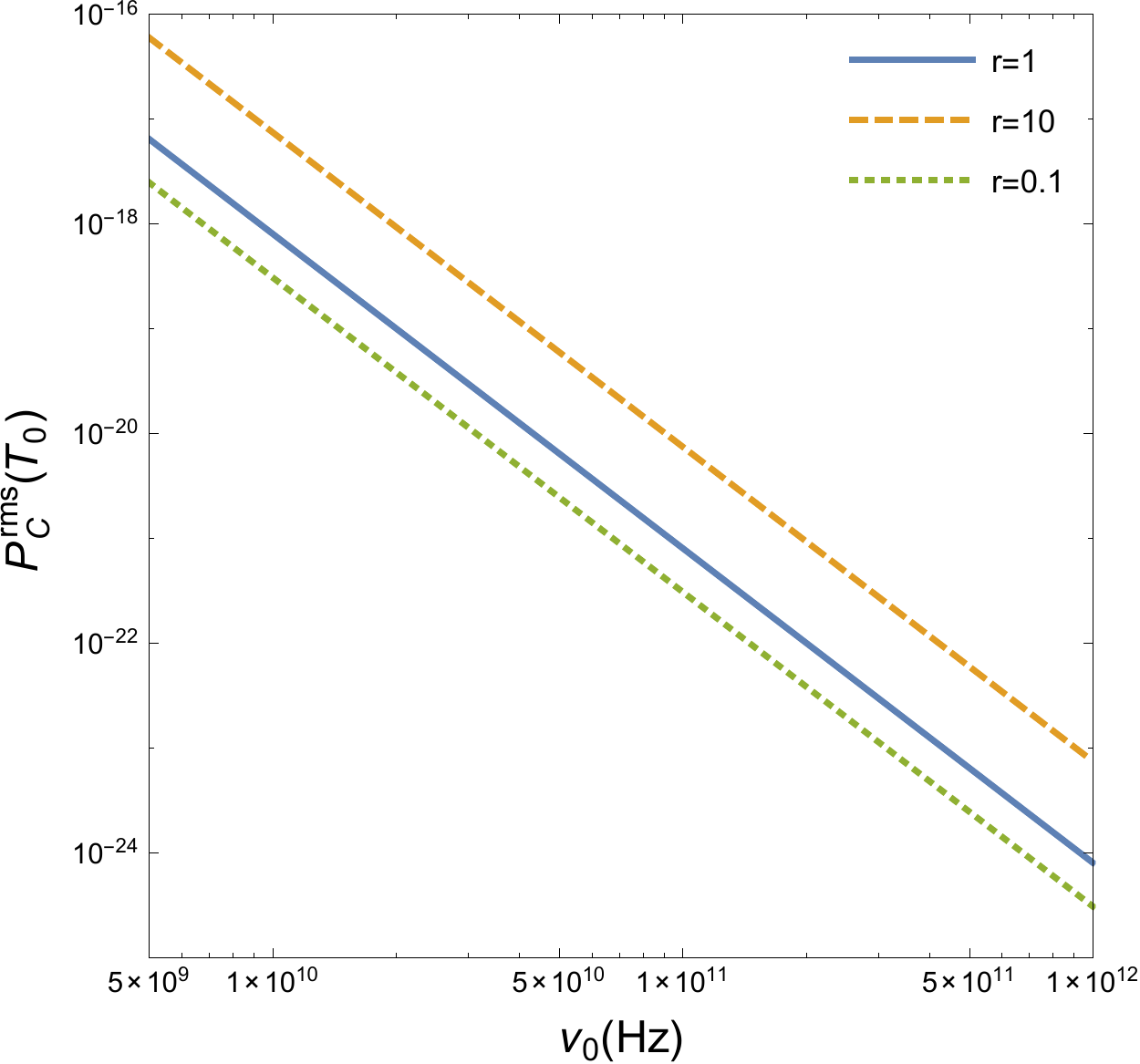}}\qquad
\subfloat[\label{fig:Fig6}]{\includegraphics[scale=0.65]{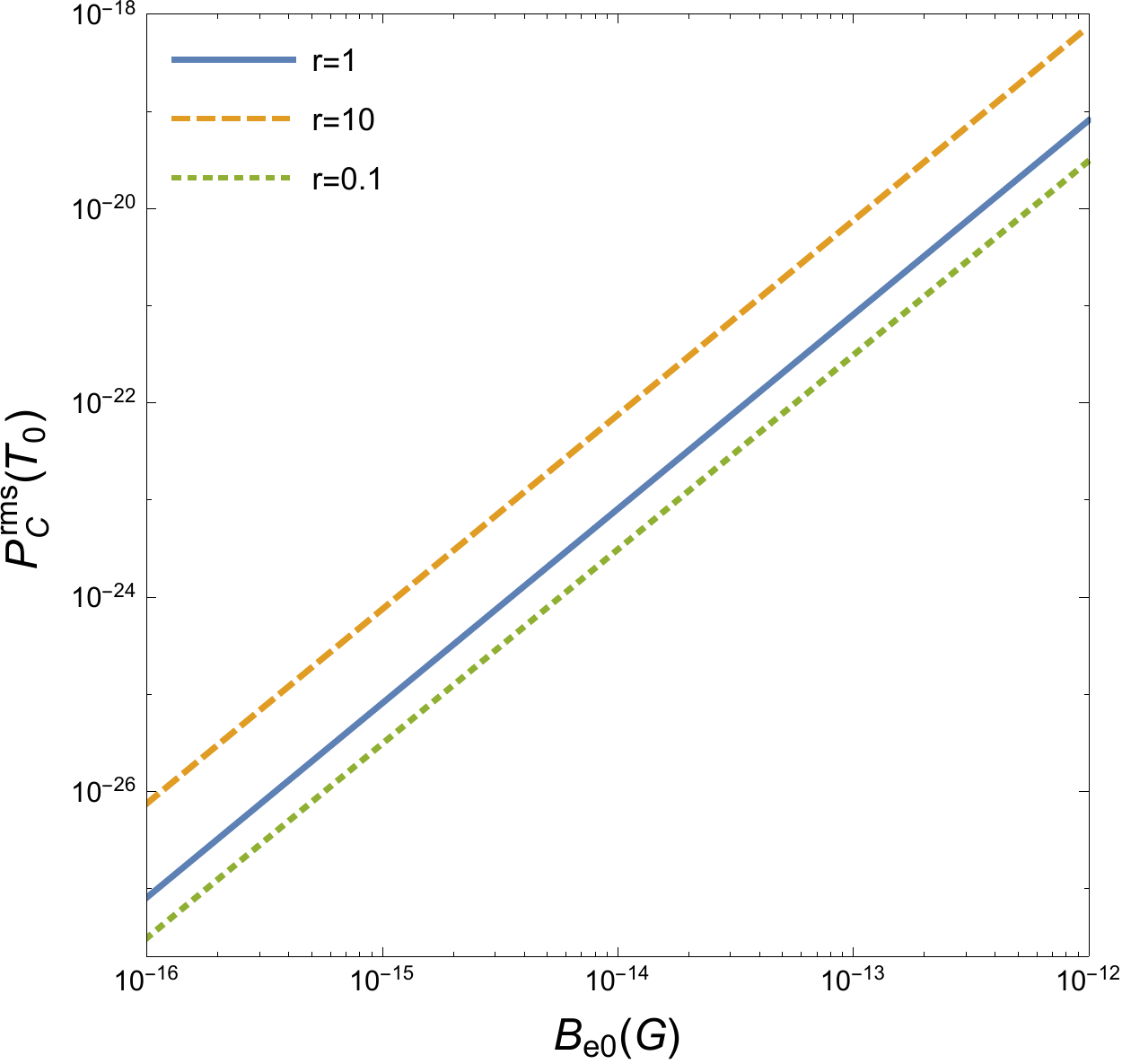}}}
\caption{(a) Logarithmic scale plots of the root mean square of the degree of circular polarization  $P_C^\text{rms}(T_0)$ as a function of the CMB frequency $\nu_0$ for a value of $B_{e0}=10^{-10}$ G and $|r|= 0.1, 1, 10$ are shown.  (b) Logarithmic scale plots of the root mean square of the degree of circular polarization  $P_C^\text{rms}(T_0)$ as a function of the magnetic field amplitude $B_{e0}$ for a value of the CMB frequency $\nu_{0}=10^{9}$ Hz and $|r|= 0.1, 1, 10$ (where $r \equiv U_i/Q_i$) are shown. In both plots we used a value of $|Q_i|= 10^{-6}$.}
\label{fig:Fig3a}
\end{figure*}

In Figs. \ref{fig:Fig2a} and \ref{fig:Fig3a} plots of the root mean square of the degree of circular polarization  $P_C^\text{rms}(T_0)$ as functions of the magnetic field amplitude $B_{e0}$ and $\nu_0$ are shown. In obtaining the plots we used the expression \ref{av-deg-pol} where we expressed $U_i=r Q_i$ with $r$ being a parameter which can have either sign and which value is not a priori known. In addition, we have chosen those values of $B_{e0}$ and $\nu_0$ that satisfy the constraints \eqref{F-C-M-cond}. Usually if the stronger constraint on the Faraday effect term is satisfied, namely the first constraint on the left hand side in \eqref{F-C-M-cond}, the remaining two stronger constraints which arise from $|\mathcal M_\text{C}(T_0)|<1$ and $|\Delta \mathcal M(T_0)|<1$ are also satisfied. We may observe from Figs. \ref{fig:Fig2a} and \ref{fig:Fig3a} that $P_C^\text{rms}(T_0)$ is usually a very small quantity where it gets bigger values for smaller values of $\nu_0$ and bigger values of $B_{e0}$. The main reason why $P_C^\text{rms}(T_0)$ is small is because we are working under the constraints \eqref{F-C-M-cond} where there is not too much choice on the values of $\nu_0$ and $B_{e0}$ which would give much large values of $P_C^\text{rms}(T_0)$. The main reason for this situation is because of the constraint $0<|\mathcal M_\text{F}(T_0)|<1$ gives very tight constraints on $\nu_0$ and $B_{e0}$.

\subsection{Case when $|\mathcal M_\text{F}(T_0)|= 0$ and $|\mathcal M_\text{C}(T_0)|< 1,|\Delta \mathcal M(T_0)|< 1$.}
\label{subsec:5.2}

In the case when $|\mathcal M_\text{F}(T_0)|= 0$ and $|\mathcal M_\text{C}(T_0)|< 1,|\Delta \mathcal M(T_0)|< 1$, the constraints on $\nu_0$ and $B_{e0}$ are much less stringent than in the previous section. In fact, for finite values of $\nu_0$ and $B_{e0}$ which interest us, the only possibility for the condition $|\mathcal M_\text{F}(T_0)|= 0$ to hold is only when $|\sin(\Theta)\sin(\Phi)|= 0$ which occurs either when $\Theta= n \pi $ or $\Phi= n \pi$ with $n\geq 0$. In both cases the direction of the magnetic field is perpendicular to the direction of photon propagation where $\mathcal M_\text{F}(T_0)=\mathcal M_\text{C}(T_0)= 0$. Consequently, the constraints in \eqref{F-C-M-cond} reduce to only the constraint $\left(\text{Hz}/\nu_0\right)^3 \left( B_{e0}/\text{G} \right)^2 < 8.35 \times 10^{-39}$ which correspond to the stronger constraint on $|\Delta \mathcal M(T_0)|<1$, namely the region within the black line in \ref{fig:Fig2}.

\begin{figure*}[h!]
\centering
\mbox{
\subfloat[\label{fig:Fig7}]{\includegraphics[scale=0.65]{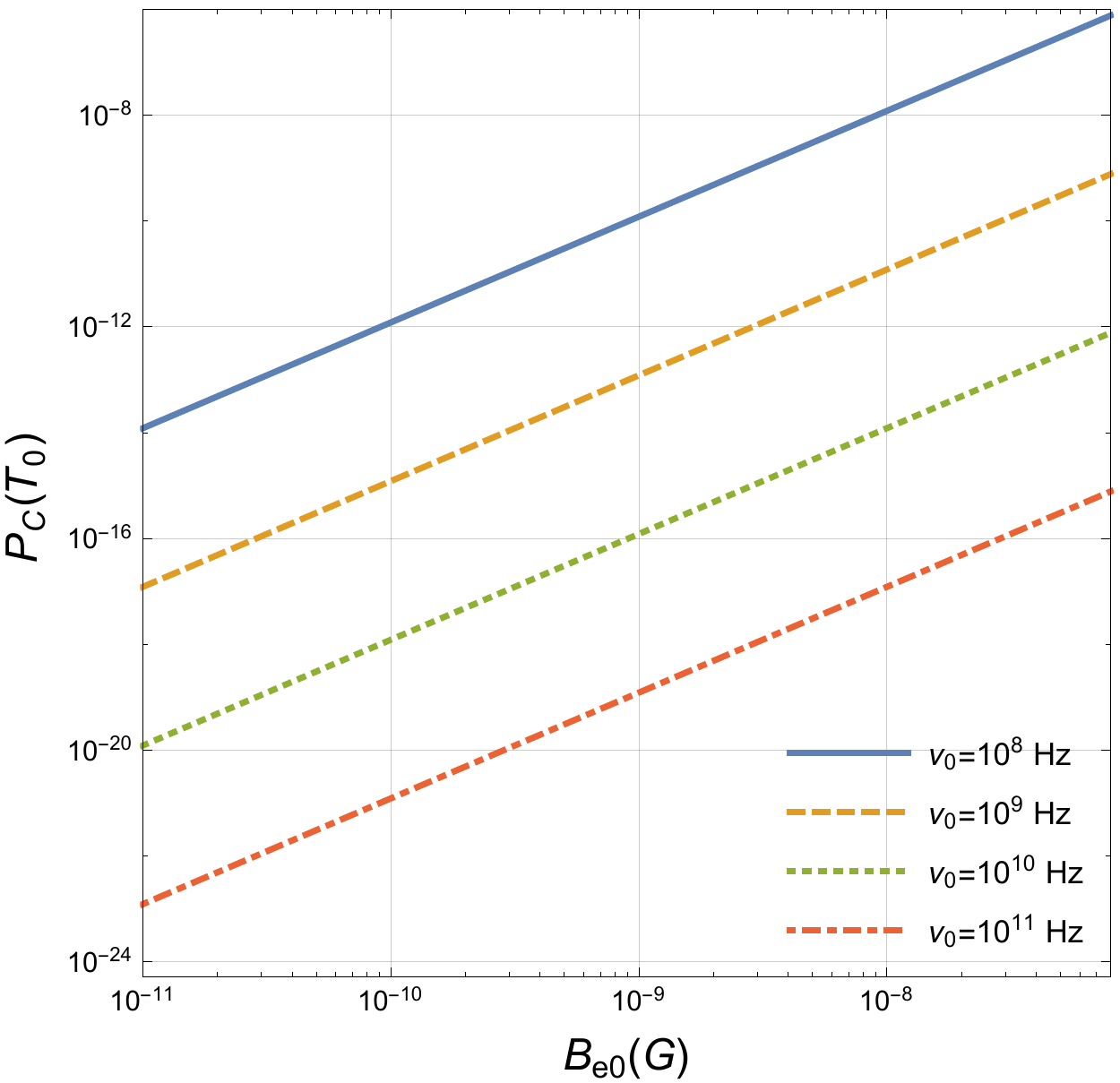}}\qquad
\subfloat[\label{fig:Fig8}]{\includegraphics[scale=0.65]{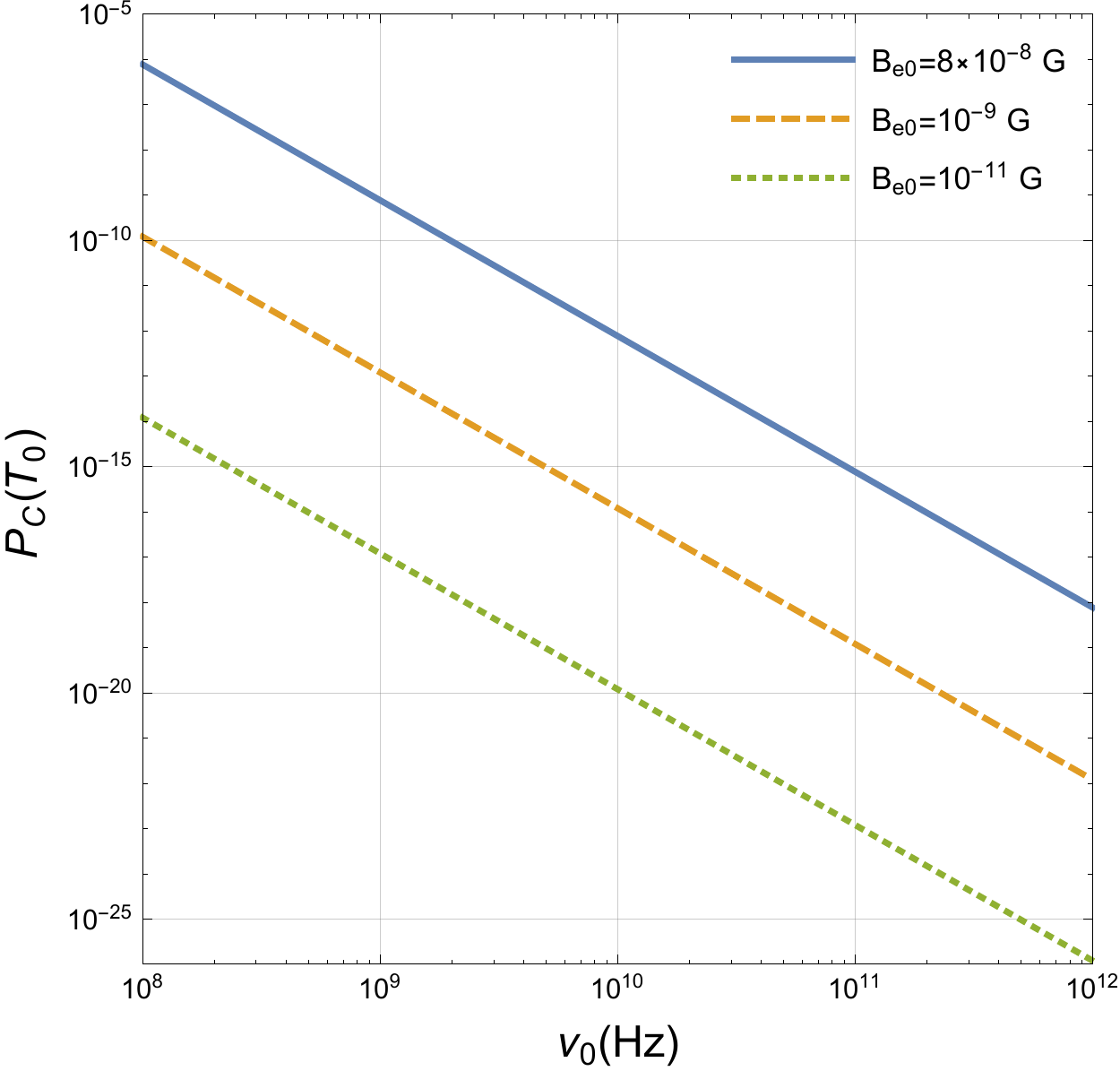}}}
\caption{ (a) Logarithmic scale plots of the degree of circular polarization  $P_C(T_0)$ as a function of the magnetic field amplitude $B_{e0}$ for values of the CMB frequencies $\nu_{0}=10^{8}$ Hz, $\nu_0=10^9$ Hz, $\nu_{0}=10^{10}$ Hz and $\nu_{0}=10^{11}$ Hz are shown. (b) Logarithmic scale plots of the degree of circular polarization  $P_C(T_0)$ as a function of the CMB frequency $\nu_0$ for values of $B_{e0}=8\times 10^{-8}$ G, $B_{e0}=10^{-9}$ G and $B_{e0}=10^{-11}$ G are shown. In both plots (a) and (b) we used a value of $|r|=1$ and $|Q_i|= 10^{-6}$ where $r \equiv U_i/Q_i$. The grey lines in both (a) and (b) are simply grid lines.}
\label{fig:Fig4a}
\end{figure*}

\begin{figure*}[h!]
\centering
\mbox{
\subfloat[\label{fig:Fig9}]{\includegraphics[scale=0.65]{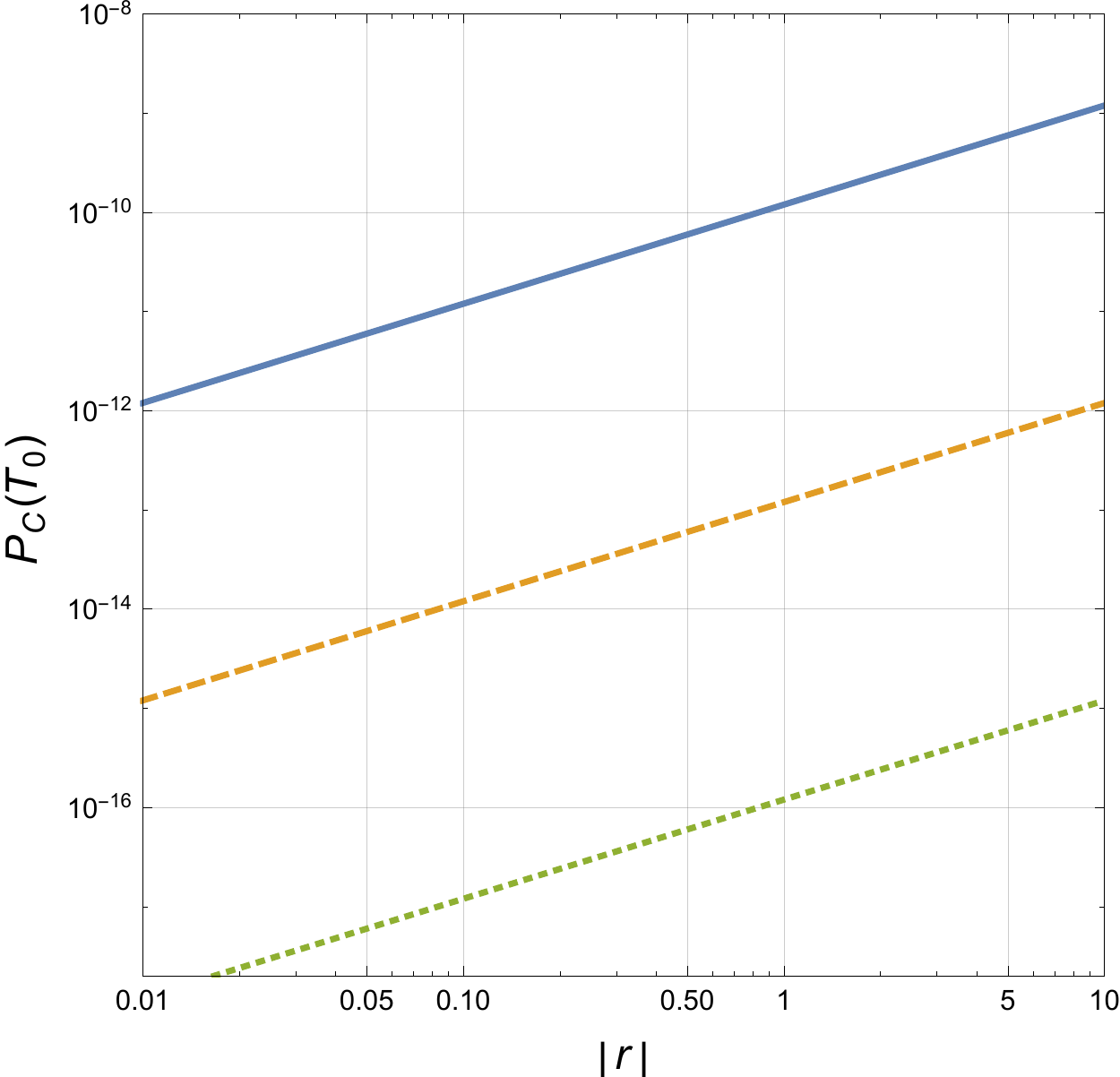}}\qquad
\subfloat[\label{fig:Fig10}]{\includegraphics[scale=0.65]{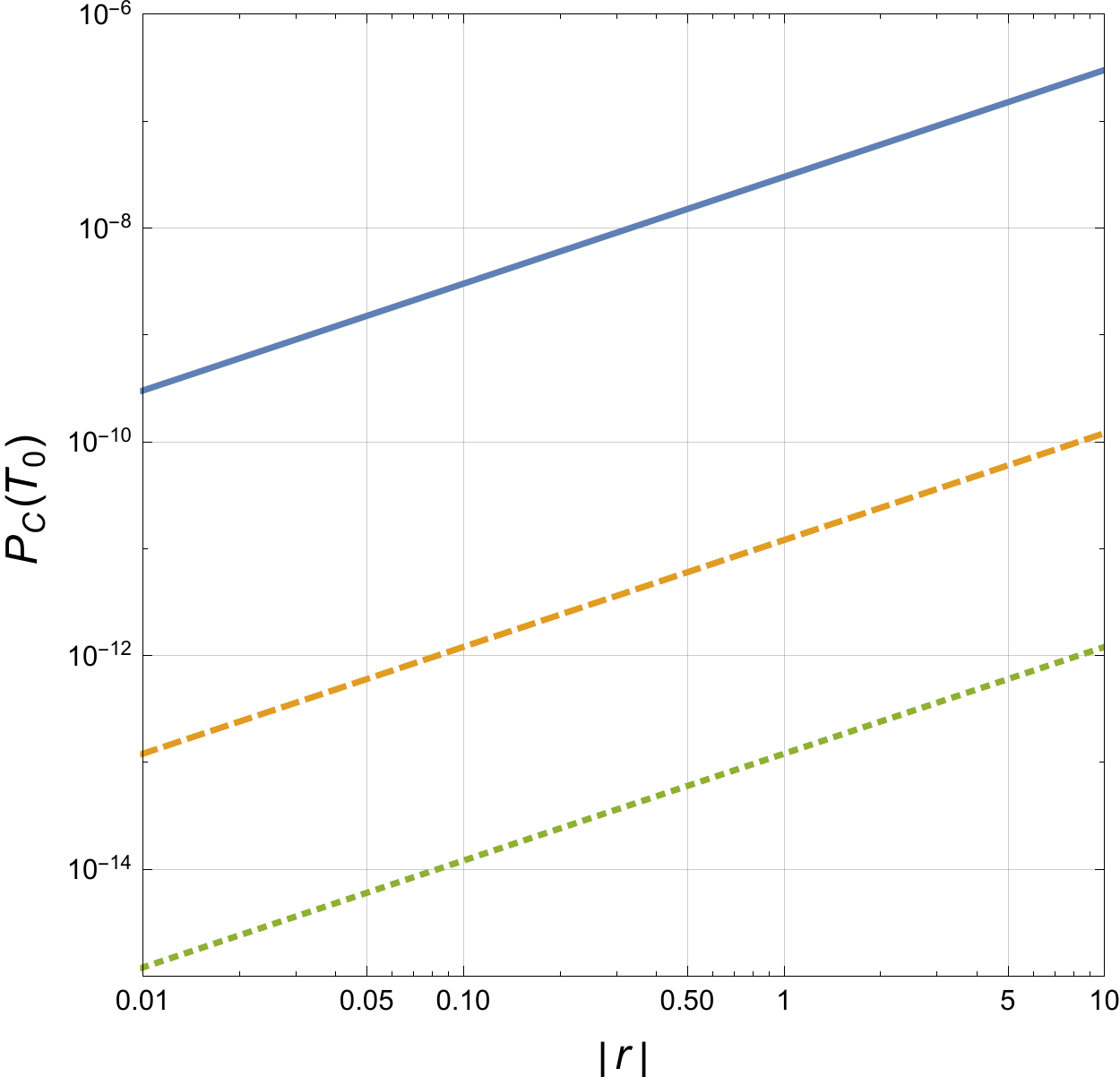}}}
\caption{(a) Logarithmic scale plots of the degree of circular polarization  $P_C(T_0)$ as a function of $|r|$ for magnetic field amplitude $B_{e0}=10^{-9}$ G and CMB frequencies $\nu_0=10^8$ Hz (continuous line), $\nu_0=10^9$ Hz (dashed line) and $\nu_0=10^{10}$ Hz (dotted line) are shown.  (b) Logarithmic scale plots of the degree of circular polarization  $P_C(T_0)$ as a function of $|r|$ (where $r \equiv U_i/Q_i$) at the CMB frequency $\nu_0=10^9$ Hz for magnetic field amplitudes $B_{e0}=8\times 10^{-8}$ G (continuous line),  $B_{e0}=10^{-8}$ G (dashed line) and $B_{e0}=10^{-9}$ (dotted line) are shown. In both plots we used a value of $|Q_i|= 10^{-6}$. For simplicity we showed also the grid lines in grey. }
\label{fig:Fig5a}
\end{figure*}

To calculate the degree of circular polarization, let us use the results obtained in Sec. \ref{subsec:4.3} that we found without any restriction on the magnitude of $\mathcal M_\text{F}(T_0)$. For absent Faraday effect $\mathcal M_\text{F}(T_0)=0$ and consequently a vanishing $\mathcal M_\text{C}(T_0)=0$, the degree of circular polarization is given by
\begin{equation}\label{deg-pol-abs-far}
P_C(T_0)=\frac{\left|\tilde V(T_0)\right|}{\tilde I(T_0)}= \left | L_4^{(0)}(T_0) Q_i -  K_4^{(0)}(T_0) U_i \right|= |\Delta \mathcal M(T_0) r Q_i| = 5.38\times 10^{38} |r Q_i| \left(\frac{\text{Hz}}{\nu_0}\right)^3\left( \frac{B_{e0}}{\text{G}}\right)^2 T_0^{-3/2} \quad (\text{K}^{3/2}).
\end{equation}
As we can see from \eqref{deg-pol-abs-far} the degree of circular polarization for transverse magnetic field depend only on $\Delta \mathcal M(T_0)$. The most important thing is that we do not have anymore the constraints on $\mathcal M_\text{(F, C)}$ but only those on $|\Delta \mathcal M(T_0)|<1$. In Figs. \ref{fig:Fig4a} and \ref{fig:Fig5a} plots of the degree of circular polarization for transverse magnetic field as a function of $\nu_0$, $B_{e0}$ and $|r|$ are shown. We may observe in Fig. \ref{fig:Fig4a} that for higher values of $B_{e0}$ and lower values of $\nu_0$, the acquired degree of circular polarization of the CMB is quite substantial and it can be comparable with that of the linear polarization for some values of the parameters. For example, as we can see from Fig. \ref{fig:Fig4a}, for $B_{e0}=8 \times 10^{-8}$ G, we get $P_C(T_0)\simeq 7.65 \times 10^{-7}$ for $\nu_0=10^8$ Hz and $P_C(T_0)\simeq 7.65 \times 10^{-10}$ for $\nu_0=10^9$ Hz. It is worth to point out that the expression \eqref{deg-pol-abs-far} can also be obtained by using the perturbative approach used in the previous section.

\subsection{Case when $|\mathcal M_\text{F}(T_0)|\geq 1$ and $|\mathcal M_\text{C}(T_0)|< 1,|\Delta \mathcal M(T_0)|< 1$.}
\label{subsec:5.3}

In the case when $|\mathcal M_\text{F}(T_0)|\geq 1$ the situation is more complicated with respect to the previous cases. One aspect is that in \eqref{F-C-M-cond} only the last two inequalities must be satisfied while the first inequality has not to be satisfied anymore. This fact tells us that the allowed region of parameters is that within the black line in Fig. \ref{fig:Fig2} and that outside the region within the dotted line. In this case the degree of circular polarization can be calculated by using the results of Sec. \ref{subsec:4.3} that we derived for arbitrary values of $\mathcal M_\text{F}(T)$
\begin{equation}\label{general-PC}
\begin{gathered}
P_C(T_0)= \left | L_4^{(0)}(T_0) Q_i -  K_4^{(0)}(T_0) U_i \right|= |Q_i|\left| \left( \int_{T_0}^{T_i} dT^\prime \left[ G_\text{C}(T^\prime) \cos[\mathcal M_\text{F}(T^\prime)] - \Delta G(T^\prime) \sin[\mathcal M_\text{F}(T^\prime)] \right] \right) -\right. \\ \left. \left( \int_{T_0}^{T_i} dT^\prime \left[ G_\text{C}(T^\prime) \sin[\mathcal M_\text{F}(T^\prime)] + \Delta G(T^\prime) \cos[\mathcal M_\text{F}(T^\prime)] \right] \right) r \right|
\end{gathered}
\end{equation}

The expression \eqref{general-PC} is valid for any value of $\mathcal M_\text{F}(T)$ and for $|\mathcal M_\text{C}(T_0)|< 1,|\Delta \mathcal M(T_0)|< 1$ even though in this section we study the case when  $|\mathcal M_\text{F}(T) | \geq 1$. The main difficulty on calculating $P_C(T_0)$ analytically stands from the fact that in all terms $\mathcal M_\text{F}, G_\text{C}$ and $\Delta G$ enters the ionization function which does not have any known analytic expression. 
To find an analytic expression for $P_C$, in this section we approximate $X_e(T)\simeq \bar X_e$ where $\bar X_e$ is the average value of $X_e(T)$ in the temperature interval $T_0\leq T\leq T_i$. Let us write 
\begin{equation}
\begin{gathered}
\mathcal M_\text{F}(T)=\mathcal A (T_i^{3/2}-T^{3/2}), \quad G_\text{C}(T)= \mathcal B T^{3/2}$, \quad $\Delta G(T)=\mathcal C T^{3/2},\\
\mathcal A\equiv (2/3) 8.71\times 10^{25} \bar X_e \sin(\Theta)\sin(\Phi)(\text{Hz}/\nu_0)^{2} (B_{e0}/\text{G}) T_{0}^{-1/2} (\text{K}^{-1}), \\
\mathcal B=6.05\times 10^{31} \bar X_e \sin(2\Theta)\cos(\Phi)(\text{Hz}/\nu_0)^{3} (B_{e0}/\text{G})^2 T_{0}^{-3/2}\, (\text{K}^{-1}),\\
 \mathcal C\equiv -1.21\times 10^{32} \bar X_e \left[\sin^2(\Theta)\cos^2(\Phi) -\cos^2(\Theta)\right]  (\text{Hz}/\nu_0)^{3} (B_{e0}/\text{G})^2 T_{0}^{-3/2}\, (\text{K}^{-1}).
 \end{gathered}
\end{equation}

Then we have that
\begin{equation}\label{integr-0}
\begin{gathered}
\int_{T_0}^{T_i} dT^\prime \, G_\text{C}(T^\prime) \cos[\mathcal M_\text{F}(T^\prime)] = \mathcal B \int_{T_0}^{T_i} dT^\prime \, T^{\prime 3/2} \cos[\mathcal A (T_i^{3/2}-T^{\prime 3/2})] = \mathcal B \cos[\mathcal A T_i^{3/2}] \int_{T_0}^{T_i} dT^\prime \,T^{\prime 3/2}\cos[\mathcal A T^{\prime 3/2}]  + \\
\mathcal B \sin[\mathcal A T_i^{3/2}] \int_{T_0}^{T_i} dT^\prime \,T^{\prime 3/2}\sin[\mathcal A T^{\prime 3/2}]
\end{gathered}
\end{equation}
where we used the identity $\cos(\alpha-\beta)=\cos(\alpha)\cos(\beta) + \sin(\alpha)\sin(\beta)$. 
Let us define $x \equiv \mathcal A T^{3/2}$ where $dT = (2/3) \mathcal A^{-2/3} x^{-1/3} dx$ and get
\begin{equation}
\begin{gathered}
 \int_{T_0}^{T_i} dT^\prime T^{\prime 3/2}\,\cos[\mathcal A T^{\prime 3/2}]  =\frac{2 \mathcal A^{-5/3}}{3} \int_{x_0}^{x_i} dx\, x^{2/3} \cos(x),\quad
  \int_{T_0}^{T_i} dT^\prime T^{\prime 3/2}\,\sin[\mathcal A T^{\prime 3/2}] = \frac{2 \mathcal A^{-5/3}}{3} \int_{x_0}^{x_i} dx\, x^{2/3} \sin(x). \nonumber
\end{gathered}
\end{equation}
Now let us focus on for simplicity on the cosine and sine integrals types and first by integrating by parts we get
\begin{equation}\label{integr-1}
\begin{gathered}
\int_{x_{0}}^{x_i} dx\, x^{2/3} \cos(x)=\frac{1}{2} \int_{x_{0}}^{x_i} dx\, x^{2/3} \left[ \exp(-ix) + \exp(i x) \right] 
= - x_{0}^{2/3}\sin(x_{0}) + x_{i}^{2/3}\sin(x_{i})  \\ 
- \frac{(-i)^{-1/3}}{3} \left( \Gamma \left[\left(\frac{2}{3}\right), i x_{0}\right] - \Gamma \left[\left(\frac{2}{3}\right), i x_{i}\right] \right) - \frac{i^{-1/3}}{3} \left( \Gamma \left[\left(\frac{2}{3}\right), -i x_{0}\right] - \Gamma \left[\left(\frac{2}{3}\right), -i x_{i}\right] \right),\\
 \int_{x_{0}}^{x_i} dx\, x^{2/3} \sin(x) = \frac{1}{2i} \int_{x_{0}}^{x_i} dx\, x^{2/3} \left[ \exp(ix) - \exp(-i x) \right] 
=   x_0^{2/3} \cos(x_0) - x_i^{2/3} \cos(x_i)  \\ 
+ \frac{(-i)^{2/3}}{3} \left( \Gamma \left[\left(\frac{2}{3}\right), i x_{0}\right] - \Gamma \left[\left(\frac{2}{3}\right), i x_{i}\right] \right) + \frac{i^{2/3}}{3} \left( \Gamma \left[\left(\frac{2}{3}\right), -i x_{0}\right] - \Gamma \left[\left(\frac{2}{3}\right), -i x_{i}\right] \right),
\end{gathered}
\end{equation}
where we used the definition of the generalized incomplete Gamma function $\Gamma(s, z_1, z_2) = \int_{z_1}^{z_2} dx\, x^{s-1} e^{-x}=\Gamma(s, z_1) - \Gamma(s, z_2)$ where the arguments $s$ and $z$ are complex numbers and $\Gamma(s, z)=\int_{z}^{\infty} dx\, x^{s-1} e^{-x}$ is the incomplete Euler Gamma function. By using \eqref{integr-1} into expression \eqref{integr-0} 
and by summing all together we get
\begin{equation}\label{integr-2}
\begin{gathered}
\int_{T_0}^{T_i} dT^\prime \, G_\text{C}(T^\prime) \cos[\mathcal M_\text{F}(T^\prime)] = \frac{2 \mathcal B}{3 \mathcal A}
T_0\, \sin(x_i-x_0) - \frac{2 \mathcal B}{3 \mathcal A^{5/3}} \frac{(-i)^{-1/3}}{3} \cos(x_i) \left( \Gamma \left[\left(\frac{2}{3}\right), i x_{0}\right] - \Gamma \left[\left(\frac{2}{3}\right), i x_{i}\right] \right) \\
- \frac{2 \mathcal B}{3 \mathcal A^{5/3}} \frac{i^{-1/3}}{3} \cos(x_i)\left( \Gamma \left[\left(\frac{2}{3}\right), -i x_{0}\right] - \Gamma \left[\left(\frac{2}{3}\right), - i x_{i}\right] \right) + \frac{2 \mathcal B}{3 \mathcal A^{5/3}} \frac{(-i)^{2/3}}{3} \sin(x_i) \left( \Gamma \left[\left(\frac{2}{3}\right), i x_{0}\right] - \Gamma \left[\left(\frac{2}{3}\right), i x_{i}\right] \right) \\ + \frac{2 \mathcal B}{3 \mathcal A^{5/3}} \frac{i^{2/3}}{3} \sin(x_i) \left( \Gamma \left[\left(\frac{2}{3}\right), - i x_{0}\right] - \Gamma \left[\left(\frac{2}{3}\right), - i x_{i}\right] \right).
\end{gathered}
\end{equation}
The integral $\int_{T_0}^{T_i} dT^\prime \, \Delta G(T^\prime) \cos[\mathcal M_\text{F}(T^\prime)] $ can be obtained from \eqref{integr-2} by simply replacing $\mathcal B \rightarrow \mathcal C$. By proceeding in the same way as we did above, we get the following expression for 
\begin{equation}\label{integr-3}
\begin{gathered}
\int_{T_0}^{T_i} dT^\prime \, G_\text{C}(T^\prime) \sin[\mathcal M_\text{F}(T^\prime)] = \frac{2 \mathcal B}{3 \mathcal A}\left[T_i -
T_0\, \cos(x_i-x_0)\right] - \frac{2 \mathcal B}{3 \mathcal A^{5/3}} \frac{(-i)^{-1/3}}{3} \sin(x_i) \left( \Gamma \left[\left(\frac{2}{3}\right), i x_{0}\right] - \Gamma \left[\left(\frac{2}{3}\right), i x_{i}\right] \right) \\
- \frac{2 \mathcal B}{3 \mathcal A^{5/3}} \frac{i^{-1/3}}{3} \sin(x_i)\left( \Gamma \left[\left(\frac{2}{3}\right), -i x_{0}\right] - \Gamma \left[\left(\frac{2}{3}\right), - i x_{i}\right] \right) - \frac{2 \mathcal B}{3 \mathcal A^{5/3}} \frac{(-i)^{2/3}}{3} \cos(x_i) \left( \Gamma \left[\left(\frac{2}{3}\right), i x_{0}\right] - \Gamma \left[\left(\frac{2}{3}\right), i x_{i}\right] \right) \\ - \frac{2 \mathcal B}{3 \mathcal A^{5/3}} \frac{i^{2/3}}{3} \cos(x_i) \left( \Gamma \left[\left(\frac{2}{3}\right), - i x_{0}\right] - \Gamma \left[\left(\frac{2}{3}\right), - i x_{i}\right] \right).
\end{gathered}
\end{equation}
Again in the same way as above, the integral $\int_{T_0}^{T_i} dT^\prime \, \Delta G(T^\prime) \sin[\mathcal M_\text{F}(T^\prime)] $ can be obtained from \eqref{integr-3} by simply replacing $\mathcal B \rightarrow \mathcal C$. While in the integrals above do appear complex valued functions, the integral in itself is real.

So far our calculations of the integrals \eqref{integr-0}-\eqref{integr-3} which enter in \eqref{general-PC} have been exact in the case when $X_e(T)\simeq \bar X_e$, namely when the ionization function is constant. At this stage in order to simplify as much as possible our results it is very convenient to find for what values of the parameters the terms proportional to $T_{0, i}\mathcal B/\mathcal A$ are larger than terms proportional to $\mathcal B/\mathcal A^{5/3}$. We have that the condition $T_{0, i}|\mathcal B/\mathcal A |> |\mathcal B/\mathcal A^{5/3}|$ is satisfied when $T_{0, i} |\mathcal A^{2/3}| >1$ or equivalently 
\begin{equation}\label{cond-8}
\left(\frac{\nu_0}{\text{Hz}}\right)^2 \left( \frac{\text{G}}{B_{e0}}\right) < 8.09\times 10^{23}\, T_{0, i}^{3/2}\, \left|\left[ \sin(\Theta)\sin(\Phi) \right]^{2/3} \right|^{3/2} \quad (\text{K}^{-3/2}), 
\end{equation}
where we must have $\sin(\Theta)\sin(\Phi) \neq 0$. Consequently, in the case when \eqref{cond-8} is satisfied, the leading terms in \eqref{general-PC} are only those of the form $\int_{T_0}^{T_i} dT^\prime \, G_\text{C}(T^\prime) \sin[\mathcal M_\text{F}(T^\prime)]$ and $\int_{T_0}^{T_i} dT^\prime \, \Delta G(T^\prime) \sin[\mathcal M_\text{F}(T^\prime)]$ and which are respectively the terms $2 T_i \mathcal B/(3\mathcal A)$ and $2 T_i \mathcal C/(3\mathcal A)$. So, in the case when \eqref{cond-8} is satisfied we get
\begin{equation}\label{first-apr}
P_C(T_0)=\left| \frac{2 Q_i T_i}{3\mathcal A} \right| \left| \mathcal C + r\mathcal B \right| = 7.57\times 10^{8} |Q_i| \left(\frac{\text{Hz}}{\nu_0}\right) \left( \frac{B_{e0}}{\text{G}}\right) \left| \frac{2 \left[-\sin^2(\Theta)\cos^2(\Phi) + \cos^2(\Theta)\right]}{\sin(\Theta)\sin(\Phi)} + r \frac{\sin(2\Theta)\cos(\Phi)}{\sin(\Theta)\sin(\Phi)} \right|,
\end{equation}
where we may note that $P_C(T_0)$ in \eqref{first-apr} does not depend on $\bar X_e\simeq 0.023$. 

\begin{figure*}[h!]
\centering
\mbox{
\subfloat[\label{fig:Fig11}]{\includegraphics[scale=0.65]{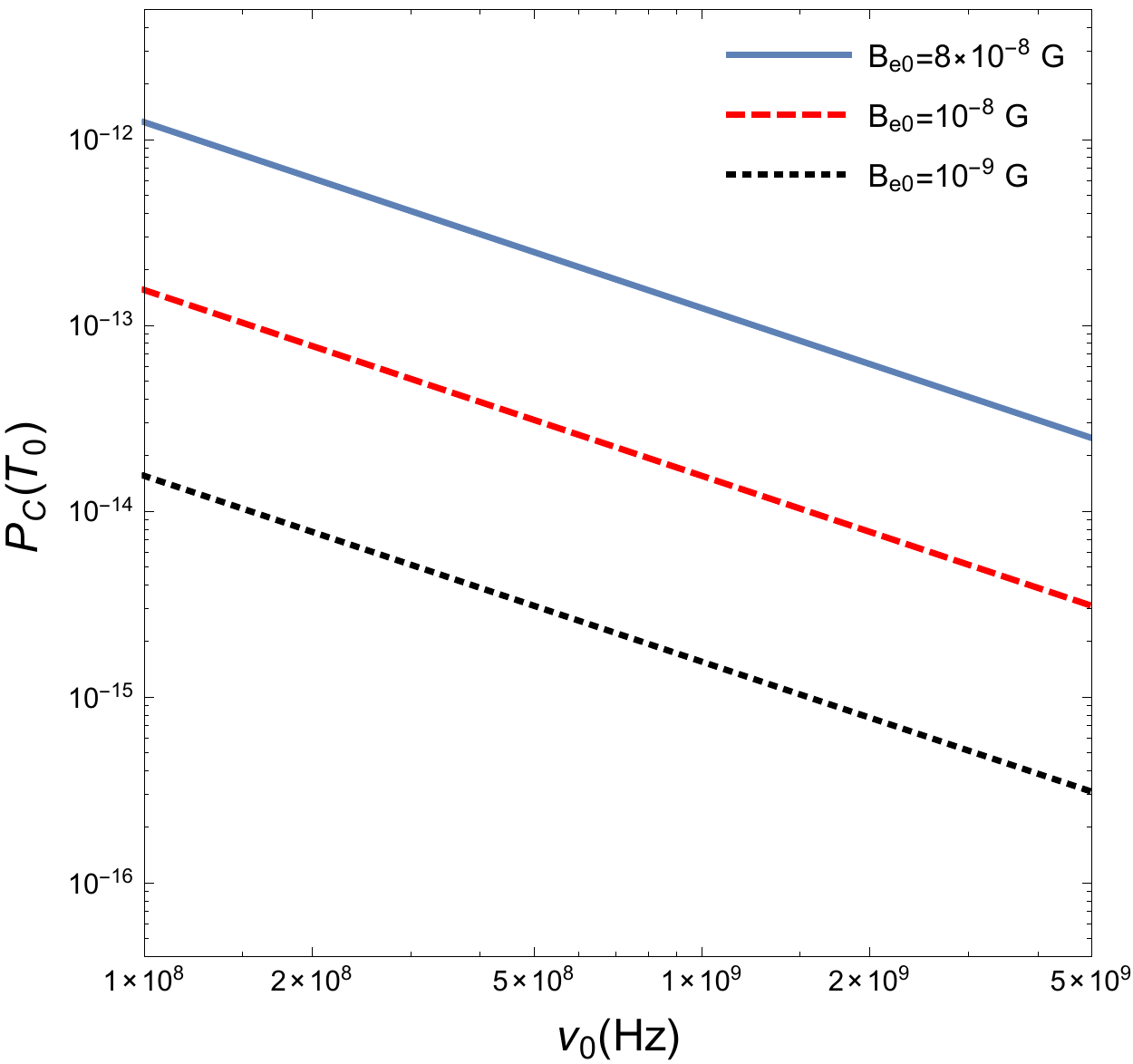}}\qquad
\subfloat[\label{fig:Fig12}]{\includegraphics[scale=0.65]{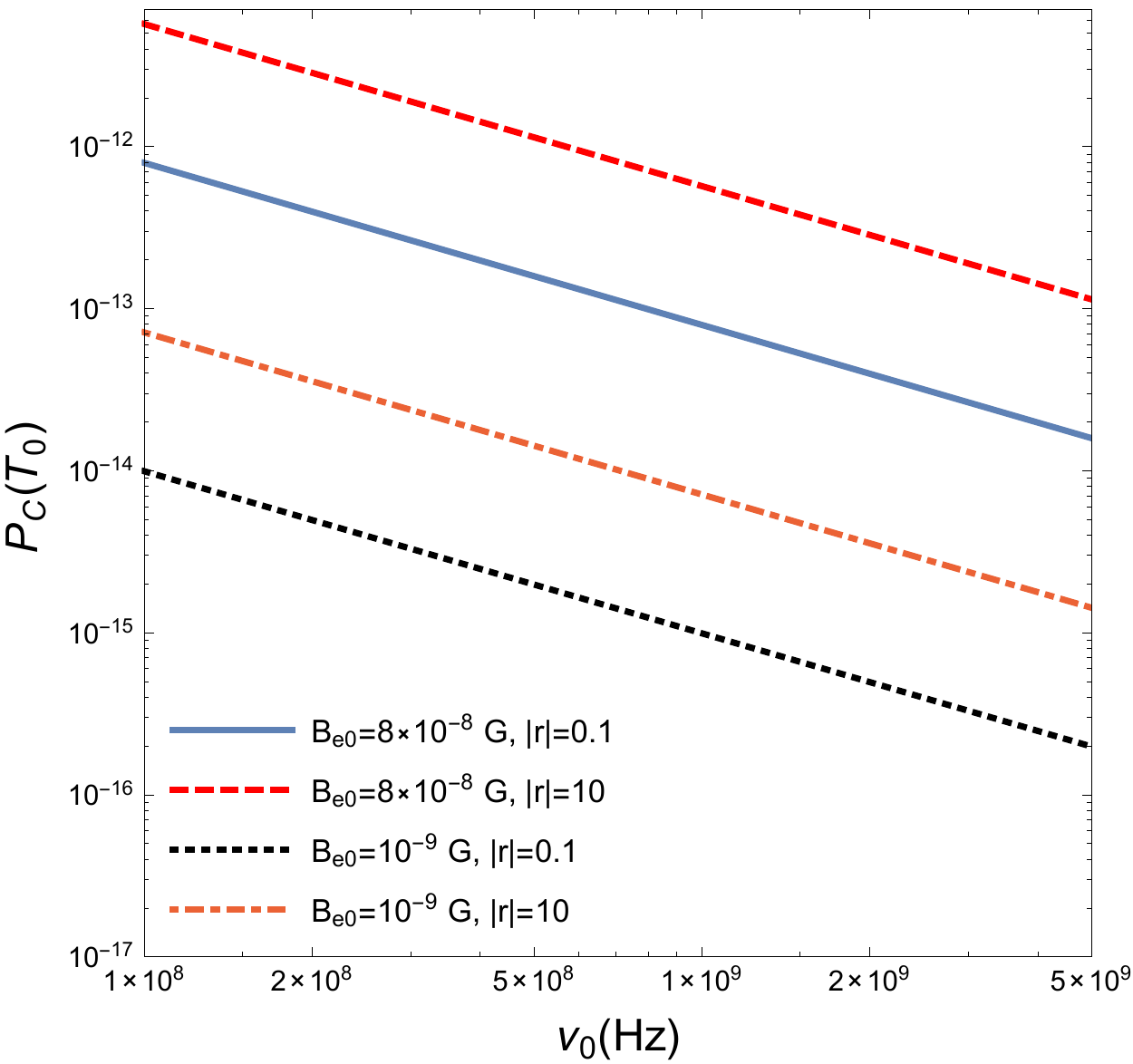}}}
\caption{(a) Logarithmic scale plots of the degree of circular polarization  $P_C(T_0)$ given in \eqref{first-apr} as a function of $\nu_0\in [10^8, 5\times 10^9]$ Hz for magnetic field amplitude $B_{e0}=8\times 10^{-8}$ G (continuous line), $B_{e0}=10^{-8}$ G (dashed line), $B_{e0}=10^{-9}$ G (dotted line) and $|r|=1$ are shown.  (b) Logarithmic scale plots of the degree of circular polarization  $P_C(T_0)$ given in \eqref{first-apr} as a function of $\nu_0\in [10^8, 5\times 10^9]$ Hz for magnetic field amplitudes $B_{e0}=8\times 10^{-8}$ G and $|r|=0.1$ (continuous line), $B_{e0}=8\times 10^{-8}$ G and $|r|=10$ (dashed line), $B_{e0}=10^{-9}$ G and $|r|=0.1$ (dotted line) and $B_{e0}=10^{-9}$ G and $|r|=10$ (dotdashed line) are shown. In all plots we used a value of $|Q_i|= 10^{-6}$ and values of $\Theta=\pi/4$ and $\Phi=\pi/3$ with $r \equiv U_i/Q_i$. In both (a) and (b) the values of the parameters have been chosen in such a way that condition \eqref{cond-8} is satisfied.}
\label{fig:Fig6a}
\end{figure*}

\begin{figure*}[h!]
\centering
\mbox{
\subfloat[\label{fig:Fig13}]{\includegraphics[scale=0.63]{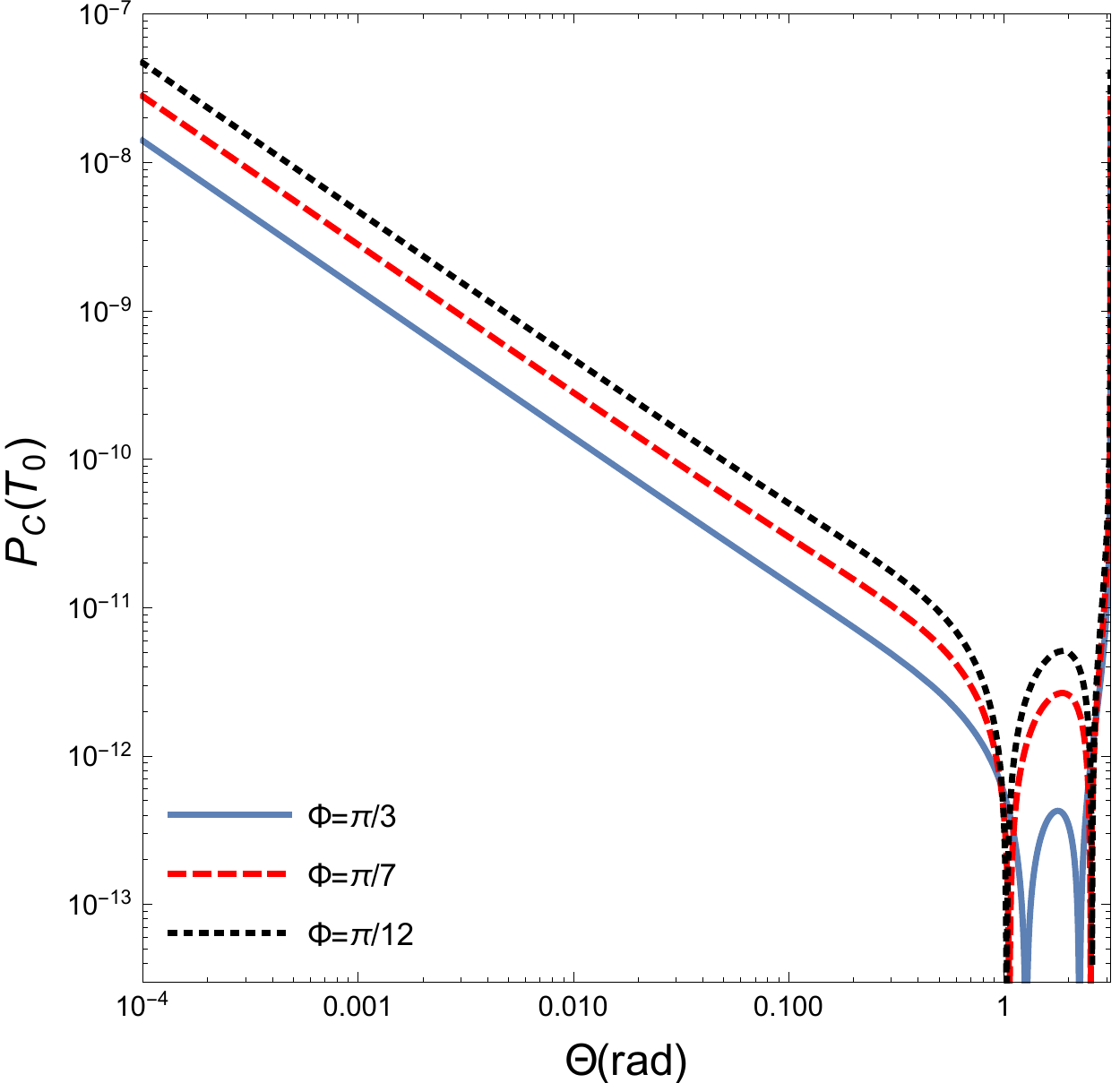}}\qquad
\subfloat[\label{fig:Fig14}]{\includegraphics[scale=0.65]{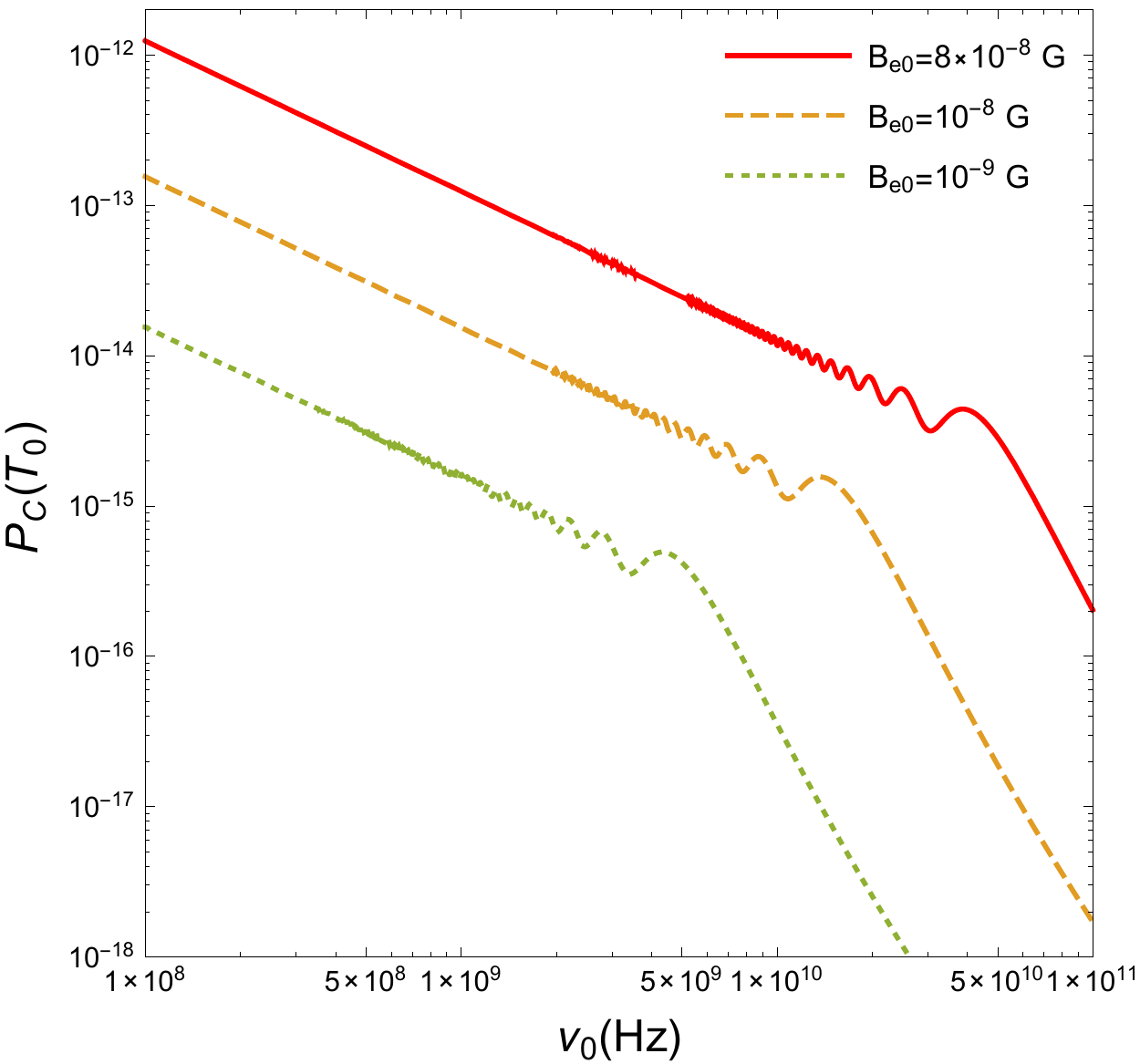}}}
\caption{(a) Logarithmic scale plots of the degree of circular polarization  $P_C(T_0)$ given in \eqref{first-apr} as a function of $\Theta\in [10^{-4}, \pi - 10^{-4}]$ (rad) for magnetic field amplitude $B_{e0}=8\times 10^{-8}$ G, frequency $\nu_0=10^{8}$ Hz for several values of the angle $\Phi$, $|r|=1$ and $|Q_i|=1$ are shown.  (b) Logarithmic scale plots of the degree of circular polarization  $P_C(T_0)$ given in \eqref{general-PC} (for $X_e(T)\simeq \bar X_e=0.023$) as a function of $\nu_0\in [10^8, 10^{11}]$ Hz for magnetic field amplitudes $B_{e0}=8\times 10^{-8}$ G (continuous line), $B_{e0}=10^{-8}$ G (dashed line), $B_{e0}=10^{-9}$ G (dotted line) are shown. In all plots in (b) we used a value of $|Q_i|= 10^{-6}$, $|r|=1$ and values of $\Theta=\pi/4$ and $\Phi=\pi/3$. In (a) the values of the parameters have been chosen in such a way that condition \eqref{cond-8} is satisfied.}
\label{fig:Fig7a}
\end{figure*}

One thing which is worth to mention is that \eqref{first-apr} is valid as far as the condition \eqref{cond-8} is satisfied and when $\sin(\Theta)\sin(\Phi) \neq 0$. The latter condition still appears because we had to divide by $\mathcal A$ in the integration procedure. Another important fact is that $P_C(T_0)$ in \eqref{first-apr} can assume zero values as far as $\sin(\Theta)\sin(\Phi) \neq 0$ and $r \sin(2\Theta) \cos(\Phi) + 2 \left[-\sin^2(\Theta)\cos^2(\Phi) +\cos^2(\Theta)\right] =0$. In Fig. \ref{fig:Fig6a} plots of the degree of circular polarization (given by expression \eqref{first-apr}) as a function of the CMB frequency $\nu_0$ for various values of the parameters are shown. The values of the parameters have been chosen in such a way that expression \eqref{cond-8} is satisfied for $T=T_i$. As we may observe from Fig. \ref{fig:Fig6a} the presence of the Faraday effect significantly reduces the degree of circular polarization by many orders of magnitude with respect to the case of absent Faraday effect. In Fig. \ref{fig:Fig13} we show the plot of $P_C(T_0)$ as a function of the angle $\Theta$ as given in \eqref{first-apr} for some given values of the angle $\Phi$. We can see that in the case when $\Theta\rightarrow 0$ the degree of circular polarization significantly increases by many orders of magnitude. In this case we recover the results of the previous section where the magnetic filed has been assumed to be purely transverse, namely $\mathcal M_\text{F}(T_0)=0$. The cusp-like behaviour that appear in the plots in Fig. \ref{fig:Fig13} correspond to those values of $\Theta$ where $P_C(T_0)= 0$ due to the trigonometric function $r \sin(2\Theta) \cos(\Phi) + 2 \left[-\sin^2(\Theta)\cos^2(\Phi) + \cos^2(\Theta)\right] =0$. However, these points are not real cusps but simply arise due to the fact that we have to take the absolute value of $\tilde V(T_0)$ in order to calculate $P_C(T_0)$. In Fig. \ref{fig:Fig14} plot of the degree of circular polarization $P_C(T_0)$ obtained by using expression \eqref{general-PC} for $X_e(T)\simeq \bar X_e=0.023$ are shown. These plots essentially correspond to the case when we include all terms which do appear in the integrals \eqref{integr-2}-\eqref{integr-3} and to their similar integral functions.  The oscillating nature of the plots arises due to the fact that for higher values of $\nu_0$ and $B_{e0}$ do contribute to the integrals the terms proportional to $\sin(x_i)$ and $\cos(x_i)$ which are fast oscillating functions of the parameters.

\section{Rotation angle of the polarization plane}
\label{sec:6}

In the previous section we studied the generation of the CMB circular polarization by calculating explicitly $P_C(T_0)$ in various regimes. In this section, we focus our attention on the rotation angle of the CMB polarization plane from the decoupling epoch until today. Apart from generating circular polarization, the CM effect also generates  linear polarization with non zero Stokes parameters $\tilde Q(T)$ and $\tilde U(T)$. At a given cosmological temperature $T$ the rotation angle of the polarization plane is given by
\begin{equation}\label{rot-angle}
\tan[2\psi(T)]=\frac{\tilde U(T)}{\tilde Q(T)},
\end{equation}
where we must have $\tilde Q(T)\neq 0$.
Let us write $\psi(T)=\psi(T_i)+\delta\psi(T)$ where $\psi(T_i)$ is the angle of the CMB polarization plane at the temperature $T_i=2970$ K corresponding to the decoupling time in the common reference frame used to study the CMB and $\psi(T)$ is the angle of the polarization plane at a temperature $T<T_i$. Here $\delta\psi(T)$ is the amount of the rotation angle of the polarization plane from the decoupling time until at the time corresponding to the temperature $T$ and it is the quantity which interests us. Since for the frequency range of interest in this work, the magnitude of the effects which we study are in general small, namely $|\mathcal M_\text{F}(T)|<1, |\Delta\mathcal M(T)|<1, |\mathcal M_\text{C}(T)|<1$, and because experimentally $\delta\psi(T_0)$ is constrained to a small quantity (in radians), we expect that the rotation angle of the CMB polarization plane from decoupling epoch until at present to be a small quantity $|\delta\psi(T)|\ll 1$. In this case by using the trigonometric identity we can write
\begin{equation}\label{exps-angle}
\tan[2\psi(T)] = \tan[2\psi(T_i) + 2\delta\psi(T)] = \frac{\tan[2\psi(T_i)]+\tan[2\delta\psi(T)]}{1-\tan[2\psi(T_i)]\tan[2\delta\psi(T)]} \simeq \frac{r + 2\delta\psi(T)}{1-2 r \delta\psi(T)} \simeq r+2\delta\psi(T)(1+r^2) + 4 r \delta\psi^2(T),
\end{equation}
where we used $\tan[2\psi(T_i)]=\tilde U_i/\tilde Q_i=r$ and $r\neq \delta\psi(T)/2$ in expression \eqref{exps-angle} where $\delta\psi(T)$ can have either sign depending on the rotation effect and on the conventions used. In \eqref{exps-angle} we used the series expansion $\tan[2\delta\psi(T)] \simeq 2\delta\psi(T)$ for $|\delta\psi(T)|\ll 1$ and the geometric series expansion $(1-2r\delta\psi(T))^{-1}\simeq 1+ 2 r\delta\psi(T) +...$ which is valid as far as $|2r\delta\psi(T)|<1$.


In the case when $ |\mathcal M_\text{F}(T)|<1, |\Delta\mathcal M(T)|<1$ and $|\mathcal M_\text{C}(T)|<1$, the expressions for the Stokes parameters up to second order in perturbation theory are given in \eqref{Stokes-1} and \eqref{M-ele}. Therefore from expressions \eqref{rot-angle}, \eqref{exps-angle} and \eqref{Stokes-1} we get for $T=T_0$ and $\tilde V_i=0$
\begin{equation}\label{pol-angle-1}
r+2\delta\psi(T_0)(1+r^2) + 4 r \delta\psi^2(T_0) = \frac{\tilde M_{32}(T_0)+r  \tilde M_{33}(T_0)}{ \tilde M_{22}(T_0)+r \tilde M_{23}(T_0)}.
\end{equation}
To calculate $\delta\psi(T_0)$ we need to calculate each matrix element in \eqref{pol-angle-1} at $T=T_0$. Consequently, we have that
\begin{equation}\label{m23}
\begin{gathered}
\tilde M_{23}(T_0)= -\mathcal M_\text{F}(T_0) + \int_{T_0}^{T_i} dT^\prime G_\text{C}(T^\prime) \int_{T^\prime}^{T_i} dT^{\prime\prime} \Delta G(T^{\prime\prime})=- \mathcal M_\text{F}(T_0) + \mathcal B \mathcal C \bar X_e^{-2} \int_{T_0}^{T_i} dT^\prime X_e(T^\prime) T^{\prime 3/2} \times \\ 
\int_{T^\prime}^{T_i} dT^{\prime\prime} X_e(T^{\prime\prime}) T^{\prime\prime 3/2}  = -\mathcal M_\text{F}(T_0) + 9.88\times 10^{12} \mathcal B \mathcal C \bar X_e^{-2} \quad (\text{K}^5),
\end{gathered}
\end{equation}
where we numerically calculated $ \int_{T_0}^{T_i} dT^\prime X_e(T^\prime) T^{\prime 3/2} \int_{T^\prime}^{T_i} dT^{\prime\prime} X_e(T^{\prime\prime}) T^{\prime\prime 3/2} =9.88\times 10^{12}$ (K$^{5}$).
We also get the following expression for 
\begin{equation}\label{m32}
\begin{gathered}
\tilde M_{32}(T_0)= \mathcal M_\text{F}(T_0) + \int_{T_0}^{T_i} dT^\prime \Delta G(T^\prime) \int_{T^\prime}^{T_i} dT^{\prime\prime} G_\text{C}(T^{\prime\prime})= \mathcal M_\text{F}(T_0) + \mathcal B \mathcal C \bar X_e^{-2} \int_{T_0}^{T_i} dT^\prime X_e(T^\prime) T^{\prime 3/2} \times \\ 
\int_{T^\prime}^{T_i} dT^{\prime\prime} X_e(T^{\prime\prime}) T^{\prime\prime 3/2}  = \mathcal M_\text{F}(T_0) + 9.88\times 10^{12} \mathcal B \mathcal C \bar X_e^{-2} \quad (\text{K}^5).
\end{gathered}
\end{equation}
On the other hand, we get the following expressions for 
\begin{equation}
\begin{gathered}\label{m22}
\tilde M_{22}(T_0)= 1- \int_{T_0}^{T_i} dT^\prime G_\text{F}(T^\prime)\int_{T^\prime}^{T_i} dT^{\prime\prime} G_\text{F}(T^{\prime\prime}) - \int_{T_0}^{T_i} dT^\prime G_\text{C}(T^\prime)\int_{T^\prime}^{T_i} dT^{\prime\prime} G_\text{C}(T^{\prime\prime})\\
= 1 - \frac{9 \mathcal A^2}{4 \bar X_e^2} \int_{T_0}^{T_i} dT^\prime X_e(T^\prime) T^{\prime 1/2} \int_{T^\prime}^{T_i} dT^{\prime\prime} X_e(T^{\prime\prime}) T^{\prime\prime 1/2} -\mathcal B^2 \bar X_e^{-2} \int_{T_0}^{T_i} dT^\prime X_e(T^\prime) T^{\prime 3/2} \int_{T^\prime}^{T_i} dT^{\prime\prime} X_e(T^{\prime\prime}) T^{\prime\prime 3/2}\\
= 1-1.6 \times 10^6 \left(\frac{9 \mathcal A^2}{4 \bar X_e^2}\right) \quad (\text{K}^3)- 9.88\times 10^{12} \mathcal B^2 \bar X_e^{-2} \quad (\text{K}^5),
\end{gathered}
\end{equation}
where we used the numerically integrated value of $\int_{T_0}^{T_i} dT^\prime X_e(T^\prime) T^{\prime 1/2} \int_{T^\prime}^{T_i} dT^{\prime\prime} X_e(T^{\prime\prime}) T^{\prime\prime 1/2}\simeq 1.6 \times 10^{6}$ (K$^3$).
The last expression to calculate is
\begin{equation}\label{m33}
\begin{gathered}
\tilde M_{33}(T_0)= 1- \int_{T_0}^{T_i} dT^\prime G_\text{F}(T^\prime)\int_{T^\prime}^{T_i} dT^{\prime\prime} G_\text{F}(T^{\prime\prime}) - \int_{T_0}^{T_i} dT^\prime \Delta G(T^\prime)\int_{T^\prime}^{T_i} dT^{\prime\prime} \Delta G(T^{\prime\prime})\\
= 1 - \frac{9 \mathcal A^2}{4 \bar X_e^2} \int_{T_0}^{T_i} dT^\prime X_e(T^\prime) T^{\prime 1/2} \int_{T^\prime}^{T_i} dT^{\prime\prime} X_e(T^{\prime\prime}) T^{\prime\prime 1/2} -\mathcal C^2 \bar X_e^{-2} \int_{T_0}^{T_i} dT^\prime X_e(T^\prime) T^{\prime 3/2} \int_{T^\prime}^{T_i} dT^{\prime\prime} X_e(T^{\prime\prime}) T^{\prime\prime 3/2}\\
= 1-1.6 \times 10^6 \left(\frac{9 \mathcal A^2}{4 \bar X_e^2}\right) \quad (\text{K}^3)- 9.88\times 10^{12} \mathcal C^2 \bar X_e^{-2} \quad (\text{K}^5).
\end{gathered}
\end{equation}

\begin{figure*}[h!]
\centering
\mbox{
\subfloat[\label{fig:Fig15}]{\includegraphics[scale=0.63]{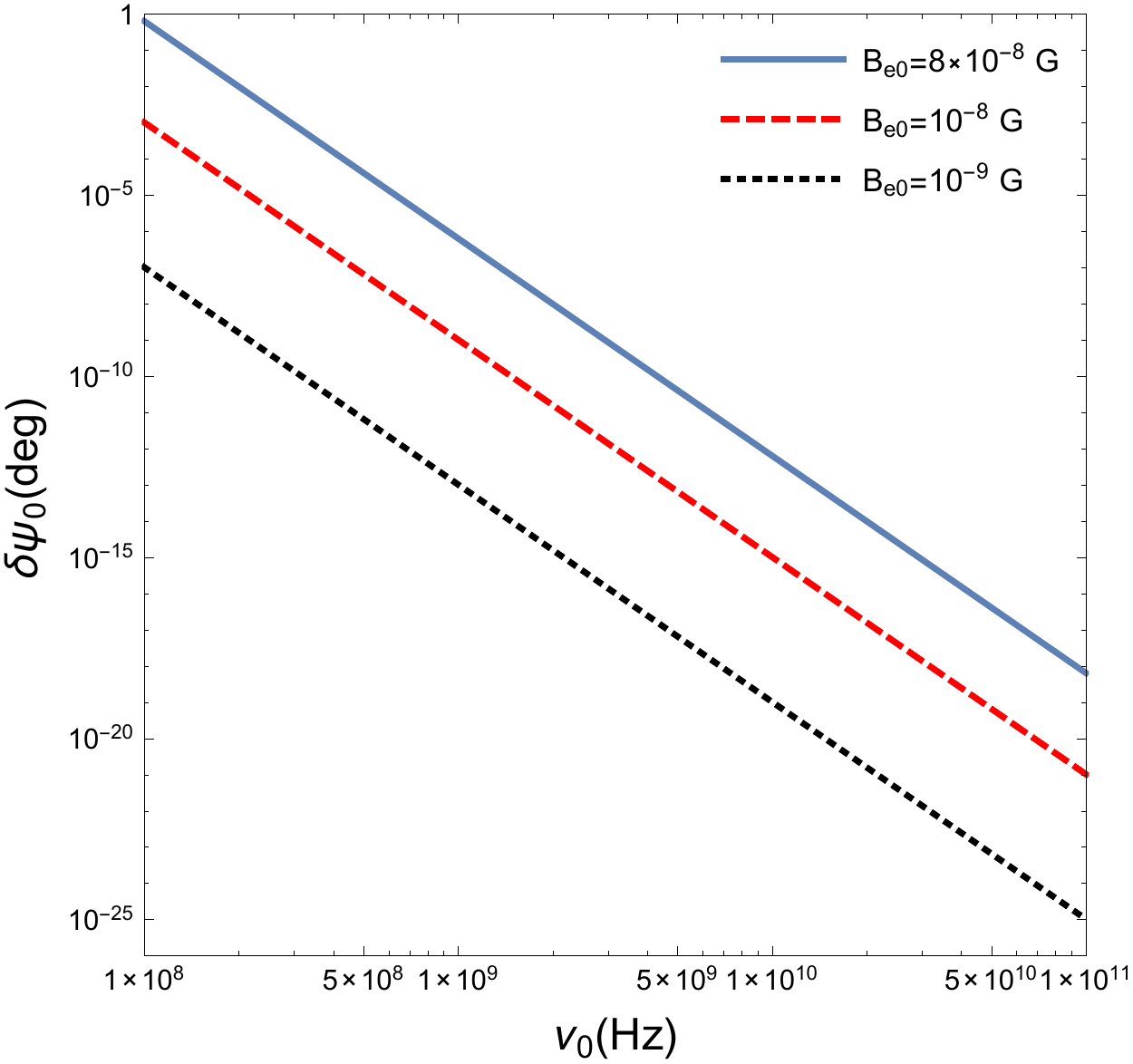}}\qquad
\subfloat[\label{fig:Fig16}]{\includegraphics[scale=0.62]{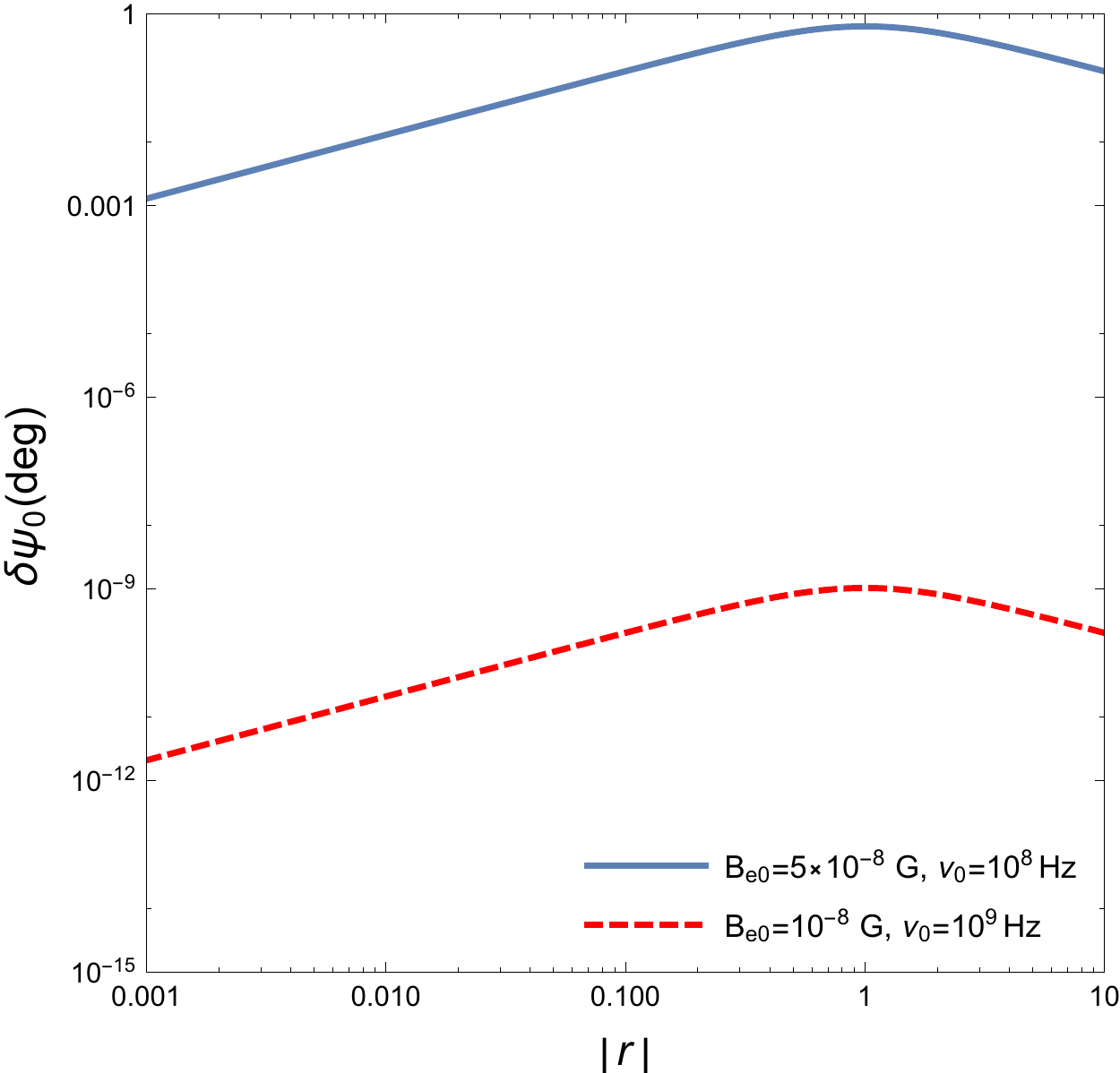}}}
\caption{(a) Logarithmic scale plots of the rotation angle of the CMB polarization plane at present $\delta\psi_0$ (the absolute value in degrees) due to the CM effect given in \eqref{abs-far-angle} as a function of the CMB frequency $\nu_0\in [10^8, 10^{11}]$ (Hz) for some specific values of the magnetic field amplitude $B_{e0}$, $\Theta=0$ and $|r|=1$ are shown. (b) Logarithmic scale plots of the rotation angle of the CMB polarization plane at present $\delta\psi_0$ (in degrees) given in \eqref{abs-far-angle} as a function $|r|\in[10^{-3}, 10]$ for some specific values of the magnetic field amplitudes $B_{e0}$ and frequencies $\nu_0$ for $\Theta=0$ are shown. In both (a) and (b) we have chosen those values of the parameters that satisfy the condition $|2 r\delta\psi(T)|<1$ that is always satisfied at $T=T_0$ for $T_0\leq T\leq T_i$ .}
\label{fig:Fig8a}
\end{figure*}

One important thing to observe about expressions \eqref{m22} and \eqref{m33} is that the second order of iteration terms must be smaller than unity because we are in the regime when $|\mathcal M_\text{F}(T)|<1, |\Delta\mathcal M(T)|<1$ and $|\mathcal M_\text{C}(T)|<1$. Consequently, one must choose the values of the parameters carefully in such way that such conditions are met. However, as far as the conditions in \eqref{F-C-M-cond} are satisfied we do not have to worry about what we said above. 
All told, let us consider first the case when $\mathcal M_\text{F}(T_0)=0$ which occurs when $\Theta=0$, namely absent Faraday effect. In this case we also have $\mathcal B=0$. From expressions \eqref{pol-angle-1}, \eqref{m22}-\eqref{m33} and dropping for simplicity the units we obtain
\begin{equation}\label{abs-far-angle}
\begin{gathered}
|2\delta\psi(T_0)/r| \simeq \left| \frac{- 9.88\times 10^{12} \mathcal C^2 \bar X_e^{-2} }{1+r^2} \right|
\end{gathered}
\end{equation}
where we neglected the sub-leading order term proportional to $\delta\psi^2$ on the left hand side in expression \eqref{pol-angle-1}. It is evident that since we are working under the condition $ |\Delta \mathcal M(T_0)|<1$ we have that the right hand side of \eqref{abs-far-angle} is less than one for any value of $r$. In Fig. \ref{fig:Fig8a} plots of the present epoch CMB rotation angle of the polarization plane $\delta\psi_0=\delta\psi(T_0)$ given by expression \eqref{abs-far-angle} are shown for various values of the parameters. We may note that substantial rotation of the polarization plane occurs only at low frequencies and for higher values of $B_{e0}$. On the other hand for higher values of the frequency and lower values of the magnetic field amplitude $\delta\psi_0$ is extremely small.

In case when $|\mathcal M_\text{F}(T)| \geq 1, |\Delta\mathcal M(T)|<1$ and $|\mathcal M_\text{C}(T)|<1$, we cannot use anymore the same expressions that we used above because of the fact that $|\mathcal M_\text{F}(T)|\geq 1$. In this case we can use the expressions for the Stokes parameters found in Sec. \ref{subsec:4.3}  up to first order in perturbation theory for arbitrary value of $\mathcal M_\text{F}(T)$. The expressions for $\tilde Q(T)$ and $\tilde U(T)$ are given by \eqref{tilde-S-sol} and read 
\begin{equation}\label{Stokes-2}
\tilde Q(T) = \cos[\mathcal M_\text{F}(T)] \tilde Q_i - \sin[\mathcal M_\text{F}(T)] \tilde U_i, \quad  \tilde U(T) = \sin[\mathcal M_\text{F}(T)] \tilde Q_i + \cos[\mathcal M_\text{F}(T)] \tilde U_i,
\end{equation}
where we may notice that there is no contribution to the linear polarization from the CM effect at the first order in perturbation theory. From \eqref{Stokes-2} we get for the rotation angle of the polarization plane 
\begin{equation}\label{FAR-ro}
\tan[2\psi(T)]=\frac{\sin[\mathcal M_\text{F}(T)] \tilde Q_i + \cos[\mathcal M_\text{F}(T)] \tilde U_i }{\cos[\mathcal M_\text{F}(T)] \tilde Q_i - \sin[\mathcal M_\text{F}(T)] \tilde U_i }=\frac{\tan[\mathcal M_\text{F}(T)] + \tan[2\psi(T_i)]}{1 - \tan[\mathcal M_\text{F}(T)] \tan[2\psi(T_i)]} = \tan[\mathcal M_\text{F}(T) + 2\psi(T_i)],
\end{equation}
where we can extract immediately $\delta\psi(T)=\mathcal M_\text{F}(T)/2$. It is important to note that, in principle, $\delta\psi(T)$ can have arbitrary values because we are not anymore under the hypothesis that $|\delta\psi(T)|\ll 1$ in \eqref{FAR-ro}. 
Thus, this result derived at the first oder in perturbation theory suggests that if\footnote{In the case when $0<|\mathcal M_\text{F}(T)|<1$, the situation is slightly more complicated and depending on the values of the parameters either the Faraday effect term or CM effect term, give the biggest contribution to $\delta\psi(T)$. In this case one can use directly the results obtained in expressions \eqref{pol-angle-1}, \eqref{m22}-\eqref{m33} in order to calculate $\delta\psi$.} $|\mathcal M_\text{F}(T)|\geq 1$ and as far as the stronger conditions for $|\Delta \mathcal M(T)|, |\mathcal M_\text{C}(T)| <1$ are satisfied, the rotation angle of the CMB polarization plane is given by $\delta\psi(T) \simeq \mathcal M_\text{F}(T)/2$ where the contribution of the CM effect is sub-leading. However, if we are interested to take simply the average value of $\delta\psi(T) \simeq \mathcal M_\text{F}(T)/2$ over $\Theta$ and $\Psi$, the Faraday effect gives on average zero contribution. So, in the case of simple average value, we also need to keep the second order terms of the CM effect in \eqref{FAR-ro}, which gives a small contribution to $\delta\psi$ but not zero. On the other hand, if we still insist on average values over the angles $\Theta$ and $\Phi$, the Faraday effect dominates when we consider the root mean square of $\delta\psi(T)$ as far as $|\mathcal M_\text{F}(T)| \gg |\Delta \mathcal M(T)|, |\mathcal M_\text{C}(T)| $. In this case we explicitly have
\begin{equation}\label{angle-pol}
\langle \delta\psi(T_0) \rangle_\text{rms} \simeq \frac{\langle \mathcal M_\text{F}(T_0) \rangle_\text{rms}}{2} = 2.36\times 10^{28} \left( \frac{\text{Hz}}{\nu_0} \right)^2 \left( \frac{B_{e0}}{\text{G}} \right) \quad (\text{rad}).
\end{equation}

\begin{figure*}[h!]
\centering
\mbox{
\subfloat[\label{fig:Fig17}]{\includegraphics[scale=0.62]{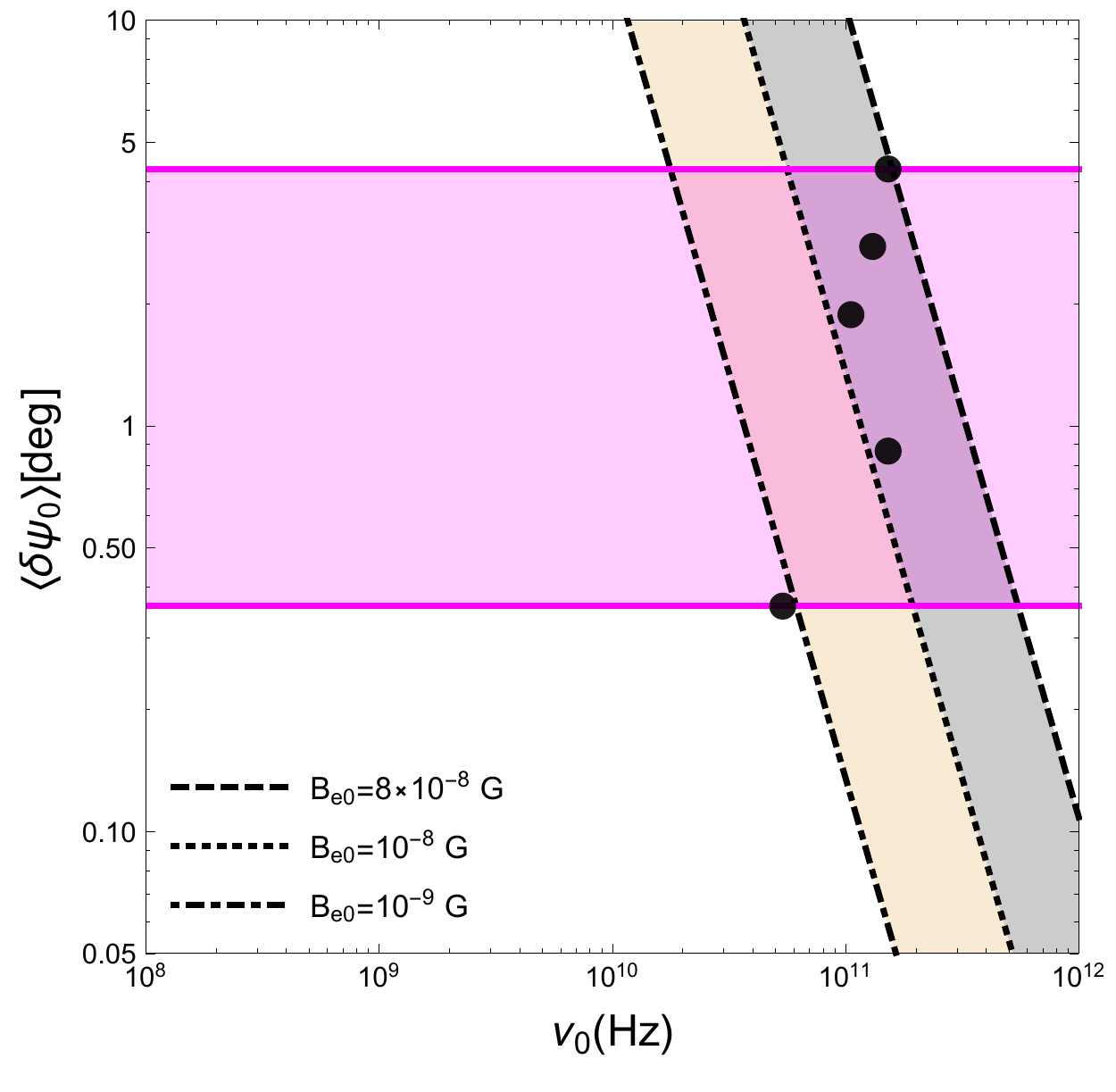}}
\qquad \subfloat[\label{fig:Fig18}]{\includegraphics[scale=0.61]{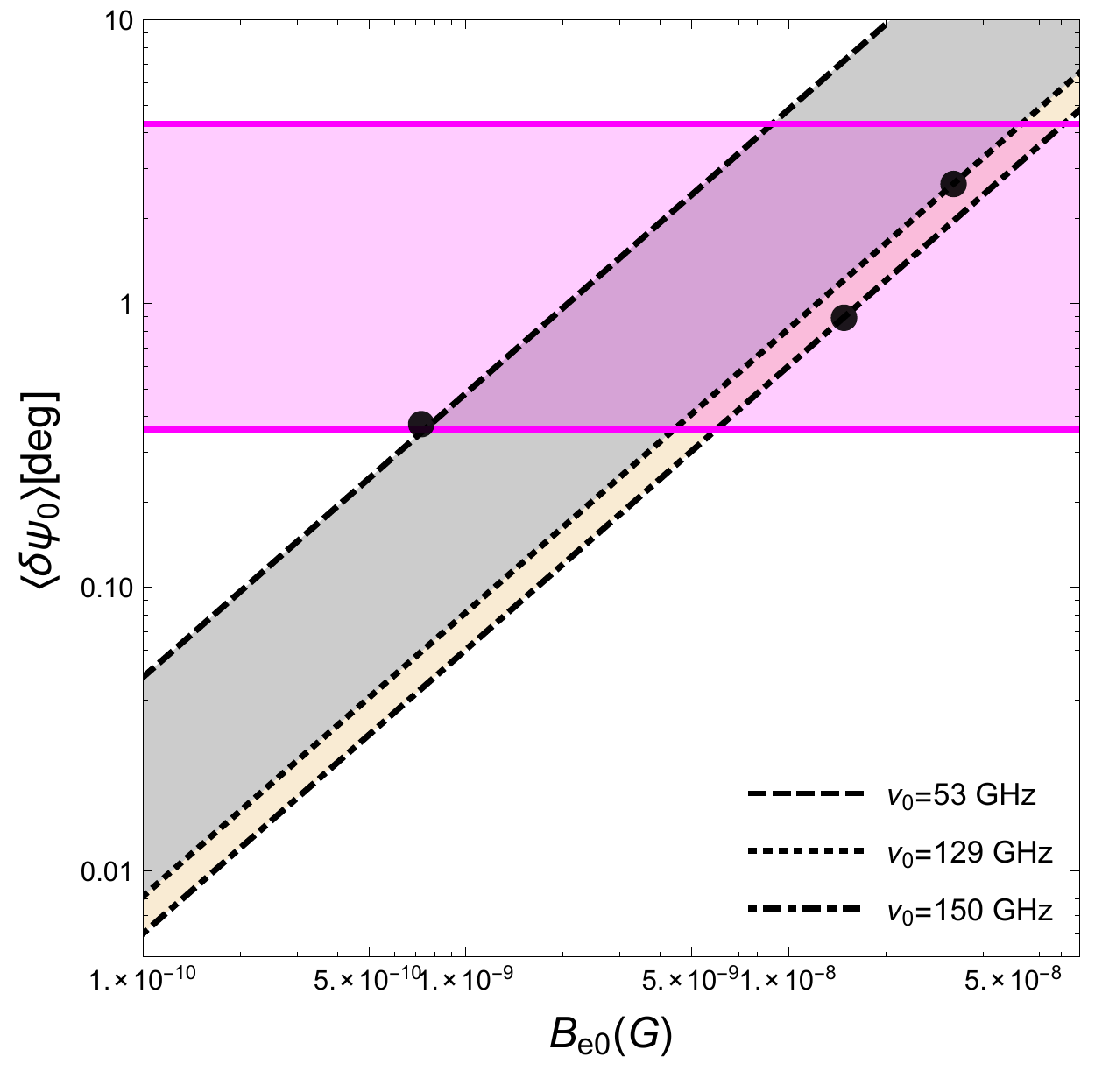}}}
\caption{(a) Logarithmic scale plots of the root mean square of the rotation angle of the CMB polarization $\langle \delta\psi_0 \rangle $ (in degrees) given in \eqref{angle-pol} as a function of the CMB frequency $\nu_0\in [10^8, 10^{12}]$ (Hz) due to the Faraday effect for some specific values of the magnetic field amplitude $B_{e0}$ are shown. (b) Logarithmic scale plots of the root mean square of the rotation angle of the CMB polarization plane $\delta\psi_0$ (in degrees) given in \eqref{angle-pol} as a function of the magnetic field amplitude $B_{e0}\in [10^{-10}, 8\times 10^{-8}]$ (G) due to the Faraday effect for some specific values of the frequency $\nu_{0}$ are shown. In both (a) and (b) the region in magenta colour between $0.36^\circ \leq \langle\delta\psi_0 \rangle \leq 4.3^\circ $ represents the region where experimentally do exist constraints on $\langle \delta\psi_0 \rangle$. In (a) the first black point from the top is the constraint found by BOOM3 experiment \cite{Pagano:2009kj} at the frequency $\nu_0=145$ GHz where $|\delta\psi_0| = 4.3^\circ$. The second point from the top is the constraint found by BICEP 1\cite{Kaufman:2013vbd} at the frequency $\nu_0=129$ GHz where $|\delta\psi_0| = 2.77^\circ$. The third and fourth points from the top are the constraint found by the QUaD collaboration \cite{Wu:2008qb} at the frequencies 100 GHz and 150 GHz where the constraints on $\delta\psi_0$ are respectively $|\delta\psi_0| = 1.89^\circ$ and $|\delta\psi_0| = 0.83^\circ$. The fifth point from the top is the constraint found by the WMAP9 collaboration \cite{Hinshaw:2012aka} at the frequency $\nu_0=53$ GHz where $|\delta\psi_0| = 0.36^\circ$, see also Ref. \cite{Galaverni:2014gca} for a general discussion on the constraints on $\delta\psi_0$. In (b) the first point from the bottom correspond to the WMPA9 collaboration \cite{Hinshaw:2012aka} constraint at $\nu_0=53$ GHz, the second point from the bottom correspond to the QUaD constraint \cite{Wu:2008qb} at $\nu_0=150$ GHz and the third one correspond to the BICEP 1\cite{Kaufman:2013vbd} at $\nu_0=129$ GHz.}
\label{fig:Fig9a}
\end{figure*}

In Fig. \ref{fig:Fig9a} plots of the root mean square of the CMB rotation angle of the polarization plane given in \eqref{angle-pol} due to the Faraday effect are shown. 
It is worth to stress that in Fig. \ref{fig:Fig9a} we have chosen the values of the parameters in such a way that the conditions $|\mathcal M_\text{C}(T_0)| <1$ and $|\Delta \mathcal M(T_0)| < 1$ are satisfied, see Fig. \ref{fig:Fig2}. 
We may observe from Fig. \ref{fig:Fig17} that most of experimental constraints\footnote{Here we consider for simplicity only the value of $|\delta\psi_0|$ without the error and assume that the magnetic field configuration is statistically the same in every direction in the sky. } on $|\delta\psi_0|$, represented by the black points, are within the grey region between the magnetic field values $10^{-8}$ G $\leq B_{e0} \leq 8\times 10^{-8}$ G. The only exception is the constraint found by WMPA9 where the magnetic field amplitude corresponding to $|\delta\psi_0|=0.36^\circ$ is by equation \eqref{angle-pol} $B_{e0}= 7.47\times 10^{-10}$ G. In Fig. \ref{fig:Fig18} plots of the root mean square $\langle \delta\psi_0 \rangle$ as a function of the CMB frequency are shown. Similarly to the Fig. \ref{fig:Fig17}, the black points represent the constraints on $|\delta\psi_0|$. 
For example, the QUaD constraint on $|\delta\psi_0|=0.83^\circ$ is consistent with $B_{e0}=1.38\times 10^{-8}$ G, while the BICEP 1 constraint on $|\delta\psi_0|=2.77^\circ$ is consistent with $B_{e0}=3.4 \times 10^{-8}$ G.

\section{Conclusions}
\label{sec:7}

In this work, we have studied the generation of the CMB circular polarization and rotation angle of the CMB polarization plane due to the CM effect in a large-scale cosmic magnetic field. We worked with the Stokes parameters and derived a system of differential equations for their evolution in an expanding universe. In the equations governing the evolution of the Stokes vector, we included all standard magneto-optic effects which manifest in a magnetized plasma which are the CM and Faraday effects. Then we looked for solutions of the equations of motion of the Stokes parameters in different regimes by using several perturbative approaches such as the regular perturbation theory and the Neumann series expansion. The equations of motion that we found in \eqref{Stokes-eq} are a generalization to the equations of motion found in Ref. \cite{Ejlli:2016avx} in the case of an arbitrary direction of $\bs B_{e}$ with respect to the photon direction of propagation. For an arbitrary direction of $\bs B_e$, the equations of motion \eqref{Stokes-eq} include two additional terms proportional to $M_\text{C}(T)$, which, would be absent in the particular case when the magnetic field $\bs B_e$ is in the same plane with the wave-vector $\bs k$. These two terms proportional to $M_\text{C}(T)$ make possible the mixing of $Q(T)$ and $V(T)$ Stokes parameters with each other.

The magnitude of the degree of circular polarization for the CM effect depends on several parameters where the most important ones are the CMB frequency $\nu_0$ and the magnetic field amplitude $B_{e}(\bs x, t_0)$. In addition, other parameters which play also an important role are the angles $\Theta$ and $\Phi$. Consequently, depending on the values of these parameters, in this work, we divided our analysis of the CM effect in three major regimes. In the regime where $|\mathcal M_\text{C,F}(T_0)|<1$ and $|\Delta \mathcal M(T_0)|<1$, the degree of circular polarization assumes the lowest values as shown in Figs. \ref{fig:Fig2a} and \ref{fig:Fig3a}, where at best its value reaches $P_C(T_0)\simeq 10^{-17}$. The reason for such low values of $P_C(T_0)$ stands from the fact that the condition $|\mathcal M_\text{F}(T_0)|<1$, drastically restricts the values of the parameters $\nu_0$ to very high frequencies and the values of $B_{e0}$ to very low ones.

In the case when the Faraday effect is completely absent, which happens when the direction of $\bs B_{e}$ is perpendicular to the direction of propagation of the CMB photons, we essentially have that $\mathcal M_\text{F}(T)=0$ and the generation of circular polarization is maximal. The absence of the Faraday effect for such specific configuration results in an enhancement of the generation of the CMB circular polarization. For such case, we have found in Sec. \ref{subsec:5.2} that the degree of circular polarization can reach values close to the CMB degree of linear polarization in the CMB low-frequency part of the spectrum. The maximum values of the degree of circular polarization are reached in the case when we concentrate at the frequency $\nu\simeq 10^{8}$ Hz, where depending on the magnetic field amplitude, the degree of circular polarization is in the range $1.19 \times 10^{-10}\lesssim P_C(T_0)\lesssim 7.65\times 10^{-7}$ for magnetic field values $10^{-10}$ G $\leq B_{e0}\leq 8\times 10^{-8}$ G. These results are plotted in Figs. \ref{fig:Fig4a} and \ref{fig:Fig5a} for different values of the parameters.

In the case when the Faraday effect is present and in particular when $|\mathcal M_\text{F}(T)|\geq 1$, the generation of the CMB circular polarizartion is strongly suppressed with respect to the case of absent Faraday effect where $\mathcal M_\text{F}(T)=0$. However, the generation of the CMB circular polarization in the case when $|\mathcal M_\text{F}(T)|\geq 1$ is usually much efficient than that in the case when $0<|\mathcal M_\text{F}(T)|< 1$ which we studied in Sec. \ref{subsec:5.1}. Even in the case $|\mathcal M_\text{F}(T)|\geq 1$ the degree of circular polarization depends on $B_{e0}$ and $\nu_0$, where in some specific range of these parameters, the degree of circular polarization scales with the frequency as $P_C(T_0) \propto \nu_0^{-1}$ and with the magnetic field amplitude as $P_C(T_0) \propto B_{e0}$, see for example the expression \eqref{first-apr}. As shown in Fig. \ref{fig:Fig12}, the degree of circular polarization can reach values in the range $ 10^{-14}\lesssim P_C(T_0)\lesssim 6\times 10^{-12}$ for magnetic field values $10^{-9}$ G $\leq B_{e0}\leq 8\times 10^{-8}$ G at the frequency $\nu_0\simeq 10^{8}$ Hz and $|Q_i|=10^{-6}$ and $|r|=1$. At the frequency $\nu_0\simeq 10^9$ Hz, the values of $P_C(T_0)$ decrease exactly by an order of magnitude since $P_C(T_0) \propto \nu_0^{-1}$ in the frequency range considered. On the other hand, $P_C(T_0) \propto |r|$, so, higher values of $|r|$ give higher values of $P_C(T_0)$ and vice-versa for smaller values of $|r|$.

Apart from generating circular polarization, the CM effect also generates linear polarization and this fact is evident in all expressions of the Stokes parameters that we found in Sec. \ref{sec:4}. In connection with linear polarization, in this work, we have studied the rotation angle of the CMB polarization plane due to the CM effect in the case when the Faraday effect is absent and in combination with the Faraday effect when the latter is present. In the case when it is present only the CM effect, the rotation angle is $\delta\psi(T_0) \propto \nu_0^{-6} B_{e0}^4$ and consequently, significant rotation of the polarization plane occurs in the low-frequency part of the CMB spectrum and for higher values of the magnetic field amplitude. We have found in Sec. \ref{sec:6} that at $\nu_0\simeq 10^8$ Hz, the rotation angle in units of degrees is in the range $10^{-3}\leq \delta\psi(T_0)\leq 1$ for magnetic field amplitude in the range $10^{-8}$ G $\leq B_{e0}\leq 8\times 10^{-8}$ G, $|r|=1$ and $|Q_i|=10^{-6}$, see Fig. \ref{fig:Fig8a}. For higher frequencies, $|\delta\psi(T_0)|$ acquires extremely smaller values which are uninteresting for any practical purpose.

In the case when the rotation angle $\delta\psi(T_0)$ is due to a combination of the CM and Faraday effects the situation slightly changes with respect to the case of absent Faraday effect. If we are interested in taking the average value of $\delta\psi(T_0)$, the Faraday effect gives null contribution while the CM effect gives in average almost the same contribution as it does in the case of absent Faraday effect. If we take the root mean square of $\delta\psi$, the Faraday effect usually dominates over the CM effect in the case when it is present unless the magnetic field is almost transverse with respect to the direction of propagation. One important aspect is that in case we take the root mean square of $\delta\psi(T_0)$, the Faraday effect generates significant rotation of the polarization plane depending on the CMB frequency and magnetic field amplitude. As shown in Fig. \ref{fig:Fig9a}, the Faraday effect generates substantial rotation of the polarization plane especially in the low-frequency part especially for $\nu_0\lesssim 10^{10}$ Hz. In the high-frequency part of the spectrum, namely for frequencies above 10 GHz, the rotation angle is still large depending on the magnetic field amplitude. An interesting fact is that most of the constraints on $\delta\psi(T_0)$ experimentally found correspond to magnetic field amplitudes in the range $10^{-8}$ G $\lesssim  B_{e0}\lesssim 8\times 10^{-8}$ G.

If we have to consider current limits on $\delta\psi(T_0)$ as a potential indicator of the existence of the large-scale cosmic magnetic field and consequently a non zero rotation angle of the polarization, these limits would allow us to make some predictions on the signal of the circular polarization due to the CM effect. Indeed, if we consider the hypothesis that the rotation angle is due to the Faraday effect only (root mean square value) and that most experimental constraints on $\delta\psi(T_0)$ would suggest a magnetic field with amplitude approximately $10^{-8}$ G $\lesssim  B_{e0}\lesssim 8\times 10^{-8}$ G, we would have that the signal of circular polarization for these values of $B_{e0}$ would be quite substantial. For these values of the magnetic field, in the case when the field is perpendicular to the photon direction of propagation, we would have a circular polarization signal at present in the range $3\times 10^{-8}$ K $\lesssim |V(T_0)|\lesssim 2 \times 10^{-6}$ K at $\nu_0\simeq 10^{8}$ Hz and a signal of $3\times 10^{-11}$ K $\lesssim |V(T_0)|\lesssim 2 \times 10^{-9}$ K at $\nu_0\simeq 10^9$ Hz, see Fig. \ref{fig:Fig4a}. In finding these values we used $|V(T_0)|=P_C(T_0)I(T_0)$ with $I(T_0)=T_0$ due to most CMB physics conventions. In the case when the magnetic field is not perpendicular, the signal of the circular polarization is reduced by many orders of magnitude and is in the range $2.7\times 10^{-14}$ K $\lesssim |V(T_0)|\lesssim 1.6\times 10^{-11}$ K at $\nu_0=10^8$ Hz and depending on the angles $\Theta$ and $\Phi$, see Fig. \ref{fig:Fig6a}.

Based on the arguments presented so far, it seems quite plausible that the CM effect is probably the most substantial effect in generating CMB circular polarization. However, the strongest signal of the circular polarization is located in the CMB frequency range $10^{8}$ Hz $\lesssim \nu_0\lesssim 10^9$ Hz. In the high-frequency range the signal of circular polarization due to the CM effect is much smaller than in the low-frequency part of the spectrum, but still, the signal is not negligible and can be comparable with the vacuum polarization circular polarization signal in a cosmic magnetic field. If we assume that there is not any major difficulty in arranging an experiment aiming to detect the circular polarization in the low-frequency part of the CMB spectrum, then it is quite logical to concentrate the attention in this frequency part of the spectrum where the signal is the strongest and more likely to be detected in a relatively short time.

\vspace{+1cm}

 \appendix
 \section{Second order solutions of equations of motion for arbitrary $|\mathcal M_\text{F}(T)|$ and $|\mathcal M_\text{C}(T)|<1, |\Delta \mathcal M(T)|<1$.}
 \label{app:1}

In Sec. \ref{subsec:4.3} we presented a solution of the equatons of motion for arbitrary $\mathcal M_\text{F}(T)$ and  
$|\mathcal M_\text{C}(T)|<1, |\Delta \mathcal M(T)|<1$ up to the first order in perturbation theory. However, there may be some specific cases when it is necessary to have the solutions up to the second order in perturbation theory. A typical example would be the case when the terms proportional to $\cos[\mathcal M_\text{F}(T)]$ and $\sin[\mathcal M_\text{F}(T)]$ in $Q(T)$ and $U(T)$ assume very small values for some values of $\mathcal M_\text{F}(T)$ and it might be necessary to see also the contribution of the CM effect which does appear at the second order in perturbation theory. So, let us follow exactly the same notations as in Sec. \ref{subsec:4.3} and extend expression \eqref{bar-S-sol} to the second order Neumann series
\begin{equation}\label{bar-S-sol-1}
\begin{gathered}
\bar S(T)= \left[ \bs I_{4\times 4} -\int_{T}^{T_i} dT^\prime \bar M(T^\prime) + \int_T^{T_i} \int_{T^\prime}^{T_i} dT^\prime dT^{\prime\prime} \bar M(T^\prime) \bar M(T^{\prime\prime}) - ...\right] \bar S(T_i) \\
\simeq 
\begin{pmatrix}
1 & 0 & 0 & 0\\
0 & 1- L_4^{ (0)}(T)L_4^{ (0)}(T^\prime)  & L_4^{ (0)}(T)K_4^{ (0)}(T^\prime) &  -L_4^{ (0)}(T) \\
0 & K_4^{ (0)}(T)L_4^{ (0)}(T^\prime)  & 1 - K_4^{ (0)}(T)K_4^{ (0)}(T^\prime) & K_4^{ (0)}(T)\\
0 & L_4^{ (0)}(T) & -K_4^{ (0)}(T) & 1 - L_4^{ (0)}(T)L_4^{ (0)}(T^\prime) - K_4^{ (0)}(T)K_4^{ (0)}(T^\prime)
\end{pmatrix} \bar S(T_i),
\end{gathered}
\end{equation}
where we have defined 
\begin{equation}\nonumber
\begin{gathered}
L_4^{ (0)}(T)L_4^{ (0)}(T^\prime) \equiv \int_T^{T_i} dT^\prime \left[ G_\text{C}(T^\prime) \cos[\mathcal M_\text{F}(T^\prime)] - \Delta G(T^\prime) \sin[\mathcal M_\text{F}(T^\prime)] \right] \\ \times \int_{T^\prime}^{T_i} dT^{\prime\prime} \left[ G_\text{C}(T^{\prime\prime}) \cos[\mathcal M_\text{F}(T^{\prime\prime})] - \Delta G(T^{\prime\prime}) \sin[\mathcal M_\text{F}(T^{\prime\prime})] \right] \\
K_4^{ (0)}(T)K_4^{ (0)}(T^\prime) \equiv \int_T^{T_i} dT^\prime \left[ G_\text{C}(T^\prime) \sin[\mathcal M_\text{F}(T^\prime)] + \Delta G(T^\prime) \cos[\mathcal M_\text{F}(T^\prime)] \right] \\ \times \int_{T^\prime}^{T_i} dT^{\prime\prime} \left[ G_\text{C}(T^{\prime\prime}) \sin[\mathcal M_\text{F}(T^{\prime\prime})] + \Delta G(T^{\prime\prime}) \cos[\mathcal M_\text{F}(T^{\prime\prime})] \right], \\ 
L_4^{ (0)}(T)K_4^{ (0)}(T^\prime) \equiv \int_T^{T_i} dT^\prime \left[ G_\text{C}(T^\prime) \cos[\mathcal M_\text{F}(T^\prime)] - \Delta G(T^\prime) \sin[\mathcal M_\text{F}(T^\prime)] \right] \\ \times \int_{T^\prime}^{T_i} dT^{\prime\prime} \left[ G_\text{C}(T^{\prime\prime}) \sin[\mathcal M_\text{F}(T^{\prime\prime})] + \Delta G(T^{\prime\prime}) \cos[\mathcal M_\text{F}(T^{\prime\prime})] \right], \\
 K_4^{ (0)}(T)L_4^{ (0)}(T^\prime) \equiv \int_T^{T_i} dT^\prime \left[ G_\text{C}(T^\prime) \sin[\mathcal M_\text{F}(T^\prime)] + \Delta G(T^\prime) \cos[\mathcal M_\text{F}(T^\prime)] \right] \\ \times \int_{T^\prime}^{T_i} dT^{\prime\prime} \left[ G_\text{C}(T^{\prime\prime}) \cos[\mathcal M_\text{F}(T^{\prime\prime})] - \Delta G(T^{\prime\prime}) \sin[\mathcal M_\text{F}(T^{\prime\prime})] \right].
 \end{gathered}
\end{equation}
Now by using \eqref{bar-S-sol-1}, we can return back to the components of $\tilde S(T)$ through the relation $\tilde S(T) \equiv \exp\left[-\int_T^{T_i} dT^\prime B_1(T^\prime) \right] \bar S(T)$, we get the following expressions for the components of $\tilde S(T)$
\begin{equation}\label{second-order-sol}
\begin{gathered}
\tilde I(T)=I_i,\\
\tilde Q(T) = \left[ \cos[\mathcal M_\text{F}(T)] \left( 1- L_4^{ (0)}(T)L_4^{ (0)}(T^\prime) \right) -  \sin[\mathcal M_\text{F}(T)] \left( K_4^{ (0)}(T) L_4^{ (0)}(T^\prime) \right) \right]Q_i + \\
\left[ \cos[\mathcal M_\text{F}(T)] L_4^{ (0)}(T) K_4^{ (0)}(T^\prime) -  \sin[\mathcal M_\text{F}(T)] \left( 1 - K_4^{ (0)}(T) K_4^{ (0)}(T^\prime) \right) \right] U_i - 
 \left[ \cos[\mathcal M_\text{F}(T)] L_4^{ (0)}(T) + \sin[\mathcal M_\text{F}(T)] K_4^{ (0)}(T) \right] V_i, \\
 \tilde U(T) = \left[ \cos[\mathcal M_\text{F}(T)]  \left( K_4^{ (0)}(T) L_4^{ (0)}(T^\prime) \right)   +  \sin[\mathcal M_\text{F}(T)] \left( 1- L_4^{ (0)}(T)L_4^{ (0)}(T^\prime) \right) \right] Q_i + \\
\left[ \cos[\mathcal M_\text{F}(T)]  \left( 1 - K_4^{ (0)}(T) K_4^{ (0)}(T^\prime) \right)  +  \sin[\mathcal M_\text{F}(T)] L_4^{ (0)}(T) K_4^{ (0)}(T^\prime) \right] U_i - 
 \left[ \sin[\mathcal M_\text{F}(T)] L_4^{ (0)}(T) - \cos[\mathcal M_\text{F}(T)] K_4^{ (0)}(T) \right] V_i, \\
 \tilde V(T) =  L_4^{ (0)}(T) Q_i - K_4^{ (0)}(T) U_i + \left[ 1 - L_4^{ (0)}(T)L_4^{ (0)}(T^\prime) -K_4^{ (0)}(T) K_4^{ (0)}(T^\prime)  \right] V_i.
   \end{gathered}
\end{equation}

\end{document}